\documentclass[twocolumn,times,tighten]{aastex63}
\usepackage{color}
\usepackage{enumitem}

\usepackage{hyperref}
\usepackage{verbatim}
\usepackage{amsmath,amstext,mathrsfs}
\usepackage[all]{hypcap} 
\usepackage{afterpage}    
\usepackage{marginnote}
\usepackage{makecell}
\usepackage{subfigure}
\usepackage{xfrac}
\defcitealias{LP00}{LP00}
\defcitealias{LP12}{LP12}
\defcitealias{GS95}{GS95}
\defcitealias{LV99}{LV99}
\defcitealias{KLP16}{KLP16}
\defcitealias{KLP17a}{KLP17}
\defcitealias{PaperI}{Paper I}

\newcommand*{\kh}{} % This is KH's color

\DeclareFontFamily{OMS}{oasy}{\skewchar\font48 }
\DeclareFontShape{OMS}{oasy}{m}{n}{%
         <-5.5> oasy5     <5.5-6.5> oasy6
      <6.5-7.5> oasy7     <7.5-8.5> oasy8
      <8.5-9.5> oasy9     <9.5->  oasy10
      }{}
\DeclareFontShape{OMS}{oasy}{b}{n}{%
       <-6> oabsy5
      <6-8> oabsy7
      <8->  oabsy10
      }{}
\DeclareSymbolFont{oasy}{OMS}{oasy}{m}{n}
\SetSymbolFont{oasy}{bold}{OMS}{oasy}{b}{n}

\DeclareMathSymbol{\smallleftarrow}     {\mathrel}{oasy}{"20}
\DeclareMathSymbol{\smallrightarrow}    {\mathrel}{oasy}{"21}
\DeclareMathSymbol{\smallleftrightarrow}{\mathrel}{oasy}{"24}

\defcitealias{PaperI}{PaperI}

\graphicspath{{./}{}}

\shorttitle{Gradient technique and magnetic field strength}
\shortauthors{Lazarian, Yuen \& Pogosyan}
\received{\today}

\submitjournal{ApJ}

\begin{document}

\title{Gradient Technique Theory: Tracing magnetic field and obtaining magnetic field strength }

\author[0000-0002-7336-6674]{Alex Lazarian}
\affiliation{Department of Astronomy, University of Wisconsin-Madison, USA}
\affil{Centro de Investigación en Astronomía, Universidad Bernardo O’Higgins, Santiago, General Gana 1760, 8370993, Chile}
\email{alazarian@facstaff.wisc.edu}

\author[0000-0003-1683-9153]{Ka Ho Yuen}
\affiliation{Theoretical Division, Los Alamos National Laboratory, Los Alamos, NM 87545, USA}
\email{kyuen@lanl.gov}

\author[0000-0002-7998-6823]{Dmitri Pogosyan}
\affiliation{Department of Physics, University of Alberta, Edmonton, Canada}
\affiliation{Korea Institute for Advanced Studies, Seoul, Republic of Korea}
\email{pogosyan@ualberta.ca}

\begin{abstract}
The gradient technique is a promising tool with theoretical foundations based on the fundamental properties of MHD turbulence and turbulent reconnection. Its various incarnations use spectroscopic, synchrotron, and intensity data to trace the magnetic field and measure the media magnetization in terms of Alfven Mach number. We provide an analytical theory of gradient measurements and quantify the effects of averaging gradients along the line of sight and over the plane of the sky.  We derive analytical expressions that relate the properties of gradient distribution with the Alfven Mach number $M_A$. We show that these measurements can be combined with measures of sonic Mach number or line broadening to obtain the magnetic field strength. The corresponding technique has advantages to Davis-Chandrasekhar-Fermi way of obtaining the magnetic field strength. 
\end{abstract}

\keywords{Interstellar magnetic fields (845); Interstellar medium (847); Interstellar dynamics (839);}

\section{Introduction}
\label{sec:intro}

The magnetic field is an important player in most astrophysical processes (e.g., \citealt{1956MNRAS.116..503M,2006ApJ...647..374G,2006ApJ...646.1043M,2007IAUS..243...31J}) and therefore it is really important to have ways of studying magnetic fields from observations. In terms of magnetic field tracing, the polarization arising from aligned dust (see \citealt{2015ARA&A..53..501A}) as well as synchrotron polarization (see e.g.\citealt{LY18b}) provide the main avenues of research (see \citealt{2010ApJ...710..853C}). The development of new ways of studying magnetic fields is essential for understanding the complex dynamics of multiphase interstellar media with ubiquitous magnetic fields. Therefore, several techniques, e.g.~\citeauthor{1981ApJ...243L..75G} effect (\citeyear{1981ApJ...243L..75G}), that uses polarization from molecular lines and the one employing ground state alignment (GSA, \citealt{YanL06, YanL07,YanL08}) have been put forward. 

In view of astrophysical flows with large Reynolds numbers, the magnetic fields are turbulent (see \citealt{2004ARA&A..42..211E,MO07,spectrum}). The evidence of turbulent magnetic field is coming from observations of density structure of the interstellar medium (e.g. \citealt{1995ApJ...443..209A, CL09,instability,2024arXiv240307976H}, see review from \cite{filament_review}) and velocity fluctuation studies \citep{1981MNRAS.194..809L,2004ApJ...615L..45H,2010ApJ...710..853C,2015ApJ...810...33C}, synchrotron polarization studies \citep{2020NatAs...4.1001Z,2024arXiv240517985P}. The magnetic field makes the turbulence anisotropic, and this was proposed as a way to study the magnetic field \cite{2002ASPC..276..182L,2005ApJ...631..320E,2008ApJ...680..420H,LP12}.

Recently, based on the theory of MHD turbulence and turbulent reconnection (\citealt{GS95}, henceforth GS95, \citealt{LV99}, henceforth LV99, see also a monograph by \citealt{BeresnyakLazarian+2019}) a Gradient Technique of magnetic field tracing was introduced. This technique employs the fact that in MHD turbulence, the gradients of the velocity and magnetic field amplitudes are directed perpendicular to the local direction of the magnetic field. The technique was developed to study velocity gradients \citep{GCL17, YL17a,YL17b, LY18a,LY18b} but later expanded to studies of gradients of different nature, most notable branches being Synchrotron Intensity Gradients (SIGs, \citealt{LYLC}), Synchrotron Polarization Gradients (SPGs, \citealt{LY18b}) as well as Intensity Gradients (IGs, \citealt{YL17b, IGvHRO,2023MNRAS.520.3857H}). In addition, Faraday Rotation Gradients (FRGs) and Dust Polarization Gradients (DPGs) were also discussed in \cite{LY18b}. In this paper, we provide a formal mathematical theory of these gradients and compare the properties of gradient tracing with that of polarisation measurements. In addition, in \cite{dispersion,YL20,curvature}, it was demonstrated that the properties of the gradient distribution can be successfully employed to measure the sonic and Alfv{\'e}n Mach number, key input parameters that determine the properties of magnetized turbulence. 

The subsequent papers successfully applied the gradient technique to studies of magnetic fields both in diffuse media and molecular clouds (see  \citealt{Hu_Snez21, Hu_mapping23}). The velocity gradients have also demonstrated their power in predicting the foreground polarization from galactic dust \citep{2020ApJ...888...96H,2020MNRAS.496.2868L}. However, progress was primarily achieved through empirical studies of gradients, which always have disadvantages. This motivates our present formal mathematical study of the properties of gradients in this paper. In particular, it is important to quantify how the ability of gradients to trace magnetic fields and measure magnetization changes with the averaging along the line of sight procedure and the change of the sub-block size. The corresponding theoretical advances can help to develop reliable magnetic field study procedures. However, the major focus for the practical applications of the theory in this paper is using gradients to obtain the strength of the magnetic field from the properties of the gradient distribution in the sub-blocks. In other words, while the study in \cite{dispersion} provides an empirical way to obtain $M_A$, which is of the order of the ratio of the magnetic field fluctuation to its mean value, the theory can provide analytical predictions for this dependence. When combined with line width broadening or other ways of evaluating the sonic Mach number of the turbulence, one gets a new way of obtaining magnetic field strength.

Traditionally, by the magnetic field is measured by combining the polarization and spectral linewidth.
The corresponding technique proposed in  \cite{1951PhRv...81..890D} and \cite{CF53} is termed Davis-Chandrasekhar-Fermi technique (henceforth DCF technique). It uses dust polarization to trace the direction othe f magnetic field and spectral linewidth to measure the velocity dispersion corresponding to the dispersion in magnetic field directions.  The DCF technique was later studied numerically and improved in several of subsequent publications (e.g. \citealt{2001ApJ...561..800H,2004Ap&SS.292..225C,2004ApJ...616L.111H,2006Sci...313..812G,Fal08}, \cite{CY16} ). These improvements did not change the essence of the technique. Based on the measurements of dispersions of both magnetic field and velocities, the DCF technique is affected by large-scale shear velocities and the variations of magnetic field directions not directly related to the Alfv{\'e}nic motions. 

An attempt to use the local measures of magnetic field and velocities to obtain the detailed distribution of magnetic field strength and remove the irrelevant large-scale contribution is made in \citeauthor{ch5} (\citeyear{ch5}, {\bf {\bf Paper I}}).
The approach, named the Differential Measure Analysis (DMA), employs the theory of MHD turbulence to obtain the magnetic field strength by combining the polarization data with spectroscopic data. 

This paper explores the possibility of obtaining the magnetic field strength without using polarization data while using gradients instead. This study capitalizes on our earlier studies of media magnetization using gradients. For instance, Velocity Channel Maps provide a way of obtaining both the maps of magnetic field directions using a new Velocity Channel Gradient (VChG) technique \citep{LY18a} and the Alfv{\'e}n Mach number \citep{dispersion}. The latter can be obtained on the scale of the data sub-block employed for calculating a single gradient vector within the gradient technique. Using this magnetization, obtaining the magnetic field strength using the same spectroscopic data is possible. Naturally, this approach is synergistic with the DCF and the DMA and has several advantages, which are discussed in this paper. 

The magnetization can be successfully obtained with velocity gradients and other types of gradients \citep{2020ApJ...905..130C}. This opens a way to obtain magnetic field strength with the different data sets that have never been employed for such purposes follows, {\kh \S \ref{sec:grad_intro} discusses the essence of the gradient technique. We discuss our numerical setup in \S \ref{sec:sim}. In \S \ref{sec:grad_obs}, we review the different observables used in the gradient technique.  In \S \ref{sec:Grad_cblock_theory} we formulate the analytical theory gradient statistics dispersions, and introduce a quantity $J_2$ mimicking circular dispersion in polarization statistics that is shown to be connected to the underlying turbulence structures.
In \S \ref{sec:J2_from_P}, we employ the actual MHD turbulence model to give an estimate of $J_2$ as functions of Alfv{\'e}nic Mach number. In \S \ref{sec:gradient_orientation} we discuss the measures that can be obtained from observations, and the corresponding statistical error estimates. In \S \ref{sec:J2_MHD} we present the numerical results verifying the analytical relations predicted in \S \ref{sec:gradient_orientation}. In \S \ref{sec:newopp}, we discuss the new opportunities based on the development of the analytical theory of gradients. In \S \ref{sec:discussion} we discuss the implication of our results, and in \S \ref{sec:conclusion} we conclude our paper.

\section{ Alfv{\'e}nic component of MHD turbulence and Gradient Technique}
\label{sec:grad_intro}

MHD turbulence in sub-Alfv{\'e}nic and trans-Alfv{\'e}nic regimes can be presented as a superposition of cascading Alfven, slow and fast modes \citep{GS95,2001ApJ...562..279L,2002PhRvL..88x5001C,CL03,2020PhRvX..10c1021M,2021ApJ...911...53H,2024NatAs.tmp...71Z,2024ApJ...962...89Z,2024arXiv240517985P}. The small-scale Alfv{\'e}nic fluctuations are marginally affected by their interaction with fast and slow modes \citep{2002PhRvL..88x5001C}. The Alfv{\'e}nic motions are most important for the Gradient Technique and, in what follows, we focus on the Alfv{\'e}nic turbulence.

\subsection{Scaling of eddies in local system of reference}

The Gradient Technique (GT) is a technique that employs the properties of MHD turbulence to study magnetic field from observational data. These properties include the anisotropy of velocity and magnetic field fluctuations in the MHD cascade as well as the alignment of these fluctuations with the {\it local} direction of magnetic field. The latter property is absolutely crucial for the ability of the GT to trace dynamically important magnetic field.\footnote{If magnetic field is not dynamically important, it can be advected by velocity motions and passively reflect the statistics of velocity fluctuations. Thus the velocity gradients can trace the magnetic field even in this case. However, we do not consider this situation in the present paper.} 

The well-accepted hydrodynamic Kolmogorov turbulence \citep{1941DoSSR..30..301K} can be understood using a mental picture of an eddy cascade. Traditionally, MHD turbulence was considered a wave-type phenomenon, similar to acoustic turbulence \citep{1963AZh....40..742I,1965PhFl....8.1385K} with Alfv{\'e}nic waves having isotropic distribution. This was shown to be an incorrect picture (\citealt{1981PhFl...24..825M}, \citetalias{GS95}). Note that further, we focus on the Alfv{\'e}nic turbulence cascade only, as this cascade determines the properties of gradients we deal with. The separation of this cascade from the other two MHD motions is possible because the back-reaction of slow and fast modes of the Alfv{\'e}nic cascade is marginal in the strong Alfv{\'e}nic turbulence regime (\citetalias{GS95}, \citealt{2002ApJ...564..291C}, \cite{leakage}, see also appendix of \cite{2024arXiv240517985P}). Thus, the scaling properties of Alfv{\'e}nic cascade stay unaltered in the compressible media in good agreement with numerical calculations \citep{CL03}.

The anisotropic distribution in \citetalias{GS95} picture corresponds to Alfv{\'e}nic wave vectors getting perpendicular to magnetic field. The corresponding motions are equivalent to eddies according to \citetalias{LV99}. 
The eddy motions were considered impossible in the presence of dynamically important magnetic field. However, the theory of turbulent reconnection (\citetalias{LV99}, see \citealt{2020PhPl...27a2305L} for a review) predicts that the magnetic field reconnection happens within one eddy turnover time. This enables unconstrained eddy motions in the direction perpendicular to the magnetic field. As a result, the Kolmogorov picture of turbulence is revived, but with the restriction that the eddy cascade is expected to involve only in the eddies with rotation axes aligned with the direction of magnetic field that surrounds the eddies.

Naturally, the magnetic field in question is those involved in the Alfv{\'e}nic motion, i.e. the field that percolates the eddies. This is the {\it local} direction of magnetic field, which is an important advance in understanding Alfv{\'e}nic cascade. In other words, due to turbulent reconnection,  the turbulent motions perpendicular to the magnetic field are not constrained by the back-reaction of magnetic field. This presents the favorable way of turbulent cascading with most energy concentrated in eddies rotating perpendicular to the {\it local} direction of the magnetic field.\footnote{A common misconception about the MHD theory is related to the fact that the concept of the local direction of the magnetic field is not a part of the original \citetalias{GS95} idea. This  concept of local magnetic field direction was introduced and proven numerically in \cite{CV00} and subsequent studies, e.g., \citep{MG01, 2002PhRvL..88x5001C}.}  Dealing exclusively with Alfv{\'e}nic turbulence simplifies our further treatment as velocities and magnetic field are related through the Alfv{\'e}nic relations. Further, in particular in \S 9 we discuss complicated situations involving non-Alfv{\'e}nic turbulence.

Accepting the natural constraint that eddies perform the motions that minimize the magnetic field bending, i.e. that due to fast reconnection they easily mix up magnetic field around them in the perpendicular direction, one can get the Kolmogorov-type condition for such perpendicular Alfv{\'e}nic motions 
\begin{equation}
v_l^2/t_{casc, l}=const,
\label{casc}
\end{equation}
where $v_l$ is the velocity of perpendicular to magnetic field eddies and the scale $l_{\bot}$ and $t_{casc,l}$ is the energy cascading time, i.e. time of the energy transfer from a scale $l_{\bot}$ scales to a smaller eddy. We use the symbol $\bot$ to explicitly reflect the fact that the cascade involves "perpendicular" eddies. For unconstrained energy transfer from scale to scale, the cascading time is similar to the Kolmogorov turbulence
\begin{equation}
t_{cas, l}\approx \frac{l_{\bot}}{v_l}
\label{t_casc}
\end{equation}
The eddies, however, can change their parallel scale $l_\|$. Indeed, the magnetic field mixing with the period $l_{\bot}/V_l$ induces a wave with a period $l_\|/V_A$, where $V_A$ is the Alfv{\'e}n velocity. The latter condition, i.e.
\begin{equation}
  l_{\bot}/V_l\approx  l_\|/V_A,
  \label{crit}
\end{equation}
was termed {\it critical balance} in the \citetalias{GS95} theory of MHD turbulence.\footnote{The critical balance assumes that in Eq. (\ref{crit}) it is $l_\|$ rather than the injection scale $L_{inj}$. If this is not true, the cascade is modified. The eddies cannot freely mix and the turbulence in the {\it weak MHD turbulence} regime with $l_\|=L_{inj}$ and perpendicular to magnetic field motions scaling $v_l\sim l^{1/2}$ (\citetalias{GS95}, \citealt{2000JPlPh..63..447G}. See also \citealt{2015MNRAS.449L..77M,2018PhRvX...8c1066M} for proofs of critical balance.). This does not change the picture for the gradient studies as the motions are still perpendicular to magnetic field.}
However, in the latter study the critical balance was derived for the corresponding parallel and perpendicular scales obtain in relation to the {\it mean} magnetic field rather than to the {\it local} magnetic field. The subsequent studies \citep{CV00,MG01, 2002PhRvL..88x5001C} testified that the critical balance and the relations between the $l_{\bot}$ and $l_{\|}$ that follow from combining Eqs. (\ref{casc},\ref{t_casc}) and (\ref{crit}), i.e. 
\begin{equation}
    l_{\|} \sim l_{\bot}^{2/3},
    \label{l_par}
\end{equation}
are valid only if $\|$ and $\bot$ scales are calculated in terms of the {\it local direction of magnetic field}. It follows naturally from our presentation of Alfv{\'e}nic cascade that the scaling of eddy velocities is Kolmogorov-type (see Eq. (\ref{casc}), i.e.
\begin{equation}
    v_l\sim l^{1/3}~{\rm and}~ b_l\sim l^{1/3}
    \label{kolm}
\end{equation}
where $b_l$ is the fluctuation of magnetic field.

Apart from Alfv{\'e}nic motions, slow and fast modes are present in MHD turbulence. The slow modes are present even in the limit of incompressible turbulence and they follow the scaling of Alfv{\'e}n modes (\citetalias{GS95},\citealt{2002PhRvL..88x5001C,CL03,2001ApJ...562..279L}). The fast modes are present in compressible MHD turbulence and it was demonstrated in \cite{2002PhRvL..88x5001C,CL03} that that provide an isotropic cascade similar to the acoustic turbulence. For magnetic field tracing with the GT, the properties of Alfv{\'e}n and slow modes are employed. The energy of those usually dominates the energy in the fast modes (see \citealt{2002PhRvL..88x5001C}). For subsonic turbulence slow modes dominate the formation of density fluctuations and therefore $\rho_l \sim l_{\bot}^{1/3}$ and are elongated with the axis ratio scaling given by Eq.~(\ref{l_par}). The situation is more complicated, however, for supersonic turbulence where the shocks distort significantly the density structure within MHD turbulence (see \citealt{2005ApJ...624L..93B,2007ApJ...658..423K}).

\subsection{Dependence on magnetization}

All the considerations above apply to sub-Alfv{\'e}nic, i.e., the magnetic field fluctuations $\delta B$ that are less than the underlying regular magnetic field $B_0$. For Alfv{\'e}nic motions, one can define the Mach number 
\begin{equation}
    M_A\equiv \frac{\delta B}{B_0} \approx \frac{V_{inj}}{V_A},
    \label{Alfven_Mach}
\end{equation}
where $V_{inj}$ is the injection velocity of the Alfv{\'e}nic cascade.
One identifies $M_A<1$ with sub-Alfv{\'e}nic turbulence and $M_A=1$ with the trans-Alfv{\'e}nic turbulence. If turbulence is super-Alfv{\'e}nic, i.e. $M_A>1$, at the injection scale $L_{inj}$ the motions are hydrodynamic-like down to the scale
\begin{equation}
    l_A=L_{inj} M_A^{-3},
\end{equation}
where the turbulence transfers to the magnetic field-dominated regime. All our arguments about trans-Alfv{\'e}nic turbulence are applicable provided that use $l_A$ as the effective injection scale of turbulence. 

For scales larger than $l_A$, the magnetic fields are moved by hydrodynamic flows and the gradients of velocities perpendicular to magnetic fields (see \citealt{LYLC}, Hu et al. 2024). For scales less than $l_A$, the velocity gradients are similar to the case of trans-Alfv{\'e}nic turbulence. For this type of turbulence and $M_A<1$ turbulence, the gradients are also perpendicular to magnetic field \citep[see][]{dispersion}. In the latter case, this is the consequence of the Alfv{\'e}nic turbulence anisotropy (\citetalias{GS95}; \citeauthor{KLP16} \citeyear{KLP16,KLP17a}, henceforth KLP16, KLP17). 

The properties of MHD turbulence depend on   
magnetization that is measured using the Alfv{\'e}n Mach number $M_A$ which is the ratio of the turbulent injection velocity $V_{inj}$
to the Alfv{\'e}n velocity $V_A$, i.e. $M_A=V_{inj}/V_A$. For sub-Alfv{\'e}nic turbulence (i.e. $M_A<1$) at scales smaller than
\begin{equation}
l_{tr} = L_{inj} M_A^2
\end{equation}
the velocity scaling is given by \citetalias{LV99}:
\begin{equation}
    v_l\approx V_{inj} \left(\frac{l_\bot}{L_{inj}}\right)^{1/3} M_A^{1/3},
    \label{eq:vscal}
\end{equation}
where $l_{\bot}$is measured in the local system of reference.  \footnote{At separations larger than $L_{inj} M_A^2$ as we mentioned earlier, the turbulence scaling follows the so-called weak cascade (See Santos-Lima et al. 2021), $v_l\approx V_{inj} (l_\bot/L_{inj})^{1/2}$. The gradient studies, however, usually employ small scales as weak turbulence often occupies only a small portion in MHD simulations.} For this case the eddies get more elongated as it is obvious from the exact expression for $M_A<1$ turbulence scaling at small separations \citepalias{LV99}:
\begin{equation}
    l_{\parallel} \sim L_{inj} \left(\frac{l_{\perp}}{L_{inj}}\right)^{2/3} M_A^{-4/3}, 
\label{eq:scaling_smallma}
\end{equation}
For larger magnetization,  $M_A$ gets smaller and the eddies get more and more elongated. The distribution of the direction of eddies also changes and the corresponding changes can be traced by studying the distribution of the gradient directions. For instance, the empirical study of gradients distribution in \cite{dispersion} revealed that the dispersion of velocity gradients measured within the sub-block employed for averaging changes as a power law of $M_A$. In this paper we relate this and other measures of velocity and magnetic field gradients with the analytical expectations.

\subsection{Properties of gradients and media magnetization}

Later in the paper we derive rigorously the properties of gradients according to the statistical theory of MHD turbulence. For readers' reference, it is advantageous to discuss the gradient properties that directly follow from the MHD turbulence scaling.

For Alfv{\'e}nic motions of the magnetic and velocity fluctuations are symmetric. The corresponding eddies in magnetized turbulence, as we discussed above, are elongated along the local direction of magnetic field and rotate about this direction.  The fact that both velocity and magnetic fluctuations exhibit Kolmogorov-type scaling given by Eq.~(\ref{kolm}) ensures that the gradients of the amplitudes of fluctuations are largest at the smallest scales resolved. For instance, for velocities $\nabla|v_l|\sim v_l/l_{\bot}\sim l_{\bot}^{-2/3}$. The fact that both velocity and magnetic field fluctuations take place and are elongated in respect to the {\it local} direction of magnetic field means that gradients of individual eddies reflect the direction of the magnetic field at the eddy location. Taken together this means that the Alfv{\'e}nic gradients are tracing magnetic fields similar to the way that aligned grains trace interstellar magnetic fields. 

If one measures gradients within a single magnetized eddy, a distribution of gradient directions is obtained. The maximal amplitude of gradients is in the direction perpendicular to the local magnetic field, while the minimal value corresponding to the direction parallel to the magnetic field. The width of the gradient distribution depends on the media magnetization. Indeed, the eddy elongation increases with the media magnetization, i.e. with the decrease of the Alfv{\'e}n Mach number $M_A$ (see Eq.~\ref{eq:scaling_smallma}). As a result, the distribution of the gradients measured for an eddy is getting narrower as $M_A$ decreases. In addition, the ratio of the maximal to minimal values of the measured gradients increases with $M_A$ decrease. 

The properties of gradients arising from Alfv{\'e}nic turbulence play the major role in the Gradient Technique (GT). Through sub-block averaging in \cite{YL17a} one identifies the direction of the maximal gradient amplitude, which is perpendicular to the magnetic field. Measuring the properties of the distribution of the gradients within the sub-block, e.g. the width of the distribution, provides a way to obtain $M_A$ \citep{dispersion}. 

As Alfv{\'e}n modes impose their scaling on the slow modes, the properties of gradients that correspond to slow modes are similar to those of Alfv{\'e}nic eddies. On the contrary, the fast modes have different type of scaling and the anisotropy that arises from those modes was shown in \cite{dispersion} to be orthogonal to that of Alfv{\'e}n and slow modes. In most cases, the fast mode contribution to gradients is subdominant\cite{LY18a,LY18b} and therefore we focus in this paper on exploring of the gradients arising from Alfv{\'e}n and slow modes.

The properties of density gradients vary with the sonic Mach number of the media $M_s$. At low $M_s$ the density can be passively moved by Alfv{\'e}n velocity fluctuations and therefore the gradients of density will be similar to the velocity gradients. At the same time, at higher $M_s$ shocks are being produced. Compressions arising from shocks tend to change the directions of gradients and complicate the interpretation of the density gradients in terms of underlying magnetic field\footnote{It was shown in \cite{2005ApJ...624L..93B}  that even in high $M_s$ media the low contrast density fluctuations follow the Alfv{\'e}nic \citetalias{GS95} scaling. }

The analytical study of the relation between the underlying properties of turbulence and the observables was started with the statistics of velocity available via channel maps (\citealt{LP00}, henceforth LP00). The analytical description of the relation of $M_A$ and the anisotropy of the observable statistics was obtained in \citeauthor{LP12} (\citeyear{LP12}, henceforth LP12) for the case of synchrotron emission fluctuations. Later studies in \citetalias{KLP16} and \citetalias{KLP17a} provided the relation of the anisotropies in spectroscopic channel maps and velocity centroids with $M_A$. These theoretical studies provide the foundations for obtaining the analytical expressions relating $M_A$ with the statistics of the gradients of the observable fluctuations. Importantly,  they relate both the isotropy degree and quadropole-to-monopole ratio to the Alfv{\'e}nic Mach number based on the analytical calculations of \citetalias{LP00}, suggesting that the Alfv{\'e}nic Mach number can be obtained from observations by studying the statistics of velocity observables.

\section{Numerical simulations}
\label{sec:sim}
The numerical simulations are originated from three numerical codes: ZEUS-MP/3D \citep{2006ApJS..165..188H,2010ApJS..187..119C}, Athena++ \citep{2020ApJS..249....4S}, and an incompressible MHD spectral code "MHDFlows" \citep{2022zndo...8242702H}. ZEUS-MP/3D approach uses the 2nd order staggered-grid discretization while Athena++ has multiple options, which we are using 3rd order PPM in space and VL2 in time. The difference of discretization method decides how the numerical viscosity scales as a function of grid size $\Delta x$. We summarize the simulations in Table~\ref{tab:sim} in Appendix~\ref{app:sim}. There, incompressible run was done with MHDFlows, models labeled $Ms??Ma??$ use Athena++, and the rest, which are the majority of our simulations, is based on ZEUS-3D.

Our data cubes are three-dimensional, triply periodic, isothermal MHD simulations with periodic force driving via {\it direct spectral injection} unless specified. Simulations have the injection scale which is one-half of the size of the cube, so that we only have injected eddies at scales $L_{inj}/L_{box}=1/2$. The driving force is taken to be solenoidal, giving incompressible driving.

The parameters that are employed are $V_{inj}$ is the injection velocity, $V_A$ and $V_s$ that are the Alfv\'en and sonic velocities respectively. The parameters give two dimensionless combinaations,  namely, the Alfv\'en $M_A=V_{inj}/V_A$ and sonic  $M_s=V_{inj}/V_s$ Mach numbers.
The results of isothermal MHD simulations can be easily transformed to arbitrary units keeping dimensionless parameters $M_A,M_s$ fixed. The chosen $M_A$ and $M_s$ are listed in Table \ref{tab:sim}. For the case of $M_A<M_s$, the simulations correspond to thermal pressure smaller than the magnetic pressure, i.e. plasma with $\beta/2=V_s^2/V_A^2<1$. Similarly, the case 
$M_A>M_s$ corresponds to plasma where the {thermal} pressure dominates, i.e. $\beta/2>1$.  The simulations are referred in Table \ref{tab:sim} by their model name. The ranges of $M_s, M_A, \beta$ encompass possible scenarios of astrophysical turbulence.  

Simulations exhibit the self-similar turbulent cascade which stops at the dissipation scale that is determined by numerical viscosity and resistivity. The existence of this dissipation scale $d_{diss}$ corresponds to the rapid decrease of turbulent velocities. Our theoretical results should only be referred to turbulence within the inertial range and we will defer our discussion on viscous turbulence in later papers.

\section{Examples of observational data to apply gradients}
\label{sec:grad_obs}

{\bf Spectroscopic data: Velocity Gradients}
Spectroscopic data of line emission (e.g. 21~cm HI line) in frequency direction reflects, through the Doppler effect,  
the line-of-sight velocities of the emitters.  Speaking about velocity gradients we assume the use of the spectroscopic data.
If we denote $\rho ({\bf X}, v)$ the density of emitters in Position-Position-Velocity space
as a function of Plane-Of-Sky (POS) direction ${\bf X}=(x,y)$ and the velocity $v$, the simplest quantity that can be analyzed is the intensity within the velocity channel centered at $v_c$ :
\begin{equation}
    I({\bf X},v_c)\sim \int_{v_c-\delta v/2}^{v_c+\delta v/2} \rho ({\bf X}, v) dv, 
    \label{chan}
\end{equation}
where $\delta v$ is the width of the channel. The instrument's spectral resolution determines the narrowest available width, but one can also consider synthesized channels and vary the
width above the resolution limit. Thermal broadening effectively adds to the width of a channel,
blurring the effect of turbulent velocities (see \citetalias{LP00} for the exact criterion, and \citealt{YL20} for the case when the thermal width is larger than turbulent velocity). One should
note that it is the thermal velocity 
$\beta_{\text{T}}^{1/2}\equiv \sqrt{k_BT/m}$, $m$ being the mass of atoms, $T$ being the temperature and $k_B$ being the Boltzmann constant, 
rather than temperature \textit{per se} that matters. Thus, lines from heavier species exhibit less thermal broadening than the ones from lighter species
at the same temperature.

When the effective channel width exceeds the typical turbulent velocities, the channel maps are
dominated by column density fluctuations. The effect of velocities is enhanced by using velocity centroids, which can be generalized to {\it reduced centroids} \citep{LY18a} defined as
\begin{equation}
C_{red}(\mathbf{X}) \propto \int_{\delta v} dv v \rho(\mathbf{X},v),
%\label{subcentroid}
\end{equation}
where the choice of $\delta v$ can be arbitrary.  Conventional (here, un-normalized, following \citealt{2003ApJ...592L..37L}) 
centroids correspond to $\delta v$ encompassing the full range of the velocities in the line.
In this case centroid correlation statistics are not sensitive to thermal broadening \citepalias{KLP17a} (See Appendix of \cite{VDA} for discussions).

If the spectral resolution of the instrument is higher than the thermal velocity, one can get additional information from {\it sub-thermal centroids} defined as
\begin{equation}
C_{subT}(\mathbf{X}) \propto \int_{\beta_T^{1/2}} dv v \rho(\mathbf{X},v) .
\label{subcentroid}
\end{equation}
These types of centroids can deliver the information on turbulent velocities at small scales that are hidden by the thermal broadening. 

From the spectroscopic data, both channel maps and centroids one can construct the measures that are related to the spacial changes of the velocity value in the neighboring points, i.e. with "velocity gradients". In a series of papers \citep{GCL17,YL17a,YL17b,LYLC,LY18a} we introduced velocity gradients as a way of tracing magnetic field and prove them as an efficient tracer. 

We stress again that the velocity gradients, as other types of gradients that we consider further trace the  "local magnetic field around the eddies". It is also important the gradients of velocity amplitude scale as $v_l/l_{\bot}\sim l^{-2/3}$, meaning that the maximal gradients are produced by the smallest resolved eddies. Due to this scaling, regular shearing motions do not affect the velocity gradient measurements.
 
Velocity gradients present a possibility of 3D studies if different molecular lines are used. Indeed, different molecules are produced and survive at different depth in molecular clouds. This opens a possibility of studying magnetic fields in molecular clouds at different depths \citep{YL17b,velac} In addition, galactic rotation provides a way to probe magnetic field at different distances from the observer \citep{GL18}. Note, that due to the galactic rotation, velocity gradients can sample magnetic fields in many more clouds in the galactic disc compared to far infrared polarimetry. For the latter, the confusion of emission from different clouds along the line is sight is detrimental. 

As we discussed earlier, the gradients in practice should be calculated over a sub-block as proposed in \cite{YL17a}. This provides the necessary averaging that is required to reliably determine the gradient direction. The distribution of gradients over the sub-block was identified in \cite{dispersion} as a source of information about media magnetization. 

Gradients of velocity-related quantities (e.g. constant density channels, velocity centroids, velocity channels processed by \cite{VDA}) generally have better tracing power on magnetic field than that of density-related quantities (e.g. column density maps). Therefore, both for the successful magnetic field tracing and studying media magnetization, it is advantageous to separate the PPV fluctuations that arise from velocity caustics, i.e. due to the velocity crowding, from the PPV fluctuations arising from the density fluctuations. This problem can be dealt with using the Velocity Decomposition Algorithm (VDA) introduced in \cite{VDA}.

In the presence of gravity, it was reported in \cite{YL17b} that the direction of velocity gradients measured by velocity centroids flips 90 degrees becoming parallel to magnetic field. The same effect was also demonstrated in \cite{LY18a} for velocity channel maps. A procedure to account for this effect and successfully trace magnetic field in the star formation regions was introduced in \cite{velac}. However, the application of the VDA shows that the observed flip of the gradients obtained with centroids and velocity channel maps is related to the density effects. The actual 90 degree flip of the velocity gradients takes place on scales that are smaller than those resolved in observations. All in all, even in the presence of gravity, the velocity gradients can be successfully used to trace magnetic field.

{\bf Synchrotron Intensities}.
Similar to the gradients of velocities, the gradients of magnetic field can trace the magnetic field direction. The gradients of synchrotron intensities and synchrotron polarization boil down to magnetic field gradients. 

The synchrotron emission depends both on the distribution of relativistic electrons 
\begin{equation}
N_e({\cal E})d{\cal E}\sim {\cal E}^{\alpha} d{\cal E},
\end{equation}
 with intensity of the synchrotron emission given by the line-of-sight integral
\begin{equation}
I_{sync}({\bf X}) \propto \int dz B_{\perp}^\gamma({\bf X},z)
\end{equation}
where $B_{\perp} = \sqrt{B_x^2 + B_y^2} $ is the magnitude of the magnetic field component perpendicular to the LOS $z$. In general, $\gamma=0.5(\alpha+1)$ is
a fractional power, a problem successfully addressed in \citetalias{LP12}.

Other types of gradients include gradients of Zeeman measure and  Faraday rotation gradients \cite{dispersion}. The synergy of different types of gradients opens new avenues for studying magnetic fields in multi-phase ISM. For instance, Faraday rotation gradients map the magnetic fields in dense ionized gas, while Synchrotron Intensity gradients have more tenuous halo media along the same lines of sight. The statistics of the aforementioned types of gradients are similar. Thus, we will discuss only gradients of centroids and gradients of synchrotron intensities.    

\section{Gradient Theory: Plane of Sky Observables}
\label{sec:Grad_cblock_theory}

In this section we develop a theoretical description of the simplest statistics of 2D gradients of projected observables on the sky, namely, the covariance tensor of gradient components
$\sigma_{ij}$, in relation to the properties of 3D turbulent volume from which the observed signal originates. In particular,
we'll show that eigendirections of $\sigma_{ij}$ matrix give a localized direction of the projected magnetic field, while
the level of anisotropy in $\sigma_{ij}$ can be used to measure the Alfv\'enic Mach number $M_A$. 

\subsection{Statistics of 2D gradients}

Let us discuss the mathematical foundations of the gradient methods in application to study of the direction of the magnetic field. We also refer to Appendix~\ref{app:grad_theory} for details.

Observing emission from the turbulent media, one constructs the sky maps of different observables that describe the emission. First of all, this is the intensity of the emission in PPV space, where the position-position $\mathbf{X}$ POS 2D vector and the velocity $v$ coordinate determine position of individual intensities $\rho(\mathbf{X},v)$, and the related integrated quantities, such as the total intensity  and velocity centroids which for optically thin lines are $I_c(\mathbf{X}) \propto \int dv I(\mathbf{X},v)$ and $C(\mathbf{X}) \propto \int dv v \rho(\mathbf{X},v) $ respectively (See \S \ref{sec:grad_obs}). The maps represent random fluctuating fields. For general exposition, let us denote this 2D signal map as $\Phi(\mathbf{X})$.

The main local statistical measure of the gradient of a 2D field $\Phi(\mathbf{X})$ is the gradient covariance tensor 
\begin{equation}
\label{eq:gradcov}
\sigma_{i j} \equiv \left\langle \nabla_i \Phi(\mathbf{X}) \nabla_j \Phi(\mathbf{X}) \right\rangle
= \nabla_i \nabla_j D(\mathbf{R})|_{\mathbf{R} \to 0}
\end{equation}
where ${\bf R}$ is the separation between the point at the POS plane. Notice that $\sigma_{ij}$ is the zero separation limit of the second derivatives
of the field structure function 
\begin{equation}
D(\mathbf{R})\equiv \frac{1}{2} \left\langle\left( \Phi(\mathbf{X+R}) - \Phi(\mathbf{X})\right)^2\right\rangle ~.
\end{equation}

For a statistically isotropic field, the covariance of the gradients is isotropic, $\sigma_{\nabla_i \nabla_j} = \frac{1}{2} \delta_{ij} \Delta D(R) | _{R\to 0}$. In the presence of the magnetic field, $D(\mathbf{R})$ of the signal becomes orientation dependent, depending on the angle between $\mathbf{R}$ and the projected direction of the magnetic field. This anisotropy is retained in the limit $\mathbf{R} \to 0$ and results in non-vanishing traceless part of the gradient covariance tensor
\begin{eqnarray}
&&\sigma_{ij} - \frac{1}{2} \delta_{ij} \sum_{l=1,2} \sigma_{l l } = \\
&&= 
 \frac{1}{2}\left(
\begin{array}{cc}
\left(\nabla_x^2  - \nabla_y^2\right)  D(\mathbf{R})
& 2 \nabla_x \nabla_y  D(\mathbf{R}) \\
2 \nabla_x \nabla_y  D(\mathbf{R}) &
-\left(\nabla_x^2 - \nabla_y^2\right)  D(\mathbf{R})\\
\end{array}
\right)_{\mathbf{R} \to 0} 
\ne 0
\nonumber
\label{eq:dd2}
\end{eqnarray}
As shown in \cite{2020MNRAS.496.2868L} and Appendix~\ref{sub:techinique}, the eigen-directions of this covariance tensor coincide with the direction of local statistical anisotropy of the motions which in MHD turbulence is expected to be given by the local projected direction $\theta_H$ of the magnetic field. 
\begin{align}
(\nabla_x^2 - \nabla_y^2) D(\mathbf{R}) & = 
I_2 \cos 2 \theta_H  \\
2 \nabla_x \nabla_y D(\mathbf{R}) & = I_2 \sin 2 \theta_H
\label{eq:lu_direction}
\end{align}
where the magnitude of the anisotropic part $I_2$ in relation to the isotropic trace 
$I_0 = (\nabla_x^2 + \nabla_y^2) D(\mathbf{R})$ is given by the ratio of the
quadrupole to monopole moments of the structure function
\begin{equation}
\frac{I_2}{I_0} = 2  \lim_{R \to 0} \frac{D_2(R)}{D_0(R)} 
\label{eq:I2I0}
\end{equation}
with $D_m(R)=\frac{1}{2\pi}\int d\phi D(R,\phi) e^{-im\phi}$.
Thus, the information about magnetic field direction can be recovered by measuring the distributions of the gradients over local patches of
2D sky and finding the eigendirections of $\sigma_{ij}$ in each of them.

The square of $I_2/I_0$ in Eq.~(\ref{eq:I2I0}) is related to the ratio of the two rotational invariants of the covariance tensor, that we will denote $J_2$,
\begin{equation}
J_2 = \frac{ (\sigma_{xx}-\sigma_{yy})^2 + 4 \sigma_{xy}^2}{(\sigma_{xx} + \sigma_{yy})^2} = 
4 \left( \lim_{R \to 0} \frac{D_2(R)}{D_0(R)} \right)^2.
\label{J2}
\end{equation}
The parameter $J_2$ determines the level of dispersion of the directions of the gradients relative
to the eigen-directions of $\sigma_{ij}$. This is the quantity that carries information about the level of anisotropy of turbulence processes, and, therefore, on the parameters of the turbulence.

A similar to $J_2$ parameter is the ratio of the quadrupole $D_2$ to monopole $D_0$ moments of the structure function
\begin{equation}
  F=\frac{D_2(R)}{D_0(R)} 
  \label{F}
\end{equation}
This ratio has appeared in \citetalias{LP12} as a convenient measure of turbulence anisotropy.  It was analytically examined for Alfv{\'e}n, slow, fast modes of MHD turbulence as well for some mixtures of MHD modes for both the synchrotron intensity fluctuations and the measures of fluctuations of intensities within spectroscopic data cube statistics \citepalias{LP00,KLP16,KLP17a}. These studies demonstrated that both $M_A$ and the composition of turbulent modes can be studied using the  ratio $F$. 

The difference with gradients is that the
quadrupole to monopole ratio is evaluated at vanishing lag, i.e. $R \rightarrow 0$.
We note that for both Alfv{\'e}n and slow MHD modes $D_2(R)$ and $D_0(R)$ have the same power-law dependence on the separation $R$  (see Eq. (\ref{kolm}) and \cite{CL03}). Therefore, if these modes dominate, $J_2$ dependence on $R$ is very weak.\footnote{The exact scaling of fast modes is still the issue of debates. It is suggestive that it follows the acoustic cascade $E(k)\sim k^{3/2}$ for low $M_s$ while the spectrum gets steeper, i.e. $\sim k^{-2}$ as $M_s$ increases and shocks are formed \citep{CL03,2010ApJ...720..742K,2020PhRvX..10c1021M}.} This makes it possible to relate $J_2$ to factor $F$ at finite lag
\begin{equation}
  J_2(\text{Alfven, Slow}) {\approx} 4 F^2 
\end{equation}
This equation relates earlier approach of \citetalias{LP12} based on the structure function multipole ratio and the new one based on measuring the gradients.

\subsection{Spatial localization of the gradient information}

Strictly local statistical measures, such as $\sigma_{ij}$ at a given sky position, are not accessible in observations, since one needs to average over some region of the sky to simulate the ensemble averaging. In case of a statistically homogeneous measure, the ergodicity principle shows that increasing the volume for averaging allows for ever better determination of statistical values.  However, in our case the direction of the magnetic field varies on the sky, therefore statistics of the gradients is not homogeneous. In this situation, determining gradient covariance over the finite patch gives the direction of the projected magnetic field locally averaged over the patch. When choosing the patch size, one needs to balance the loss of resolution and information when increasing the patch size versus the increase of errors due to insufficient statistics if the patch is too small. The choice of the optimal patch size becomes an important parameter of the analysis.

The situation can be modelled by a patch-background split, where within a local patch of scale $L_p$ centered at position $\mathbf{X}_p$, termed {\bf sub-block} in \cite{YL17a}, the signal can be decomposed into short-wave Fourier modes with $ K > K_p \approx 1/L_p$ that have approximately homogeneous but anisotropic distribution with the power spectrum $P(\mathbf{K},\hat\Lambda(\mathbf{X}_p))$, while the direction of the magnetic field $\hat\Lambda(\mathbf{X}_p)$ has long-wave variations between the patches. Within a patch the formalism of the homogeneous axis-symmetric turbulence can be applied (\citealt{PhysRevE.56.2875}, \citetalias{LP12}) to the gradients, giving the local direction of the magnetic field via the eigen-direction of the gradient covariance tensor $\theta_H(\mathbf{X}_p)$ and the now local parameter $\tilde J_2(\mathbf{X}_p)$ expressed via the short-wave part of the power spectrum as
\begin{equation}
J_2(\mathbf{X}_p) = \left(\frac{ \int_{K_p}^\infty K dK K^2  P_2(K,\hat\Lambda_{\mathbf{X}_p}) W_b(K/K_b)}
{\int_{K_p}^\infty K dK K^2 P_0(K,\hat\Lambda_{\mathbf{X}_p}) W_b(K/K_b)}\right)^2
\end{equation}
where $P_2(K)$ and $P_0(K)$ are the quadrupole and the monopole moments of the power spectrum determined within the patch at position $\mathbf{X}_p$, and $\hat\Lambda_{\mathbf{X}_p}$ is the unit vector in the direction of the mean magnetic field within the sub-block.
The range of scales $K_p < K < K_b$ is bounded by the patch size, $K_p \approx 1/L_p$,  and by the beam of the telescope or low-pass post-processing 
window, $W_b$, that smooths out fluctuations at very short modes $ K > K_b$.  
Importantly, due to extra $K^2$ weight, the gradient covariance tensor is primarily determined by the shortest wavelengths $\sim 1/K_b$, which justifies
the patch-background split.

One should stress that low pass smoothing of the signal by $W_b$ is critical for practical applications
of the gradient techniques. It assures that the limit in Eq.~(\ref{J2}) is convergent and, correspondingly, that
the measurements of the gradients are stable. In addition, the low-pass filtering should be isotropic
as not to modify $J_2$ by introducing artificial anisotropy.  Observations inevitably have low-pass filtering due to finite beam  which usually does not resolve turbulent motions on scales less than the turbulence dissipation scale. Numerical simulations, one the other hand, also have intrinsic low-pass filtering corresponding to the numerical dissipation scale $K_d$, where the spectrum $P(\mathbf{K},\hat\Lambda(\mathbf{X}_p))$ is exponentially suppressed for $K>K_d$. Despite that, it is advantageous in the analysis to deconvolve, as much as possible, this ``instrumental'' effects and 
smooth the data with a synthetic, preferably well behaved Gaussian low-pass filter, to maintain the maximum control
over the scale and isotropy of the filtering procedure.

Due to different averaging along the line of sight in different patches, two dimensional projected spectra and, therefore $J_2(\mathbf{X}_p)$ may  vary significantly from patch to patch. We discuss this issue in the next section.

\section{Relating gradient statistics to average turbulence power spectrum}
\label{sec:J2_from_P}
\subsection{Accumulation of gradients along the line of sight}

Projected 2D gradients are results of the line-of-sight (LOS) integration over depth $L$ of 3D signal 
\begin{equation}
\nabla_R \Phi(\mathbf{R}) = \nabla_R \int_0^L d z \; \phi ( \mathbf{R},z)
\end{equation}
where LOS coordinate is chosen to be $z$. 
Analysing a patch  we shall consider only modes with 2D wavelengths on the sky shorter than $L_p$. Correlation and spectral properties of the gradient, nevertheless, include scales extending past $L_p$ along the LOS. To study how signal accumulates when we integrate over LOS distance $L \ge L_p$, we start by considering a cubical volume of magnetized fluid $L_p\times L_p \times L_p$ which we will term {\bf c-block}. Within c-block we assume that the turbulence is axis-symmetric with the mean direction of the magnetic field given by the vector ${\bf \lambda}$. Therefore, the analytical description in \citetalias{LP12}  applies to a c-block magnetic field. Over larger distances $\hat\lambda(z)$ may vary according to long-wave behaviour of the magnetic field. Let us see how the statistic of gradients is determined in this picture.

To use the MHD turbulence description in global system of reference (see \citetalias{LP12}), one should consider scales that are sufficiently small either in comparison with $L_{inj}$ or $l_A$. In fact, in most cases we want to get as high resolution as possible and deal with sub-blocks much less than any of the two scales (see \citealt{YL17a}}). As a result, for our theoretical description, we take the patch size $L_p$ to be sufficiently smaller than either $L_{inj}$ for sub-Alfv\'enic case or $l_A$ for super-Alfv\'enic one. In a c-block of volume $L_p^3$ one can approximate 3D turbulent signal $\phi$ as anisotropic with a fixed 3D anisotropy direction $\hat{\lambda}$ and described by the power spectrum $P_c(k, \mathbf{\hat{k}} \cdot \hat{\lambda})$. Note that $P_c$ is defined locally to a particular c-block, thus label ``c''.

First, we shall establish how the gradients are correlated along the line of sight. We are interested only in scaling, so we shall neglect the anisotropic nature of the spectrum, taking it in the model form 
\begin{equation}
P_c(k) \approx \langle \phi^2 \rangle k^{-3} (k L_{inj})^{-m} 
\label{eq:model}
\end{equation}
(See \citealt{LP06}) where $m$ is the spectral index, e.g., $m=-2/3$ for Kolmogorov scaling. We compute the trace of the correlation for the gradients smoothed with a Gaussian beam at the beam scale $L_b$ in approximation $L_p \gg L_b$:
\begin{align}
\xi_{\nabla\nabla}(z) &= \pi \int_{1/L_p} K^2 d^2 K e^{-K^2 L_b^2} \int d k_z P_c(K,k_z) e^{i k_z z}
\nonumber \\
& \propto \frac{\langle \phi^2 \rangle}{L_b^2} \left(\frac{L_b}{L_{inj}}\right)^{m}
\widetilde{\xi}_{\nabla\nabla}\left(z/L_b\right)
\end{align}
where   the direct integration gives
\begin{align}
&\widetilde{\xi}_{\nabla\nabla}(y=z/L_b)  \approx
\sqrt{\frac{\pi}{2}}  \cos \left( \frac{m}{2}\pi \right) \Gamma[-1-m] \times
\nonumber \\
& \times \left( y^{2+m} + 2^m \left( 2m - y^2 \right)
e^{\frac{1}{4} y^2} \Gamma\left[1+\frac{m}{2},\frac{1}{4} y^2\right] \right) .
\end{align}
This shows that the correlation decreases as $(L_b /z)^{-2+m}$ at LOS distances $z > L_b$. Thus, the gradients are added up in uncorrelated fashion during integration over the depth exceeding the smoothing scale $L_b$.

This is reflected in the variance of the gradients projected over c-block depth $L_p$
\begin{equation}
\begin{aligned}
\sigma_{ij}(L_p) & \equiv \left\langle \nabla_{R_1} \Phi(\mathbf{R}_1) \cdot  \nabla_{R_2} \Phi(\mathbf{R}_2)  \right\rangle_{R_2 \to R_1} 
\nonumber \\
& = - L_p  \int_{-L_p}^{L_p} dz \xi_{\nabla\nabla}(z)\\
&\propto  \frac{L_p}{L_b} \langle \phi^2 \rangle (L_b/L_{inj})^{m}
\end{aligned}
\end{equation}
accumulating only linearly with the c-block depth $L_p$, since $L_p \gg L_b$. Here we point out that $L_p$ appears in the gradient variance only as the depth of a c-block along the LOS.
From the point of view of orthogonal scales, gradients have sufficient small scale power to be determined predominantly by
the beam scale $L_b$. Notably, 
$\langle \phi^2 \rangle (L_b/L_{inj})^{m}$ can be viewed as a variance of differences of quantity
$\phi$ at separations $L_b$, i.e. $\langle \Delta \phi(L_b)^2 \rangle$.

\begin{figure}
\includegraphics[width=0.4\textwidth]{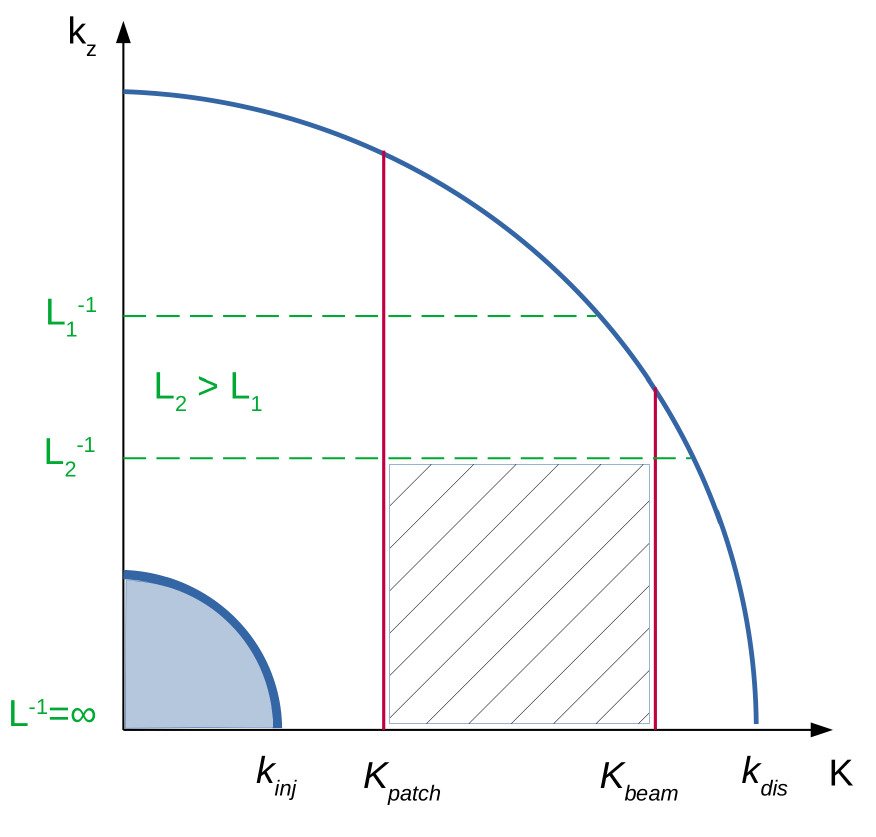}
\caption{The dashed area is the part of the phase space that contributes to the variance of the gradients, after the line of sight integration is
done through the depth $L_2=L_p$ of the c-block.  Here $K_{inj} \sim 1/L_{inj}$, 
while $K_{patch} \sim 1/L_p$ and $K_{beam} \sim 1/L_b > K_{patch}$. We see that the wavenumber of
relevant modes is dominated by sky component $K > k_z$ and modes with wavelengths approaching the injection scale (blue shaded region) have little contribution to the gradient variance.  Extending the line-of-sight
interation range shrinks the height of the dashed region, ever improving $k_z=0$ approximation.
%Top: in the presence of patch smoothing.  Bottom: in the absence of patch smoothing
}
\label{fig:boundaries}
\end{figure}

Let us now consider anisotropic aspects of the projected gradient variance tensor in a single c-block. This tensor structure in the Fourier space is 
\begin{align}
\sigma_{ij}(L_p) &  = L_p  \int_{-L_p}^{L_p} dz \times \\
\times & \int_{1/L_p}^\infty d^2 K K_i K_j e^{-K^2 L_b^2}
\int_0^\infty \!\! d k_z P_c(k, \hat{k} \cdot \hat{\lambda}) e^{i k_z z}
\nonumber
\end{align}
Line of sight integration through the c-block suppresses all the modes with $k_z > 1/L_p$.
Thus, as illustrated in Fig.~\ref{fig:boundaries}, only $ K > 1/L_p$ and $k_z < 1/L_p$ contribute, and one can
effectively set $k_z$ to zero in the result 
\begin{equation}
\sigma_{ij}(L_p) \approx L_p  \int_{1/L_p}^{\infty} d^2 K K_i K_j P_c(K, \hat{k} \cdot \hat{\lambda}, k_z=0) 
\label{eq:sigma_block}
\end{equation}
(degenerate case when this may not be accurate is when the anisotropy direction is close to the line of sight).

Subsequent c-blocks along z-axis can be considered as having different direction of the magnetic field $\hat\lambda(z)$, but uncorrelated gradient signal between the blocks, thus the total variance of projected gradients becoming sum of variances of individual c-blocks, $\sigma_{ij} \approx \sum_c \sigma_{ij}(L_p)$
\begin{align}
\sigma_{ij} &\approx \int d^2 K K_i K_j \left[ \sum_c L_p P_c(K, \hat{k} \cdot \hat{\lambda}(z), k_z=0) \right] 
\nonumber \\
 & \approx L \int d^2 K K_i K_j \overline{ P(\mathbf{K})}
\label{eq:sigma_block_sum}
\end{align}
where overline designates averaging over the directions $\hat{\lambda}$ 
takes over the integration depth $L$, 
\begin{equation}
\overline{ P(\mathbf{K})}= \frac{1}{L}  \int_0^L dz P_c(K, \hat{k} \cdot \hat{\lambda}(z), k_z=0)
\label{eq:Pc_integrated}
\end{equation}
i.e the effective 2D spectrum of projected gradients is the average
of individual c-block spectra over the directions of the local to c-block magnetic field. Note that $P_c$ is defined only within a c-block where we can assume \citetalias{LP12} description.  The result in 
Eqs.~(\ref{eq:sigma_block_sum},\ref{eq:Pc_integrated}) can be
also described using polarization analogy. Covariance $\sigma_{ij}$
has the same rotational properties as the linear polarization tensor. Introducing the effective Stokes parameters for a c-block as $Q=\sigma_{xx} - \sigma_{yy}$ and $U = 2 \sigma_{xy}$, we see that
these $Q$ and $U$ are simply additive as we sum c-blocks along the line of sight.

\subsection{Averaging of power spectrum along the LOS in strong turbulence regime }

For sub-Alfv\'enic turbulence, the directions of magnetic field in individual  c-blocks along the line of sight are correlated over the turbulence injection scale $L_{inj}$. Within this scale, 
both the regime of weak at $l > l_{tr}=L_{inj} M_A^2$ and strong at $ l < l_{tr}$ sub-Alfv\'enic turbulence is present. On scales $l \gg L_{inj}$ the magnetic field fluctuations are not correlated and add in a random walk fashion.  The rms angular difference of the magnetic field directions between c-blocks
separated by more that $L_{inj}$ is of the order of the Alfv\'en Mach number $M_A$. 

For super-Alfv{\'e}nic turbulence $M_A>1$, the correlation length of magnetic field is determined by a smaller scale $l_A=LM_A^{-3}$. The magnetic field on scales larger than $l_A$ is uncorrelated and the variation of the magnetic field direction from one volume $l_A\times l_A \times l_A$ to another similar volume is of the order of unity. 

For sub-Alfv\'enic turbulence in strong regime,  the correlated rms deviations of c-block field directions $\lambda$ over the distance $l_\perp$ measured perpendicular to the mean magnetic field is characterized by the angle $\alpha$:
\begin{equation}
    \tan \alpha\approx  \frac{l_{\bot}}{l_\|} \approx \left(\frac{l_{\bot}}{L_{inj}}\right)^{1/3}M_A^{4/3},
    \label{eq:alphastrong}
\end{equation}
which can be seen from Eq.~(\ref{eq:scaling_smallma}).  This scaling proceeds up to the transitional scale between the weak and strong cascades $l_{tr}=L_{inj} M_A^2$ \citep{Lazarian06}.\footnote{At the scale $t_{tr}$ the scaling of turbulent motions changes and, as a result, the change of the angle behaves as
\begin{equation}
    \tan\alpha \approx \frac{v_l}{V_A}\approx \left(\frac{l_\bot}{L_{inj}}\right)^{1/2} M_A,
    \label{eq:alphaweak}
\end{equation}
where we used the scaling of the weak turbulence $v_l\approx M_A V_A (l_\bot/L_{inj})^{1/2}$.}

To estimate the LOS averaging of a c-block spectrum in Eq.~(\ref{eq:Pc_integrated}) one need a specific model for the power distribution $P_c$ in a c-block. Following \citetalias{LV99}, in the strong turbulence regime at scales $k L_{inj} > M_A^{-2}$,  the peak of power distribution lies at wave angles to the local direction of the magnetic field  given by $\cos\theta_k = \hat k \cdot \hat \lambda \approx v_k/V_A$ where $v_k$ is the rms turbulent velocity at the scale $k=1/l$ (note that $v_k/V_A$ can be understood as a local Afv\'en Mach number). Relating $v_k$ to the velocity $v_{inj}$ at the injection scale $k\approx L_{inj}^{-1}$ using Eq.~(\ref{eq:vscal}), we have $\cos\theta_k \sim (k L_{inj})^{-1/3} M_A^{4/3}$ where it is taken into account that strong turbulence regime is replaced by weak turbulence at $ 1 < k L_{inj} < M_A^{-2}$. Thus, the power is strongly peaked for wavevector directions almost perpendicular to the magnetic field, $|\cos\theta_k|< M_A^2$, though strictly perpendicular waves are somewhat disfavoured. The shorter the wavelength, the more concentrated the power is in the perpendicular direction. At the level of individual c-block, the dispersion of directions
will be dominated by the longest wavelengths present, i.e,. the ones of the scale of the c-block $k \sim L_p^{-1}$. Thus, 
 similar to \citetalias{LP12} we can expect that in the regime of strong turbulence, 
\begin{equation}
P_c(\mathbf{k}) \propto k^{-11/3} \exp\left(-\frac{(L_{inj}/L_p)^{1/3}}{M_A^{4/3}} | \widehat{\mathbf{k}} \cdot \widehat{\mathbf{\lambda}} | \right)
\label{eq:power_spectrum}
\end{equation}
provides an adequate description for power distribution between differently oriented Alfv\'en and slow modes. Note, however, that to use this scaling, the size of the c-block should be sufficiently small, $L_p < L_{inj} M_A^2$.  

Adding contributions from different c-blocks along the line of sight amounts to convolution
of the angular part of the $k_z=0$ sky projection c-block power $P_c(\mathbf{k})$ with
distribution of the 3D direction of the magnetic-field $\widehat{\lambda}$ around the
global direction of the magnetic field $\widehat{\lambda}_0$. We will model the
distribution of $\lambda$ direction by assuming that the perturbations in the magnetic
field  $\delta \mathbf{b}$ are Gaussian and distributed in 3D isotropically around global $\lambda_0$ direction with
$ \langle \delta b^2 \rangle /B_0^2 = M_A^2$, which gives

\begin{equation}
\mathcal{P}(\theta_{\widehat{\lambda}})
 = \frac{1}{2\pi} 
\left( 1 + \frac{2 \cos^2\theta_{\widehat{\lambda}}}{M_A^2} \right)
\exp\left[-\frac{ \sin^2\theta_{\widehat{\lambda}}}{ M_A^2}  \right]
\label{eq:pdf_exp}
\end{equation}
where $\theta_{\widehat{\lambda}} \in (0,\pi/2) $  is the angle measuring the deviation 
of the headless vector $\widehat{\lambda}$ from $\widehat{\lambda}_0$ axis.
Let us note that in the isotropic limit $M_A \to \infty$, $\langle \cos^2 \theta_{\widehat{\lambda}} \rangle \to 1/3$ with respect to an \textit{arbitrary} axis. In the opposite limit,
$M_A \to 0$, $\mathcal{P}(\theta_{\widehat{\lambda}}) \to \delta_D(\theta_{\widehat{\lambda}})$
and $\widehat{\lambda} \to \widehat{\lambda}_0 $ with $\langle \sin^2(\theta_{\widehat{\lambda}}) \rangle \sim M_A^2$.

Combining Eq.~(\ref{eq:power_spectrum}) and Eq.~(\ref{eq:pdf_exp}) gives the following expression for the angular part of the line-of-sight averaged power spectrum
\begin{eqnarray}
\overline{P(\mathbf{K})} & \propto & \frac{1}{2\pi} \int d \Omega_{\widehat{\lambda}}
\exp\left(-\frac{(L_{inj}/L_p)^{1/3}}{M_A^{4/3}} | \widehat{\mathbf{K}} \cdot \widehat{\mathbf{\lambda}} | \right)
\nonumber \\
& \times & 
\left(1 + 2 \frac{(\widehat{\lambda} \cdot \widehat{\lambda}_0)^2}{M_A^2} \right)
\exp\left(-\frac{1-(\widehat{\lambda} \cdot \widehat{\lambda}_0)^2}{M_A^2}\right)
\label{eq:Pk_convolution}
\end{eqnarray}
For $M_A < 1$ and $L_p/L_{inj} < M_A^2$ the first exponential term
is narrower than the second and can be treated as a Dirac $\delta$-function. The resulting distribution of $\mathbf{K}$ directions is then largely determined by the local magnetic field's direction variations $\hat \lambda$ itself.
To approximate the integral over the directions of $\hat\lambda$ we notice that the maximum contribution comes from $\hat\lambda$'s that are both
perpendicular to $\widehat{\mathbf{K}}$ and lie in the plane spanned by $\widehat{\mathbf{K}}$ and $ \widehat{\mathbf{\lambda}}_0$ vectors. 
For such $\hat\lambda$'s we have $ \left(\widehat{\mathbf{\lambda}} \cdot \widehat{\mathbf{\lambda}}_0\right)^2 = 
1 - \cos^2\psi_k \sin^2\gamma$ where $\psi_k$ is the angle between $\mathbf{K}$ and 2D projection of the mean field direction  $\widehat{\mathbf{\Lambda}}_0$, so that $\cos\psi_k =\widehat{\mathbf{K}} \cdot \widehat{\mathbf{\Lambda}}_0 $. Substituting this in Eq.~(\ref{eq:Pk_convolution}) gives
\begin{equation}
\label{eq:PK_approx}
\overline{P (\mathbf{K})} 
\approx
\exp\left(-\frac{(\psi_k \pm \pi/2)^2 }{(M_A/\sin\gamma)^2} \right) 
\end{equation}
in the approximation of small $\cos\psi_k$ that has two solutions $\cos^2\psi_k \approx (\psi_k \pm \pi/2)^2$. Our numerical integration suggests that Eq.~(\ref{eq:PK_approx}) is an excellent approximation for $M_A/\sin\gamma < \pi/8$.

\begin{figure}
\centering
\includegraphics[width=0.45\textwidth]{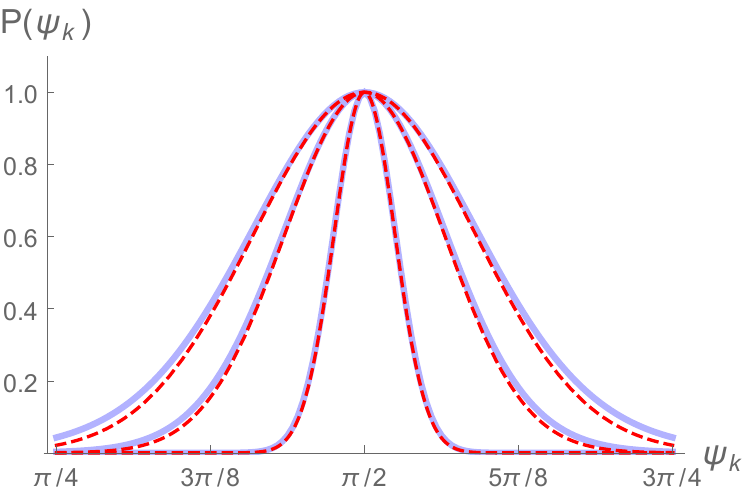}
\caption{Normalized angular dependence of the 2D spectrum of the gradients as given by Eq.~(\ref{eq:Pk_convolution}) (blue, solid) versus the analytical fit of Eq.~(\ref{eq:PK_approx}) (red, dashed). Three sets of curves correspond to $M_A=0.1, \gamma=\pi/2$ (the most narrow), $M_A=0.1,\gamma=\pi/8$ and $M_A=0.4,\gamma=\pi/2$ (the widest, which starts showing some deviations of the fit). $L_p=0.01 L_{inj}$ for all cases. 
%(Right) The numerical result from 5 numerical simulations. 
}
\label{fig:Pk_fit}
\end{figure}

\begin{figure}
\centering
\includegraphics[width=0.45\textwidth]{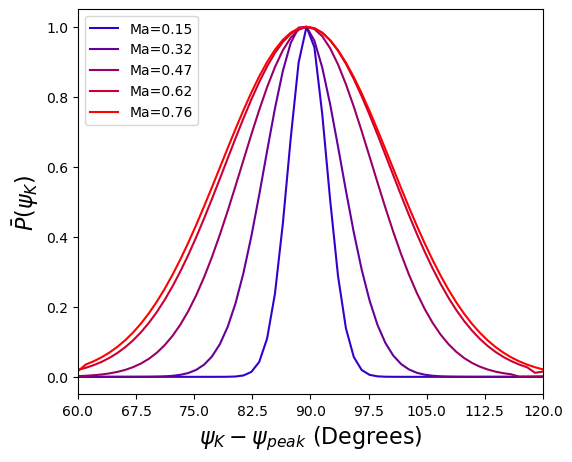}
\caption{ Angular dependence of the 2D power spectrum $P(\psi_K)$ as a function of $\psi_K$ (recentred with respect to $\psi_{center}$) in the projected maps from five selected simulation data (huge-0 to huge-4) with only Alfven modes. Due to the noises of the angular spectrum, the presented curves are the Gaussian fits of the angular spectrum instead of the raw data itself.
}

\label{fig:Pk_fit_num}
\end{figure}

In Fig.~\ref{fig:Pk_fit} we plot the comparison of Eq.~(\ref{eq:PK_approx}) with Eq.~(\ref{eq:Pk_convolution}) as a function of $M_A/\sin\gamma$. We see that the approximation of Eq.~(\ref{eq:PK_approx}) has not much visible differences from the exact form in Eq.~(\ref{eq:Pk_convolution}).  In Fig.~\ref{fig:Pk_fit_num} we show normalized angular power distribution obtained from the synthetic observational maps from pure Alfven modes extracted in numerical simulations ``huge-0'' to ``huge-4'' with different $M_A$.
We can see that Fig.~\ref{fig:Pk_fit} and Fig.~\ref{fig:Pk_fit_num} has very similar trend as $M_A$ grows.

\subsection{Theoretical approximation for $J_2$ parameter}

To summarize, we have demonstrated that the covariance matrix of projected gradients in a $L_p \times L_p$ sub-block on a sky is determined by the spectrum that has a concentration of power at wavemodes orthogonal to the mean projected direction $\widehat{\mathbf{\Lambda}}_0$ of the magnetic field within the line-of-sight cube based on the sub-block. Respectively, the eigendirection of the covariance matrix with the largest eigenvalue is orthogonal to $\widehat{\mathbf{\Lambda}}_0$ as well. Following this eigendirection change from sub-block to sub-block, one maps the magnetic field over wider region of the sky.

\begin{figure}
\includegraphics[width=0.4\textwidth]{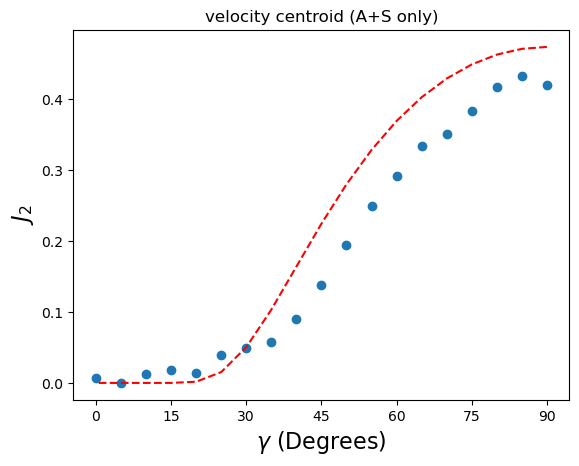}
\caption{$J_2$ in centroid maps created from synthetic simulations with equal contributions of Alfven and Slow modes as a function of $\gamma$. The theoretical approximation in Eq.~(\ref{eq:J}) is given by the red dash line. $M_A \approx 0.6$ . }
\label{fig:J2}
\end{figure}

The variance of the directions of gradients at individual pixels is determined by the $J_2$ parameter of the covariance matrix. 
The $J_2$ parameter is determined by the properties of the turbulence and the LOS angle of the mean magnetic field. In approximation of Eq.~(\ref{eq:PK_approx}) for strong turbulence it is easily calculated to be
\begin{equation}
J_2 \approx \exp\left( -2 (M_A/\sin\gamma)^2 \right) 
\label{eq:J}
\end{equation}
while complete form of the functional dependence of $J_2$ on $M_A$ and $\sin\gamma$ 
can be determined by numerically integrating the $\sigma_{ij}$ components with the power spectrum given in Eq.~(\ref{eq:Pk_convolution}). Fig.~\ref{fig:J2} shows our result from the numerical calculations as compared to the theoretical expectation in Eq.~(\ref{eq:J}). We can see that the numerical data (from synthetics, blue points) and the theoretical points (red curve) have very similar trends.

\subsection{The susceptibility of our method to noise level}

\begin{figure}
\includegraphics[width=0.4\textwidth]{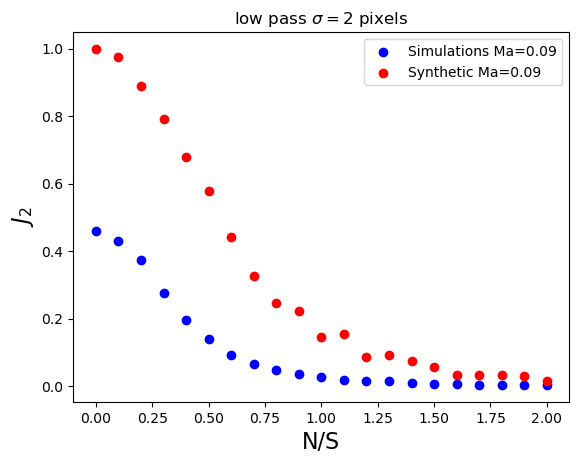}
\caption{ Two figures showing how the introduction of noises changes the measured $J_2$ in both synthetic (red) and the Alfven mode fluctuations of simulation data (blue) at  $M_A=0.09$.  }
\label{fig:noise}
\end{figure}

In realistic observations, we will have noise, and the $J_2$ parameter is naturally smaller since noise introduces more gradient isotropy to the data set. As a result, we would like to evaluate how noise affects our measurement of $J_2$ in observations. Fig.~\ref{fig:noise} shows the variation of $J_2$ of Alfven mode velocity centroid as a function of the noise-to-signal ratio (N/S) for synthetic and realistic numerical data. To mimic observations, we apply an additional 2-pixel Gaussian smoothing kernel to our data. Again, we have to emphasize that synthetic data demonstrates the effect of beam size and dissipation scales in realistic data. We can see from Fig.~\ref{fig:noise} that the $J_2$ parameter indeed decreases dramatically as $N/S$ increases for both synthetic (red points in Fig.~\ref{fig:noise}) and realistic MHD data (blue points in Fig.~\ref{fig:noise}).

\section{Estimation of the mean direction of the field and its variance on the sky}
\label{sec:gradient_orientation}

In the previous sections, we have shown that measurements of the gradients on the sky give information about
the local direction of the magnetic field $\Theta_H$ and the Alfven Mach number via the eigen-directions
and the anisotropy $J_2$ of the gradient covariance tensor, respectively. In this section, we analyze the statistical uncertainties if $\Theta_H$ and $J_2$ are determined locally in a $L_p \times L_p$ sub-block. To estimate the errors we assume the Gaussian distribution of the gradients.

\subsection{Distribution of the gradient orientations on the sky}

In \S ~\ref{sec:Grad_cblock_theory} we have developed a theoretical prediction for the one-point covariance tensor $\sigma_{ij}$ of the gradient field components. The anisotropy of the gradient map is determined by the $J_2$ parameter in an invariant coordinate way. At the same time, the two eigen-directions are one aligned and one perpendicular to the local direction of the LOS-averaged magnetic field. Here we present theoretical predictions for more practical statistics {\kh in the point of view of} variance of the directions of the gradients, that follows from the results of Section~\ref{sec:Grad_cblock_theory}, assuming that 2D gradients can be described as Gaussian distributed, i.e. in low $M_A$. {\kh Notice that \cite{2020MNRAS.496.2868L} points out that such assumption of Gaussian distribution on gradient orientation histogram is not correct in the case of low polarization percentage (i.e., $M_{A,LOS}$, the line of sight Alfv{\'e}n Mach number).}

Under Gaussian assumption for gradients,  {\kh the theoretical expectation of the gradient orientation distribution is given by \citep{2020MNRAS.496.2868L}}: 
\begin{align}
\label{eq:Gaussian-angles}
P(\theta) =  \frac{1}{\pi} \times \frac{ 
\sqrt{1 -J_2}}{1 - \sqrt{J_2}\cos 2\left(\theta-\theta_H\right)}
\end{align}
where $\theta_H$ is the preferred direction of the distribution that follows the eigen-direction of $\sigma_{ij}$ with the largest eigenvalue. 
This preferred direction is orthogonal to the magnetic fields for the turbulence regime that we are considering in this current section. The distribution is defined and normalized over an arbitrary interval of angles with the span of $\pi$ radians (i.e., $\theta\in [-\pi/2,\pi/2)$). We see the fundamental role the parameter $J_2$ plays in this distribution: In maximally anisotropic case $J_2=1$ the angle distribution given by Equation~(\ref{eq:Gaussian-angles}) becomes the Dirac $\delta_D$-function around $\theta_B$ with zero variance of orientation. In the opposite limit where the turbulence is isotropic, $J_2=0$, i.e. the angle distribution is flat and the \textit{rms} angle deviation $\sigma_\theta=\pi/(2\sqrt{3})\approx 52^o$.

\subsection{Statistical uncertainties of the mean field direction and the angle variance from gradients in sub-blocks}

Let us first determine the number $N$ of independent and uncorrelated samples of the gradient directions $\theta_i$ in a sub-block. We can estimate $N$
from the following considerations. Since gradients are determined by the short wave side of the spectrum, the correlation length $L_{corr}$ between gradients at neighboring pixels is determined by the largest experimental or synthetic beam scale of the map 
and the scale of dissipation that damps shortwave perturbations. We can estimate $L_{corr} \approx \mathrm{max}(W_{FWHM},L_{dis})$ where $W_{FWHM}$ is a 
full width at half maximum of the beam window. The resulting 
number of independent measurements in the patch is then 
\begin{equation}
N \approx (L_p/L_{corr})^2 ~.
\end{equation}

\subsubsection{Uncertainties of the magnetic field direction $\Theta_H$}
The simplest estimator $\widetilde{\theta}_H$ of the mean direction of the gradients in a sub-subblock 
is the average of individual independent measurements
\begin{equation}
\widetilde{\theta}_H = \frac{1}{N} \sum_{i=1}^N \theta_i
\end{equation}
The uncertainty in determining $\widetilde{\theta}_H$ is given by variance of the estimator
over the whole ensemble of gradient realizations
\begin{equation}
(\Delta \widetilde{\theta}_H )^2 \equiv
\left\langle {\widetilde{\theta}_H}^2 \right\rangle - \left\langle \widetilde{\theta}_H \right\rangle^2 =  \frac{1}{N} \langle (\theta - \theta_H)^2 \rangle
\end{equation}
To proceed further, let us have the gradient angle distribution described by
Eq.~(\ref{eq:Gaussian-angles}) in Gaussian assumption for the gradients.
With this distribution, the ensemble variance of the angles can be found in terms of special functions,
but for simplicity we will approximate it by the ``angle variance''  $\Sigma^2$, defined as
\begin{equation}
\Sigma^2 \equiv \frac{\left\langle 1-\cos 2(\theta - \theta_H) \right\rangle}{2}
= \frac{\sqrt{J_2} + \sqrt{1-J_2}-1}{2 \sqrt{J_2}}
\label{eq:Sigma}
\end{equation}
$\Sigma^2$ is close to true angle variance $\left\langle(\theta-\theta_H)^2\right\rangle$ if $\theta-\theta_H$ is small. Otherwise, it ranges from zero at zero variance to 1/2 if the distribution of directions is isotropic. The estimated mean direction is then, on average, equal to the ensemble
mean with the uncertainty that is suppressed by $\sqrt{N}$
\begin{equation}
\widetilde{\theta}_H = \theta_H \pm \frac{\Sigma}{\sqrt{N}}
\end{equation}

It is important to note, that as anisotropy level increases with decreasing $M_A/\sin\sigma$ and $\Sigma^2 \to 0$, the \textit{rms} deviation of the gradient direction decreases only as $\Sigma \sim  \sqrt{M_A/\sin\gamma}$, in contrast to \textit{rms} fluctuations of the magnetic field direction itself relative to the mean, which are $\propto M_A$.

\subsubsection{Uncertainty of the anisotropy level of the gradients}
The information on the $M_A$ is contained in the anisotropy level of the gradient direction distribution,
which in turn is determined by the variance of the gradient directions.
In line with the previous sections, to treat the periodicity of angles when they are not small, let us consider the quantity $\Sigma_i^2 = \frac{ 1-\cos 2(\theta_i - \theta_H)}{2} $ in place of $(\theta_i-\theta_H)^2$.

Then the estimator (designated by tilde) of $\Sigma^2$ in a sub-block is 
\begin{equation}
\widetilde{ \Sigma^2} = \frac{1}{N} \sum_{i=1}^N \Sigma_i^2 
\end{equation}
The mean value of this estimator is equal to ensemble-averaged $\Sigma^2$ given by Eq.~(\ref{eq:Sigma})  
\begin{equation}
\left\langle \widetilde{\Sigma^2} \right\rangle =  \frac{1}{N} \sum_{i=1}^N \langle \Sigma_i^2 \rangle = \Sigma^2
\end{equation}
while the variance of this estimator is
\begin{align}
\left\langle (\widetilde{\Sigma^2})^2 \right\rangle - \left\langle \widetilde{\Sigma^2} \right\rangle^2 
& =  \left\langle \left( \frac{1}{N} \sum_{i=1}^N  \Sigma_i^2 \right)^2 \right\rangle -   \left\langle \Sigma_i^2 \right\rangle^2
\nonumber \\
& = \frac{1}{N} \left(  \left\langle \Sigma_i^4 \right \rangle  -  \left\langle\Sigma_i^2 \right\rangle^2 \right) ~,
\end{align}
using the presupposition that $\theta_i$ are uncorrelated.

With distribution given by Eq.~(\ref{eq:Gaussian-angles}) the fourth order cumulant  is evaluated to be 
\begin{equation}
\langle \Sigma_i^4 \rangle  - \left\langle \Sigma_i^2 \right\rangle ^2  = \frac{J_2 - 1 + \sqrt{ 1 - J_2}}{4 J_2} 
\end{equation}
When angle fluctuations within sub-block are small, $\Sigma_i \approx \theta$, and $J_2 = 1 - \alpha$ with $\alpha \ll 1$, we can use the expansion
\begin{align}
\langle \Sigma_i^2 \rangle &  \approx \frac{1}{2}  \sqrt{\alpha} \left( 1 - \frac{1}{2} \sqrt{\alpha} \right) \\
\langle \Sigma_i^4 \rangle   - \langle \Sigma_i^2 \rangle^2 & \approx \frac{1}{4} \sqrt{\alpha} \left(1- \sqrt{\alpha} \right)
\end{align}
Thus we have an estimator $\widetilde{\Sigma^2}$ for the angle variance with the mean  and uncertainty 
as follows
\begin{align}
\left\langle \widetilde{\Sigma^2} \right\rangle &\approx \frac{1}{2}  \sqrt{\alpha} \\
\Delta \widetilde{\Sigma^2} = \sqrt{\left\langle (\widetilde{\Sigma^2})^2 \right\rangle - \left\langle \widetilde{\Sigma^2} \right\rangle^2} 
& \approx \frac{1}{2} \frac{\alpha^{1/4}}{N^{1/2}}
\end{align}
The relative uncertainty, therefore, is 
\begin{equation}
\frac{\Delta \widetilde{\Sigma^2} }{\left\langle \widetilde{\Sigma^2} \right\rangle } \approx \frac{1}{ \alpha^{1/4} N^{1/2}} 
\approx \sqrt{\frac{1}{2 N \Sigma^2 }}
\end{equation}
which demonstrates that the relative error on angle dispersion increases as angle dispersion becomes smaller (i.e., as the mean direction becomes better determined) but decreases with the increase in the number of independent samples of the gradients in the sub-block.

\begin{figure}
\includegraphics[width=0.4\textwidth]{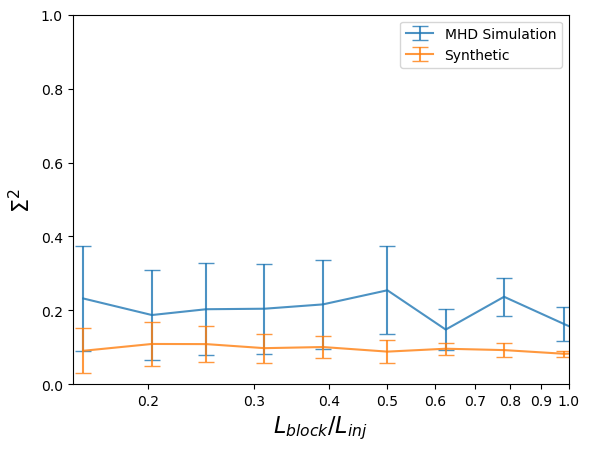}
\caption{The variation of $\Sigma^2$ and its error-bars as a function of block size for Alfv{\'e}n mode centroid at $\gamma=\pi/2, M_A=0.13$. }
\label{fig:MC_block}
\end{figure}

This is illustrated in Fig.~\ref{fig:MC_block} where we plot the  $J_2$ and its error bars as a function of block size $L_{block}$ normalized by the injection scale for both synthetic and real MHD data. We can see that the value of $J_2$ for both synthetic and real data does not change significantly. However, the numerical data has larger errors for smaller block sizes.

\subsection{Relation to projected Mach number}

From Eq.~(\ref{eq:J}), $J_2$ is related to Alfv'enic Mach number and the angle between the mean field and the line of sight, $J_2 \approx e^{-2 M_{A\perp}^2}$, i.e. $\alpha = 2 M_{A\perp}^2$, thus $\langle \theta^2 \rangle = \frac{1}{\sqrt{2}} M_{A\perp}$.
If $M_{A\perp}$ is determined from the variance of gradient angle\cite{dispersion}, the statistical error on this measurement is
\begin{equation}
\frac{\Delta M_{A\perp}}{M_{A\perp}} = \sqrt{\frac{1}{\sqrt{2} M_{A\perp} N}}
\label{eq:deltaMA}
\end{equation}
For instance,  for $M_{A\perp}=0.1$, the statistical error is $10\%$ if we have $N=1000$ \textit{uncorrelated} samples of a gradient in the patch (oversampling of a smoothing scale will not decrease the error).
 We must remember, of course, that the expansion part of our analysis is valid only when $M_{A\perp} = M_A /\sin\gamma \ll 1$, which will not hold
for too small $\gamma$. 

\begin{figure*}[th]
\includegraphics[width=0.47\textwidth]{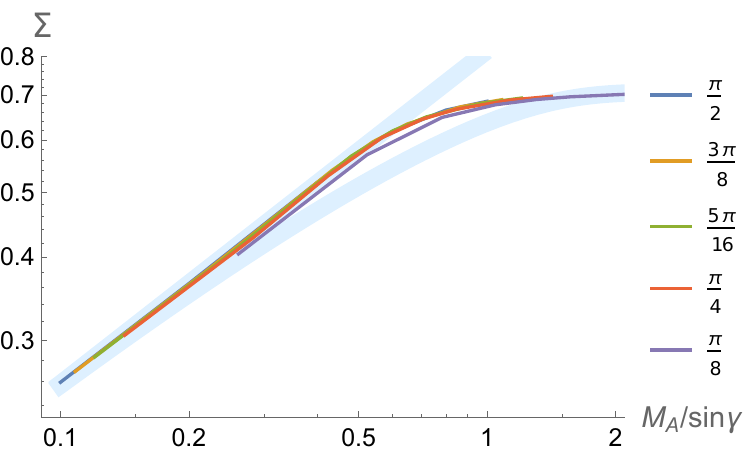}
\includegraphics[width=0.47\textwidth]{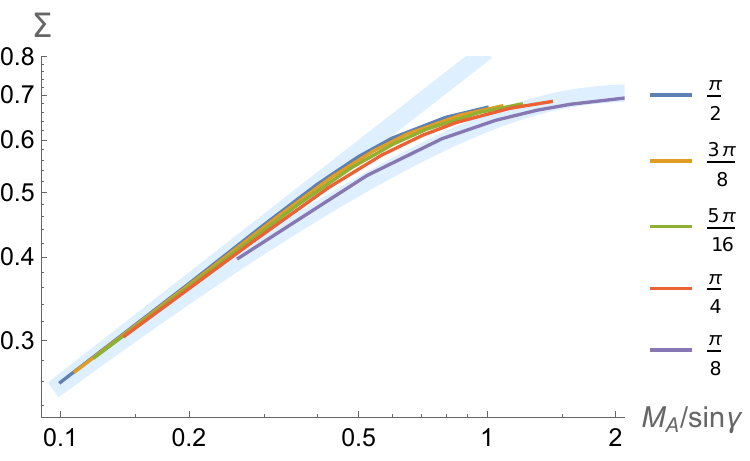}\\
\includegraphics[width=0.47\textwidth]{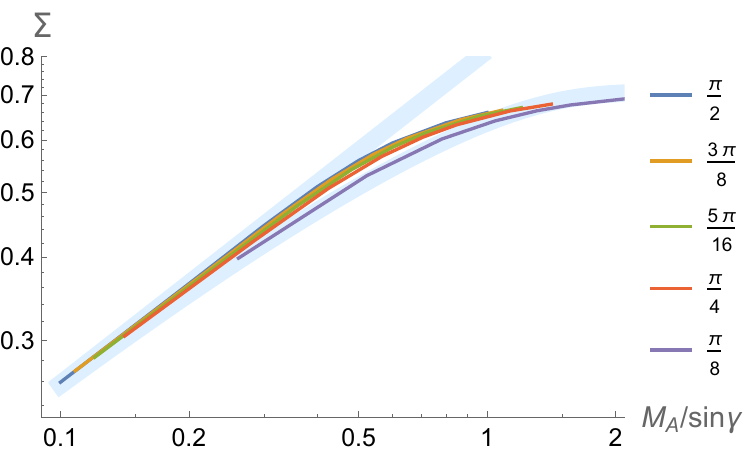}
\includegraphics[width=0.47\textwidth]{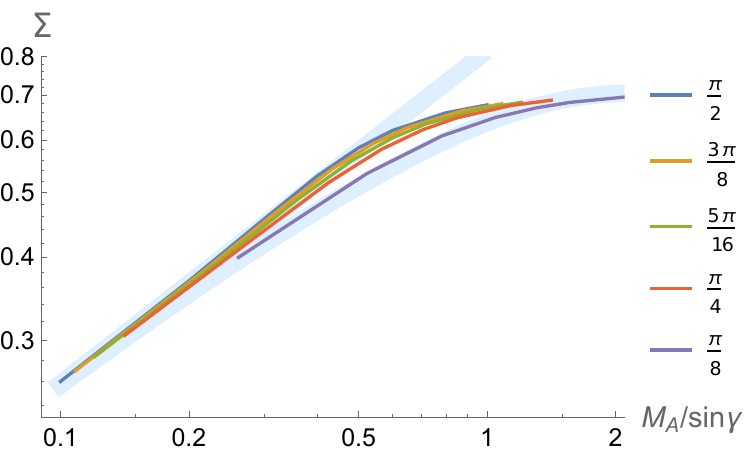}
\caption{$\Sigma$ parameter for velocity centroids (top row) and synchrotron intensity gradients (bottom row) in strong (left column) and Alfv'enic (right column) turbulence. The functions are plotted versus
perpendicular $M_{A\perp} = M_A/\sin\gamma$ argument, for several angles $\gamma$ between the mean magnetic field and the line of sight, as labeled. Curved thick light blue line corresponds to approximation $J_2=\exp(-2 M_A^2/\sin^2\gamma)$, while straight thick blue line is its expansion $J_2 \approx 1-2 M_A^2/\sin^2\gamma$}
\label{fig:J2andSigma}
%\label{fig:Sigma_sync_and_cen}
\end{figure*}

In Fig.~\ref{fig:J2andSigma}, we present theoretical curves for $\Sigma$ as functions of $M_A/\sin\gamma$ while varying the mean direction of the magnetic field within the local patch relative to the LOS.  
We consider two cases of observables, namely gradients of velocity centroids and synchrotron intensity,
and two models of turbulence. purely Alfv'enic turbulence, and strong turbulence in high-$\beta$ regime
with equal contribution of Alfv'en and slow modes. Remarkably, in all these cases,
the universal approximation $J_2 \approx \exp\left(-2 (M_A/\sin\gamma)^2 \right)$ is excellent.
One observes that for Alfv'enic turbulence, there is more sensitivity to the mean field angle $\gamma$
in the regime of $M_A/\sin\gamma > 0.5$, with the fit performing best for small $\gamma$ where magnetic
field is more along LOS. At the same time, for more perpendicular orientation, it is actually the linear
expansion of the fit $J_2 \sim 1 - 2 (M_A/\sin\gamma)^2$ that gives a better match
up to $M_A/\sin\gamma < 0.6$, i.e., until the angle variance starts reaching its saturation level at $\Sigma = 1/\sqrt{2}$.

\section{Application of gradient theory to observables of MHD turbulence}
\label{sec:J2_MHD}

\subsection{Alternative measures available from observations}

We have developed our theory in terms of $J_2$, Eq.~(\ref{J2}), which can be obtained from observations by estimating the covariance tensor of the gradients in the sky patch.
However, one can measure the anisotropy of the gradient orientations using quantities which, although mathematically related to $J_2$, may have own advantages
in convenience of the measurement or properties of the uncertainties. They are particularly useful in the regime of high anisotropy, when $J_2$ is nearly saturated close to
unity, $ J_2 \approx 1 - 2 (M_a/\sin\gamma)^2 $.  We will not study this estimators in detail here, but just enumerate some popular choices and their behavior for sub-Alfv\'enic regime
\begin{enumerate}[wide, labelwidth=!, labelindent=0pt]
\item The ``angle variance''  $\Sigma^2$, defined as
\begin{equation}
\begin{aligned}
\Sigma^2 &\equiv \frac{\left\langle 1-\cos 2(\theta - \theta_B) \right\rangle}{2}
= \frac{\sqrt{J_2} + \sqrt{1-J_2}-1}{2 \sqrt{J_2}} \\
\Sigma^2 & \approx ( 1/\sqrt{2} ) ( M_A/\sin\gamma )~,  \quad M_A/\sin\gamma < 1
%\label{eq:Sigma}
\end{aligned}
\end{equation}
$\Sigma^2 $ is close to true angle variance $\left\langle(\theta-\theta_B)^2\right\rangle$ if $\theta-\theta_B$ is small. 
Otherwise, it ranges from zero at zero variance to 1/2 if the distribution of directions is isotropic. Determining $\Sigma^2$ requires preliminary determination
of the mean direction in the patch, around which the fluctuations are measured.
\item The $p^2$ function, which is similar to degree of polarization in polarization studies,
\begin{equation}
p^2 \equiv \left\langle\cos 2 \theta\right\rangle^2 + \left\langle\sin 2 \theta\right\rangle^2
= \frac{\left(1-\sqrt{1 - J_2} \right)^2}{J_2}
\label{eq:p2}
\end{equation}
or,  its reciprocal $V=1-p^2$ with the range from zero (zero variance, i.e. maximal anisotropy) to one (isotropic):
\begin{equation}
\begin{aligned}
V  & \equiv 1-p^2 = 2 \times \frac{J_2 + \sqrt{1-J_2}-1}
{J_2} \\
V & \approx  2 \sqrt{2}  ( M_A/\sin\gamma )~,  \quad M_A/\sin\gamma < 1
\label{eq:V}
\end{aligned}
\end{equation}
\item 
The ``bottom to top ratio'' $B/T$ that we define as the ratio of peak to the lowest value of the distribution function 
\begin{equation}
\begin{aligned}
\frac{B}{T} & \equiv \frac{P(\theta = \theta_B + \pi/2)}{P(\theta = \theta_B)}
 = \frac{1- \sqrt{J_2}}{ 1 + \sqrt{J_2}} \\
\frac{B}{T} &\approx (1/2)  ( M_A/\sin\gamma )^2~,  \quad M_A/\sin\gamma < 1
 \label{eq:BT}
\end{aligned}
\end{equation}
$B/T$ ranges from zero at zero variance to one if distribution is isotropic. \cite{dispersion} presented empirical dependencies between the dispersion of the velocity gradient orientation and inverse $T/B$ (top-to-bottom) ratio with the local plane of sky Alfv{\'e}nic Mach number $M_{A,\perp}$. Employing this technique it was possible \citep{survey} to estimate the magnetization of a number of molecular clouds on the sky and obtain previously unavailable magnetic information on a high velocity cloud hid behind the galactic arm. 
\end{enumerate}

In conclusion, we see that all measures of angle variances of the gradients of any single is determined by a single parameter
$J_2$, which in turn is given by the quadrupole to monopole ratio of 2D on-sky structure function
of the signal at zero separation.  

\subsection{Gradients of Centroids}
Velocity Centroid presents the most straightforward case on our application of Eq.~(\ref{eq:J}) as there is no extra caution that we have to take care of as long as density contributions are removed properly \citep{VDA}. The projection of {\it Alfv{\'e}n mode} velocities follows the formalism of Eq.~(\ref{eq:PK_approx}) without any doubt. In this subsection, we will discuss how different modes enter Eq.~(\ref{eq:J}). For simplifications, since MHD theory predicts that the slow modes follow pretty much similar scaling as the Alfv{\'e}n mode (see, e.g. \citealt{LP04}), therefore we will limit our discussion to turbulent media with admixtures of Alfv{\'e}n and fast modes only.

\begin{figure*}
\centering
\includegraphics[width=0.96\textwidth]{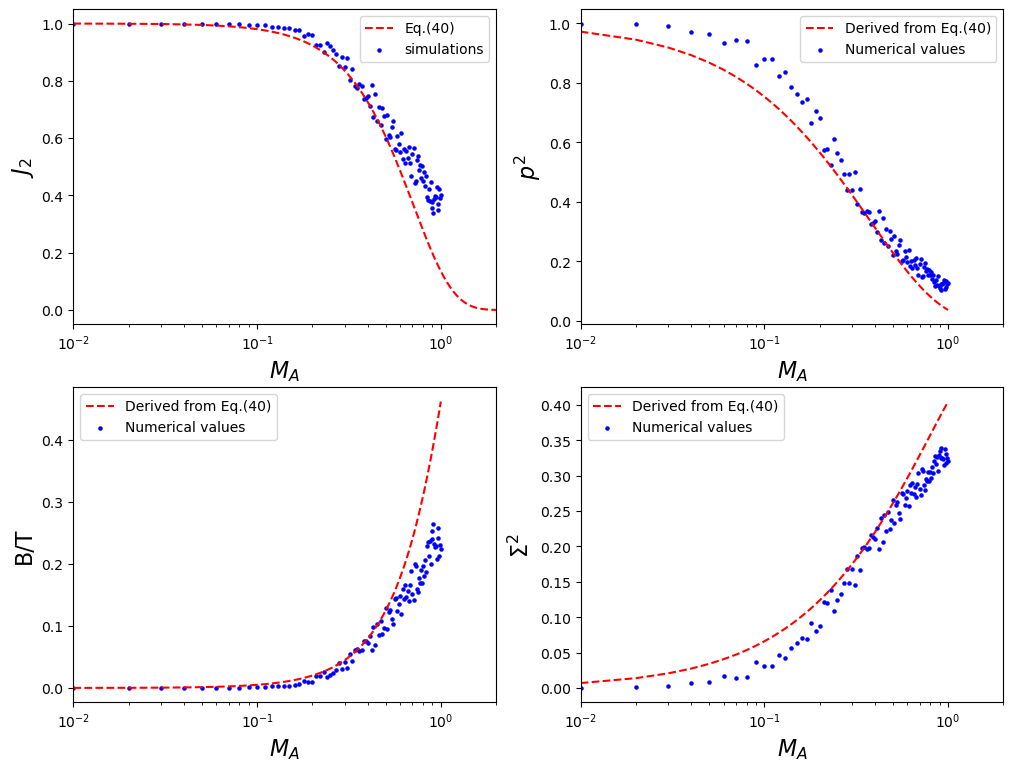}
\caption{ A figure comparing the theoretical expectation (red) to data points from synthetic simulations for velocity centroids with pure Alfven modes for all four observables as a function of $M_A$. From top left : $J_2$, $p^2$, $B/T$ and $\Sigma^2$. }
\label{fig:semi_analytical_J}
\end{figure*}

In Fig.~\ref{fig:semi_analytical_J} we show how our four observables discussed in \S \ref{sec:gradient_orientation} derived from the Alfven mode synthetic simulations are varied as a function of $M_A$. We plot the variations of these observables against the theoretical expectations that we did in \S \ref{sec:gradient_orientation}. We can see that in the case of pure Alfven modes, the synthetic data points follow the theoretical expectations very nicely. 

\begin{figure*}[th]
\includegraphics[width=0.99\textwidth]{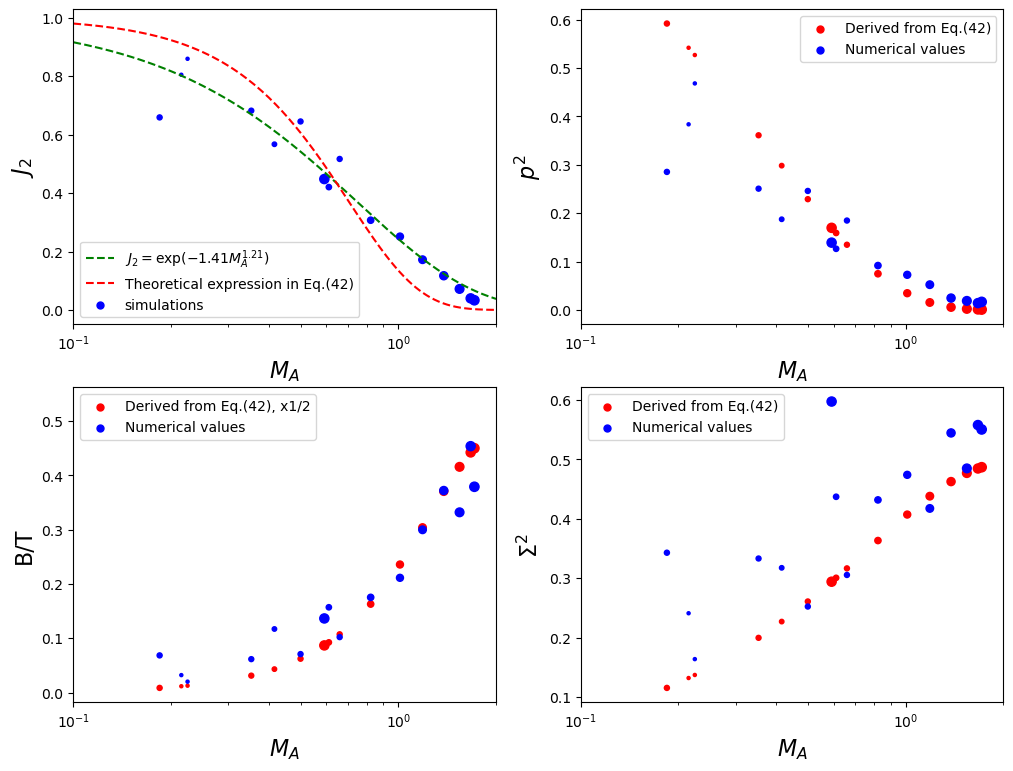}
\caption{Four panels showing our theoretical prediction (Eq.\ref{eq:J}) and its derivatives, red) relative to the actual numerical results (blue). For $J_2$ we fit the numerical results with $\exp(-CM_A^n)$ with the green curve.  }
\label{fig:parameters}
\end{figure*}

Furthermore, we show in Fig.~(\ref{fig:parameters}) on how the four parameters that we described in the current section vary as a function of $M_A$ in synthetic centroid maps from realistic numerical simulations with all non-Alfv{\'e}n mode filtered out and with $M_A<1$ and $M_s>1$. This is the condition where most molecular clouds are residing. To compare with the theoretical prediction, we plot the exact theoretical expectations following Eq.~(\ref{eq:J}). We can see that the numerical result (blue points in Fig.~\ref{fig:parameters} follows a very similar trend to the theoretical expectation.

\begin{comment}
Some challenges arise from the possible presence of the Fast modes. Fast modes, different from the Alfven modes, are only sensitive to the magnetic compressibility parameter $\beta$ instead of the Alfven Mach Number $M_A$. More importantly, the contributions of fast modes to $J_2$ are statistically orthogonal to that of Alfven modes.  This is illustrated in Fig.~\ref{fig:aspec} which shows angular dependence of the power spectrum
\begin{figure}
\centering
\includegraphics[width=0.49\textwidth]{angular_spectrum_x1.png}
\caption{The response of the angular spectrum when the Fast-to-Alfv{\'e}n Mode is varied in
realistic simulations. Left bump is due to fast modes, right peak is due to Alfb\'en modes}
\label{fig:aspec}
\end{figure}
in simulations that contain both Alfv\'en (and slow) and Fast modes.  We see clear double peak
structure of power alignment relative to magnetic field where the peak due to Fast modes is
shifted by $\pi/2$ over the peak dominated by Alfv\'en (and slow) modes.
In addition, the fast modes are less concentrated around the preferred direction.
As the result, addition of Fast modes in the areas of Alfv\'en mode dominance 
will isotropize the distribution of the gradients and
decrease $J_2$ measure, acting, to some extend, as an elevated noise contribution. Subtracting this 
contribution will require some input about the relative level of Fast modes to obtain an
unbiased estimate of the Alfv{\'e}nic Mach number $M_A$.
\end{comment}

\subsection{Gradients of Synchrotron Intensity}

For completeness, we would also like to discuss how $J_2$ varies if we consider the gradients of synchrotron intensity \citep{LYLC}. One particular difference between the gradients of centroid and the gradients of synchrotron intensities is that the former is a linear projection of the {\it line of sight} velocity fluctuations. At the same time, the latter is the quadratic projection on the plane of sky components. The tensor factor (see \citetalias{LP12}) is crucial in explaining the orientations of the gradients for the compressible modes. However, for the case of Alfven modes, the crucial factor is still the anisotropy factor we discussed in \S \ref{sec:Grad_cblock_theory}. We therefore expect the $J_2$ factor for synchrotron intensities to follow similarly to that of centroid.

\begin{figure*}[th]
\includegraphics[width=0.96\textwidth]{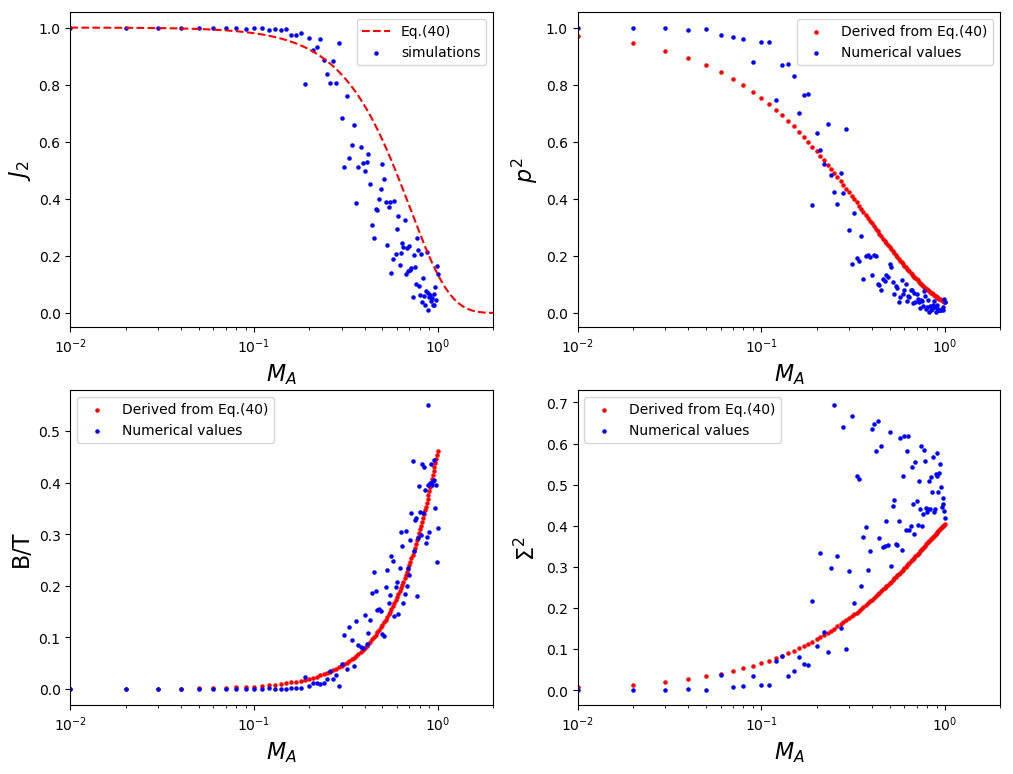}
\caption{Similar to Eq.\ref{fig:semi_analytical_J} but with synchrotron intensity gradients. From top left : $J_2$, $p^2$, $B/T$ and $\Sigma^2$.}
\label{fig:synchparameter}
\end{figure*}

To verify our idea, we perform the same 4-parameter test as in Fig.~\ref{fig:semi_analytical_J} by replacing the velocity centroid to synchrotron intensity maps from synthetic simulations. The results are shown in Fig.~\ref{fig:synchparameter}. We can see from this figure that all four parameters follow the theoretical expectations rather nicely, similar to that of the centroid.

\subsection{Variation of the $J_2$-$M_A$ relation in numerical data}

\begin{figure}
\centering
\includegraphics[width=0.48\textwidth]{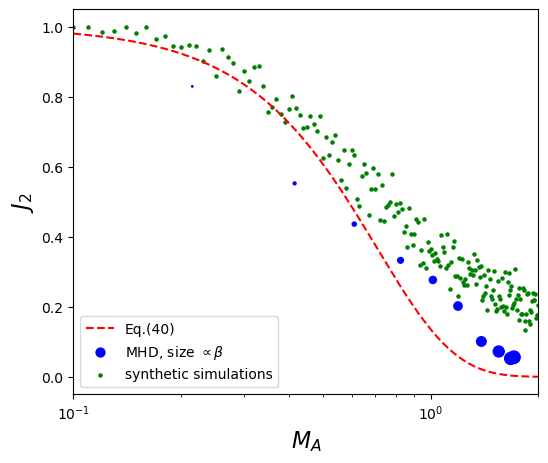}
\includegraphics[width=0.48\textwidth]{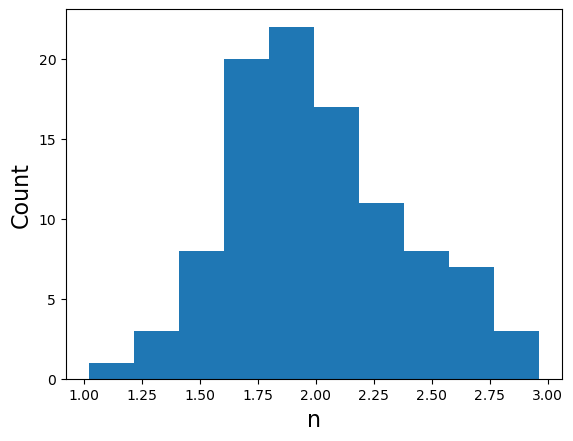}
\caption{(Top) A figure showing how the $J_2$ parameter computed from both synthetic (green points) and MHD (blue points) simulations are compared to the theoretical prediction in Eq.~(\ref{eq:J}), red dashed line). (Bottom) The Monte Carlo results in estimating the fit on synthetic data on the formula $J_2 = \exp(-CM^n)$, showing the mean value of $n$ to be $\sim2$. }
\label{fig:comparison_syn_mhd}
\end{figure}

From Figs.~\ref{fig:semi_analytical_J}--\ref{fig:synchparameter}, we observe a deviation between theory and observation even in the case of pure Alfv{\'e}nic modes. 
It could be questioned whether approximations, in particular a sample form of the power spectrum Eq.~(\ref{eq:Pk_convolution}) that led to Eq.~(\ref{eq:J}) are statistically accurate enough. To quantify the deviation between the analytical $J_2$ given by approximation of Eq.~(\ref{eq:J}) and numerics (both from synthetic models based on Eq.~(\ref{eq:power_spectrum}) and actual MHD simulations), let us assume that the actual relation between $J_2$ and $M_A$ can be parameterized as:
\begin{equation}
J_2  = \exp(-CM_A^n)
\label{eq:J2_model}
\end{equation}
for some constants $C$ and $n$. On the top panel of Fig.~\ref{fig:comparison_syn_mhd}, we plot the $J_2$ value from analytical (red dash line), synthetic (green dots), and MHD simulation data (blue dots) as a function of $M_A$. While theoretical approximation falls faster than synthetic and MHD simulations as $M_A$ approaches unity, both data resemble the analytical model of Eq.~(\ref{eq:J}). 

To quantify statistical correspondence further, we generate 100 samples of synthetic cubes at $M_A=0.1$ using Eq.~(\ref{eq:model}) and then perform an MCMC (Monte-Carlo Markov Chain) calculation on the synthetic data, assuming that Eq.~(\ref{eq:J2_model}) applies.  Each synthetic data cube gives a pair of values of $C,n$. The bottom panel of Fig.~\ref{fig:comparison_syn_mhd} shows the distribution of $n$ from the MCMC experiment, in which we select the number of bins to be the square root of the sample size (namely $10$) to satisfy the statistical criterion. The peak of the distribution is roughly at $\sim 2$, which is consistent with Eq.~(\ref{eq:J}). The dispersion of the values of $n$ is $\sim 0.4$, which provides an estimation of the accuracy of Eq.~(\ref{eq:J}) when applying to observations.

\section{Advances of theory and new opportunities}
\label{sec:newopp}

\subsection{The "kinematic" Alfv{\'e}nic Mach number under different driving cases}
\label{sec:alvenics}

Earlier in the paper, we focused on the Alfv{\'e}nic component of the MHD turbulence cascade. This treatment is feasible due to the relative independence of the Alfv{\'e}nic motions from the influence of other types of MHD fluctuations and the dominance of Alfv{\'e}nic fluctuations in determining the gradient properties. For some other applications, we must consider non-Alfv{\'e}nic types of turbulent motion. In this situation, the magnetic and velocity fluctuations are not connected through the Alfven relation, and the total injection velocity $\delta v_L$ is larger than the $V_{inj}$ that initiates the Alfv{\'e}nic cascade. 

The considerations above mean that the ratio of $\delta v_L/V_A$ is larger than the Alfven Mach number defined by Eq. (\ref{Alfven_Mach}), and this ratio defines another quantity, which we will term "kinetic" Mach number: 
\begin{equation}
    M_{A, k} \equiv \frac{\delta v_L}{V_A},
\end{equation}

A recent study in \cite{p_turb} of sub-Alfv{\'e}nic turbulence in incompressible limit revealed that the relation between $M_{A,k}$ and $M_A$ depends on whether the turbulence was driven by velocity or magnetic perturbation. In the former case, velocity components {\it parallel to the mean field} at the injection scale increase the kinetic energy of turbulent motions compared to the case of magnetic driving.  The physical effect is the natural consequence of fluid fluctuations along stochastic but stiff magnetic field lines. In the case of velocity-driven incompressible MHD turbulence, \cite{p_turb} derived that
\begin{equation}
    {\cal E}_b\approx M_A^2 {\cal E}_k ~,
    \label{energies}
\end{equation}
where ${\cal E}_b$ is the energy of magnetic turbulence and ${\cal E}_k$ is the kinetic turbulent energy. Therefore,
\begin{equation}
    M_{A}=M_{A,k}^2.
    \label{Machs}
\end{equation}

Relations similar to Eqs.~(\ref{energies}) and (\ref{Machs}) were empirically obtained in \cite{Bea22} using MHD compressible simulations, which suggests that the compressibility does not substantially the corresponding physics.\footnote{In \cite{Bea22} these expressions were interpreted as the consequence of the compressibility effects.} Thus, we expect these relations to hold approximately for various astrophysical conditions. 

If magnetic fluctuations drive the turbulence, \cite{p_turb} also showed that $M_{A,k}=M_A$ . The paper outlined a procedure for distinguishing the magnetically and velocity-driven turbulence. In many astrophysical settings, we expect the turbulence to be driven by velocity injection. 

The equations for sub-Alfv{\'e}nic turbulence in \citetalias{LV99} and subsequent studies (e.g., \citetalias{LP12}, \citealt{LP16}) assumed that the fluctuations at the injection scale are Alfv{\'e}nic, i.e., that the $M_A=M_{A,k}$ as in Eq. (\ref{Alfven_Mach}). 

\subsection{New technique in obtaining magnetic field strength independent of polarimetry: The MM2 approach}
\label{sec:VGTDMA}

Consider first the idealized  setting with pure Alfv{\'e}nic turbulence observed perpendicular to the mean magnetic field.
In this case of magnetic driving of turbulence, the magnetic field fluctuation $\delta \phi$ is directly related to the velocity Alfven Mach number:
\begin{equation}
    \delta \phi \approx M_{A}
\end{equation}
Thus, the classical Davis-Chandrasekhar-Fermi (DCF) expression \citep{1951PhRv...81..890D,CF53}  for determining magnetic field from observations can be rewritten as  
\begin{equation}
B_0 \approx \sqrt{4\pi \rho}\;\delta v M_{A}^{-1}.
\label{eq:B0viaMa}
\end{equation}

Astrophysical turbulence is a mixture of 3 cascades, fast, slow, and Alfven modes\footnote{We use the term ``modes'' rather than ``waves'', since in turbulence, the non-linear interactions are essential. These interactions make Alfv\'en and slow modes decay within one period.}, rather than pure Alvenic turbulence \citep{2002PhRvL..88x5001C,CL03}.
Accounting for different modes was done using the alternative technique of magnetic field strength study, differential measurement analysis (DMA), which can measure the distribution of magnetic field intensity (Lazarian et al. 2021). However, in this paper, we limit ourselves to the constraints of the traditional DCF, where no decomposition into modes is employed.

Let us now  eliminate the turbulent velocity term $\delta v$ using the expression of sonic Mach number $M_s = \delta v/c_s$, where $c_s$ is the sound speed,  obtaining instead of Eq.~(\ref{eq:B0viaMa})
\begin{equation}
    B_0 \approx  c_s \sqrt{4\pi \rho} M_s M_{A}^{-1},
    \label{eq:channelMM21}
\end{equation}
which is the expression obtained in Lazarian et al. (2020) within the approach termed there MM2 to denote the employment of two Mach numbers.

Our present study shows that gradient distribution directly measures from observations $M_{A, mes}=M_{A}/\sin\gamma$. Using this measure, in case of magnetic driving, one can rewrite Eq.~(\ref{eq:channelMM21}) as 
\begin{equation}
     B_\bot \approx  c_s \sqrt{4\pi \rho} M_s M_{A,mes}^{-1},
     \label{eq:MM2Bperp}
\end{equation}
where $B_\bot$ is the plane of the sky magnetic field component. 
Eq.~(\ref{eq:MM2Bperp}) can be used to find the perpendicular component of magnetic field for both velocity and magnetic field driving.

At first glance, nothing major has been achieved by substituting $M_s$ instead of $\delta v$. However, it is important to realize that $M_s$ can be measured differently. Focusing on gradients, $M_s$ can be obtained in spectroscopic channel maps measuring the amplitudes of gradients \citep{YL20}. This way of measuring magnetic field strength was used to obtain the 3D distribution of magnetic field strengths in the Milky Way using HI data, as demonstrated in 
\cite{Hu_mapping23}.  

For another example, \citet{2009ApJ...693..250B}  showed that $M_s$ can be obtained by measuring the kurtosis and skewness of the PDFs of the intensity fluctuations within the channel maps. This is important as in the galaxy, a significant line broadening arises from regular motions, e.g., differential rotation, not related to turbulence. The separation of the turbulent and regular motions may be difficult or even not possible. At the same time, the regular differential motion can distinguish the contribution along the line of sight, providing the 3D information. 

An approach similar to the one of \citet{2009ApJ...693..250B}  was employed for maps of synchrotron polarization gradients\footnote{These papers did not identify these gradients as the way of magnetic field studies. The latter step was done in \cite{LY18b}.} in \citet{2011Natur.478..214G}  and \citet{2012ApJ...749..145B}  also to constrain $M_s$. In the latter case, there is no linewidth information, but one can still obtain the value of $M_s$.

Other ways of obtaining $M_s$ from observational data include using the Tsallis statistics 
\citep{2005ApJ...631..320E,2010ApJ...710..125E}, 
genus \citep{2008ApJ...688.1021C}, with ongoing research opening up new possibilities for $M_s$ determination. Therefore, the MM2 approach utilizing two Mach numbers in Eq.~(\ref{eq:channelMM21})  provides yet another promising way of measuring the magnetic field strength. 

As discussed in \S \ref{sec:alvenics}, there are additional degrees of freedom depending on whether the turbulence is driven through magnetic or velocity fluctuations at the injection scale.  
As shown in Lazarian et al. (2024),  one can write a universal relation for magnetic field strength 
\begin{equation}
B_0 \approx \sqrt{4\pi \rho}\;\delta v M_{A,k}^{-1}.
\label{eq:B0viaMa2}
\end{equation}
where the "kinetic Alfv\'en Mach" number is employed. DCF Eq.~(\ref{eq:B0viaMa}) comes from equating  $M_{A,k}$ to $M_A=\frac{\delta B}{B}$, and obtaining the latter from 
polarization angle distribution. This is valid in case of magnetic driving. 
In the case of velocity driving, the "kinetic Alfven Mach number" $M_{A,k}$ can be determined from $M_A$ via Eq.~(\ref{Machs}) and the traditional DCF expression is modified\footnote{The empirical modification of this type was first suggested in \citet{2021A&A...647A.186S}.} to
\begin{equation}
B_0 \approx \sqrt{4\pi \rho}\;\delta v M_{A}^{-1/2}.
\label{eq:B0viaMa2_mod}
\end{equation}
Thus, if we deal with the case of velocity driving, Eq. (\ref{Machs}) should be used to relate $M_{A,k}$ to $M_{A,mes}$. The possibility of observationally determining the type of driving was discussed in \cite{p_turb}. 

\subsection{MM2 with different types of gradients}
\label{sec:variety}

Our present analytical study directly addresses two types of gradients: gradients of velocity centroids and gradients of synchrotron intensities. The extension of the theory to other types of gradients is straightforward, but is beyond the scope of our paper. The availability of analytical theory is very synergetic to our earlier attempts to gauge the properties of gradients, e.g. the relation between the statistics of gradients to the media magnetization, with numerical simulations in \cite{dispersion}. Both the analytical theory and the aforementioned gauging can be applied to a variety of data set within the MM2 approach. Several selected benefits of this are listed below.

The Synchrotron Intensity Gradients (SIGs) \citep{LYLC}  demonstrate the ability to trace the magnetic field using just the information of synchrotron intensities. Using the latter, the conventional way of obtaining the magnetic field strength is to assume the equipartition of cosmic rays and magnetic field energies. This is a very uncertain assumption, and there are many reasons why it can be violated \citep{2018MNRAS.473.4544S}. Obtaining the magnetic field with the MM2 removes this assumption and provides a unique way to explore the balance between the magnetic field and cosmic rays. 

The Synchrotron Polarization Gradients (SPGs) \citep{LY18b}  provide significant advantages compared to tracking magnetic fields with polarization directions. For instance, using the polarization measurements, the direction of the magnetic field can be obtained only after correcting for the Faraday rotation. Suppose the polarized radiation comes from an external source with no intrinsic Faraday rotation. In that case, the Faraday rotation can be accounted for using multiple polarization measurements at different wavelengths \citep[see][]{Beck15}. However, if the same volume is responsible for both the synchrotron emission and Faraday rotation, restoring the line of sight averaged magnetic field is impossible, even with multi-frequency polarization measurements. At the same time, the SPGs provide the ability to establish the POS magnetic field direction with the polarization data taken at a single frequency.\footnote{We note that the polarization data may not sample the entire volume due to the Faraday depolarization effect. Thus, the magnetic field obtained with SIGs and single-frequency SPG measurement may differ. This difference carries the information about the depolarization.} 
If, however, multi-frequency polarization measurements are available, the SPGs provide a way of restoring the 3D structure of a magnetic field, as was demonstrated in \cite{LY18polar} and \cite{HoL19polar}. As a bonus, by comparing the intensity of polarized radiation with the results obtained with MM2, one can estimate the 3D variations of the ratio of the magnetic field to the cosmic ray energies.

The particular advantages of applying the MM2 approach to different data sets
include
\begin{itemize}
\item for synchrotron intensity data, it is possible to decouple the contributions of magnetic field intensity and the relativistic electrons. 
\item for Faraday rotation data, it is possible to obtain the strength of the POS component of the magnetic field and decoupling the contribution of thermal electrons and a parallel component of the magnetic field to the rotation measures.
\item for synchrotron polarization multifrequency data, it is possible to provide the 3D tomography of magnetic field strength
\end{itemize}
The combination of the first 2 items allows to explore variations of POS magnetic field in different phases of multiphase media along the line of sight. This is synergetic with the input from item 3 in the list. 

\section{Discussion}
\label{sec:discussion}

 \subsection{Theoretical understanding of gradients}

The first major advance of our study above is the analytical descriptions of gradients.
The foundations of the GT are rooted in the theory of MHD turbulence and turbulent reconnection \citepalias{GS95,LV99} as well as in the analytical description of the statistics of the turbulence parameters available from observations \citep{LP00,LP04,LP06,LP08,LP12,LP16,KLP16,KLP17a}. The GT has already been proven to be a powerful way of tracing magnetic fields (see \citealt{YL17a,YL17b,LY18a,LY18b,dispersion, Hu_Snez21, Hu_mapping23}, The analytical theory formulated in this paper provides the solid foundations and justifications for the empirical procedures employed within the technique. It opens avenues for new approaches that increase the reliability of the gradient technique. For instance, the formulated theory allows us to understand how the distribution of the gradients over the sub-block used for obtaining individual gradient "vectors" depends on the media magnetization. Indeed, while an earlier study \cite{dispersion} empirically finds that the gradient dispersion has a power law relation to $M_A$, the power law index was not physically justified. In contrast, the present paper provides analytical predictions for the measures of gradient anisotropy  (c.f. Eqs.~(\ref{eq:J},\ref{eq:Sigma},\ref{eq:V},\ref{eq:BT})) that are available from observations.

Our present study deals with Alfven and slow mode fluctuations of sub-Alfv{\'e}nic at scales less than $LM_A^2$, at which a strong MHD turbulence model is applicable. The anisotropy related to fast modes is significantly weaker compared to Alfven modes, and fast modes' contribution is frequently subdominant in terms of energy. Thus, we do not consider the effect of fast modes within our present study.

Compared to \cite{dispersion}, our study extends the choice of measures that can be used to describe media magnetization. The ratio $F$ of the quadrupole $D_2$ and the monopole $D_0$ moments (see Eq. (\ref{F})) was calculated in \citetalias{LP12} for synchrotron intensity fluctuations and was generalized for in \citetalias{KLP16} and \citetalias{KLP17a}, respectively, for the intensity fluctuations in channel maps and velocity centroids. This ratio is directly related to the value of $J_2$ given by Eq.~(\ref{J2}) and is directly related to the measures available with gradients (see Eqs.~(\ref{eq:Sigma}), (\ref{eq:p2}), (\ref{eq:BT})). Our study demonstrates that different measures can have different advantages for practical applications. For instance, it is possible to show that for sufficiently small $M_A$, e.g., less than $0.3$, the changes of $F$ and $J_2$ are very weak with the change of $M_A$. At the same time, for this range of $M_A$, the differences in measures given by Eqs.~(\ref{eq:Sigma}), (\ref{eq:p2}), (\ref{eq:BT}) is easier to detect. 

Finally, the GT theory predicts differences in how gradients and polarization are summed up along the line of sight. In the presence of regular and random magnetic field components, the fluctuations in the polarization Stokes parameters add up in the random walk manner. In contrast, the regular magnetic field contribution increases linearly. As a result, the observed amplitude of the variations of the polarization direction decreases if the line of sight extension of the region exceeds the turbulence injection scale. In contrast, this effect is absent for the gradient magnetic field mapping. This difference stems from the fact that the regular field does not induce gradients. This is expected to induce a systematic difference between the magnetic field direction obtained with gradients and dust polarization using gradients \citep{2020MNRAS.496.2868L,2020ApJ...888...96H}. The analytical description of the gradients opens ways to quantitatively account for this difference.

\subsection{3D distribution of magnetic field strength}

The second advance achieved in the present study is the description of how gradients can provide magnetic field strength, i.e., the advancement of the MM2 technique. One of the important applications of this technique is mapping the 3D distribution of the galactic magnetic field vectors. This approach employs the galactic rotation curve as demonstrated in \cite{Hu_mapping23}. The mapping from the velocity space to the galactic coordinate is determined up to the turbulent velocity dispersion. The corresponding $\delta v\sim v_{turb}$ is the coarse grading for the line of sight $V_z$ to $z$ mapping.  In terms of molecular clouds, the galactic rotation curve opens ways of studying magnetic fields of molecular clouds in the galactic disk, i.e., the clouds for which crowding along the line of sight prevents magnetic field studies using traditional polarimetry.

The variety of species that can be used for studying magnetic fields by spectroscopic means must provide a complementary way of exploring the 3D magnetic field distribution. For instance, \cite{velac} demonstrated that the magnetic field direction and $M_A$ can be traced at different distances from the molecular cloud surface using the emission of different molecular species that form at different optical depths. A similar approach is possible for finding $M_s$ \citep{YL20}. Different species, molecules, and various ions can be employed to test the magnetic field in the ionized part of the ISM.

Measuring magnetic field with gradients using synchrotron intensity and synchrotron polarised intensity of Faraday measure input (see \S \ref{sec:variety}) provides additional synergy for the 3D magnetic field studies. For instance, synchrotron radiation from SIGs and SPGs mainly test magnetic fields in the hot and warm phases of the ISM at high latitudes (see the description of idealized ISM phases in \cite{2009ASPC..414..453D}), while Faraday rotation is biased to probing magnetic fields in the regions with higher thermal electron densities, e.g., HII regions. In addition, as described in \cite{LY18a}, measuring SPGs at different frequencies makes it possible to sample magnetic fields at different depths from the observer (see also \cite{HoL19polar}).

The MM2 approach can be applied to other non-spectroscopic data sets. For instance, the Intensity Gradients (IGs) \citep{LY18a,  HY19inten} deal with the column density data that is also affected by MHD turbulence. They can be used with emission lines without the required spectroscopic resolution. Due to the good coupling of interstellar gas and dust, they can also be used with the Far Infrared dust emission data. The IGs reflect the properties of density fluctuations in turbulent media. Those deviate significantly from the underlying turbulent velocity field in the case of supersonic MHD turbulence \citep[see][]{2005ApJ...624L..93B,2007ApJ...658..423K}. In this case, as first demonstrated by \cite{YL17a}, the IGs can trace interstellar shocks. This makes IGs an auxiliary source of information for restoring magnetic field direction and obtaining $M_A$ in combination with velocity gradients. For subsonic turbulence, IGs can be directly used to study magnetic fields within the MM2 approach.

The advances in the analytical description of turbulent statistics make the gradient approach more versatile.  In KLP18, it was shown that in the presence of dust absorption, the sampling of turbulence in the media is limited by the surface layer, the thickness of which is determined by the dust absorption. As this absorption is wavelength-dependent, it provides a way to sample deeper into the cloud using a longer wavelength. The ability to sample magnetic fields on the surface of the clouds using optical emission lines presents an advantage in itself.

\subsection{Gradients and polarization studies: advantages and synergy}

Polarization and gradient measurements present two major ways of probing POS magnetic fields. The present study illustrates the differencies and 
the synergy of the two approaches. 

Measurement of magnetic fields with gradients and polarization are similar because both trace the direction of the local magnetic field, and neither measure is sensitive to the amplitude of the magnetic field. Casting the gradient measure in terms of pseudo-Stokes parameters accumulated along the LOS \citep{2020MNRAS.496.2868L} highlights the
formal similarity.  However, there are two important differences.  First is that polarization directly reflects the mean magnetic field as well as its fluctuations. Whereas the gradients are not sensitive to constant field being directly affected only by the variable part of the field, so that the mean magnetic field direction is found from gradients only statistically.
Second difference is that polarization has a long correlation scale across the sky,
similar to magnetic field,  while the correlation length of the gradients is much shorter.

These differences affect the determination of $M_A$ from the data. The way of measuring $M_A$ with gradients was first explored in \cite{dispersion} and has shown several advantages compared to the corresponding measurements of $M_A$ using polarization. 
The most notable is that gradient measurement of $M_A$ can be achieved within an individual sub-block. Local determination of $M_A$ from polarization angles is hampered by the presence of
long correlations that requires sky coverage exceeding $L_{inj}$  to evaluate the full variance of the angles. 
Thus, applying the GT returns a spatial distribution of magnetization rather than a single value of $M_A$  from measuring dust polarization.

Additionaly, the presence of both regular and random magnetic fields along the line-of-sight decreases the dispersion of polarization angle 
$\delta \phi \sim \frac{\int^{\mathcal{L}} \delta B_\perp dz}{\overline{B}_\perp \mathcal{L}} $ by a factor $\sqrt{L_{inj}/{\cal L}}$ if the thickness of the turbulent cloud along the line of sight ${\cal L}$ is larger than the turbulent injections scale $L_{inj}$. On one hand this improves the
accuracy with which mean field direction is determined by polarization. On the other hand, this effect is not easy to quantify when analyzing the polarization angle variance, and it can significantly distort the value of $M_A$ obtained with the polarization measurements. On the contrary, the theory presented in this paper demonstrates that the gradients can return the true value of $M_A$. A comparison of the dispersion of the polarization and $M_A$ obtained with gradients can provide an estimate of $L_{inj}/{\cal L}$, which can be used to determine the turbulence injection scale $L_{inj}$.

Compared to dust polarization, the gradient  advantage is that they are unaffected by grain alignment efficiency variations and alignment direction variations that can be induced by strong radiation \citep{LazHoang07, Andersson15}. However, the outflows and gravitational collapse can switch the direction of gradients from the default perpendicular to the magnetic field to parallel to the field \citep{Huetal2020, HuL23}. The synergetic combination of polarization and gradients opens ways to decrease the uncertainties relevant to both techniques and get a unique insight into the underlying physical processes. 

A serious advantage of gradients compared to synchrotron polarization is that the gradients are unaffected by the Faraday rotation/depolarization \citep{Beck15}. By combining the gradients and synchrotron polarization, one can also get the Faraday rotation measure and, therefore, an estimate of the magnetic field component parallel to the line of sight $B_\|$ without involving troublesome multi-frequency measurements. In addition, by applying a gradient approach to different measures obtained from synchrotron polarization \citep{LYLC, ZhangL19, ZhangL20}, one can reconstruct the magnetic field distribution in 3D space \citep{HoL19polar}. 

The synergetic use of gradients and polarization is also possible for the CMB studies. The polarization gradients arising from the synchrotron foreground are aligned parallel to the foreground polarization. The polarization of cosmological origin is not expected to have this property. This can help solve the important problem of distinguishing the polarization of cosmological origin from foreground polarization.

\section{Conclusions}
\label{sec:conclusion}

This paper presents the theory of the Gradient Technique, which applies to different types of gradients originating in turbulent MHD flows. Our study applies the theory to gradients of velocity centroids and synchrotron intensities. However, this approach can be generalized to Velocity Channel Gradients (VChGs), Synchrotron Polarization Gradients (SPGs), Faraday Rotation Gradients (FRGs), etc.

The universality of the theory stems from the small size of the adopted sub-blocks of data averaging compared to the turbulence injection scale. In this setting, the effect of the correlations of the gradients along the line of sight is negligible, and the dispersion of gradients along the line of sight is determined by the dispersion of magnetic field directions on the injection scale rather than the statistics of emission that we employ to measure the gradients. The MHD turbulence cascade's Alfv{\'e}nic component determines the latter dispersion. The dispersion changes only marginally with the extent of the line of sight $L$ if $L>L_{inj}$. This is very different from the dispersion of the polarization "vectors," for which the dispersion gets narrow as $L$ gets larger compared to $L_{inj}$.  

The theory testifies that the properties of the distribution of gradients reflect the distribution of the POS component of magnetic field for $M_A/sin\gamma<1$, where $\gamma$ is the angle between the line of sight and the mean magnetic field. The deviations increase when the aforementioned ratio gets larger than unity as the fluctuating magnetic field gets larger than the POS component of the mean magnetic field. 

The observed gradient directions are dominated by the largest spacial frequencies of the image, with the gradients calculated on the smallest resolved scales dominating the resulting directions. The smallest eddies in MHD turbulence closely follow the direction of the magnetic field in their vicinity. Therefore, similar to the polarization from aligned dust, the gradients can provide the image of the magnetic field projected along the line of sight.  

In terms of practical tracing of magnetic fields and the distribution of magnetization, the theory demonstrates that 
\begin{itemize}
\item It is possible to obtain the detailed distribution of the media magnetization. 
\item The obtained magnetization value is robust and changes only marginally with the length of the line of sight $L$ when $L\gg L_{inj}$.
\end{itemize}

In addition, in the paper, we explored a new way of probing the strength of POS magnetic field by using the ratio of sonic and Alfv{\'e}n Mach numbers, i.e., $M_s$ and $M_{A,\bot}$. This technique can be applied to various data sets, including spectroscopic data, synchrotron intensity, polarization data, etc. When applied to spectroscopic data, MM2 provides
\begin{itemize}
    \item a detailed distribution of the plane of sky magnetic field strength;
    \item 3D distribution of plane of sky magnetic field galactic disk magnetic fields, if galactic rotation curve is employed;
    \item the 3D distribution $B_{\bot}$ strength in molecular clouds if a combination of emission lines from molecular species formed at different optical depths is used.
\end{itemize}
Our study ofor extending the MM2 technique for studying magnetic field strength using synchrotron intensity, synchrotron polarization gradients, and gradients, as well as density gradients.

{\noindent\bf Acknowledgments.} 
We cordially thank Jungyeon Cho, Chris Mckee and Martin Houde and  for detailed suggestions that significantly improved the manuscript. A.L. acknowledges the support the NSF AST 1816234, NASA TCAN 144AAG1967 and NASA ATP AAH7546. Flatiron Institute is supported by the Simons Foundation. This research used resources provided by the Los Alamos National Laboratory Institutional Computing Program, which is supported by the U.S. Department of Energy National Nuclear Security Administration under Contract No. 89233218CNA000001.  This research also used resources of the National Energy Research Scientific Computing Center (NERSC), a U.S. Department of Energy Office of Science User Facility located at Lawrence Berkeley National Laboratory, operated under Contract No. DE-AC02-05CH11231 using NERSC award FES-ERCAP-m4239 (PI: KHY, LANL). D.P. thanks Theoretical Group at Korea Astronomy and Space Science Institute (KASI) for hospitality. Flatiron Institute is supported by Simons Foundation. Research presented in this article was supported by the Laboratory Directed Research and Development program of Los Alamos National Laboratory under project number(s) 20220700PRD1. This research was supported in part by grant NSF PHY-2309135 to the Kavli Institute for Theoretical Physics (KITP). 

{\noindent \bf Data Availability} The data underlying this article will be shared on reasonable request to the corresponding author. The code can be found in \url{https://github.com/kyuen2/MHD_modes}

%\clearpage
\appendix

\section{Gradient variance and its dependence on $M_a$ and $\gamma$}
\label{app:grad_theory}

\subsection{Gradient methods to study magnetic field}
\label{sub:techinique}

Let us briefly summarize the mathematical foundations of the gradient methods in application to study of the direction of the magnetic field. 
The formalism was developed in Fourier space in \cite{2020MNRAS.496.2868L}. Here we complement the exposition by real space treatment, as well as add more details.

Let us denote by $f(\mathbf{X})$ a sky map of observable emission from a turbulent medium.  For example, this can be the intensity of the emission in PPV (position-position $\mathbf{X}$-velocity $v$) space $I(\mathbf{X},v)$, or the related integrated quantities, such as the total intensity $I_c(\mathbf{X}) \propto \int dv I(\mathbf{X},v)$ or velocity centroids which for optically thin lines are 
$VC(\mathbf{X}) \propto \int v dv I(\mathbf{X},v) $.

The simplest local statistical measure for the gradient of a random field 
$f(\mathbf{X})$ is the gradient covariance tensor
\begin{equation}
\sigma_{i j} \equiv \left\langle \nabla_i f(\mathbf{X}) \nabla_j f(\mathbf{X}) \right\rangle
= \nabla_i \nabla_j D(\mathbf{R})|_{\mathbf{R} \to 0}
\end{equation}
which is the zero separation limit of the second derivatives
of the field structure function 
$D(\mathbf{R})\equiv \frac{1}{2} \left\langle\left( f(\mathbf{X+R}) - f(\mathbf{X})\right)^2\right\rangle$.

For a statistically isotropic field, the covariance
of the gradients is isotropic, $\sigma_{\nabla_i \nabla_j} = \frac{1}{2} \delta_{ij} \Delta D(R) | _{R\to 0}$.
In the presence of the magnetic field, the structure function
of the signal becomes orientation dependent, depending
on the angle between $\mathbf{R}$ and the projected direction of the magnetic field. This anisotropy is retained in the limit $\mathbf{R} \to 0$ and results in non-vanishing  traceless part of the gradient covariance tensor
\begin{equation}
\sigma_{ij} - \frac{1}{2} \delta_{ij} \sum_{l=1,2} \sigma_{l l } = \frac{1}{2}\left(
\begin{array}{cc}
\left(\nabla_x^2  - \nabla_y^2\right)  D(\mathbf{R})
& 2 \nabla_x \nabla_y  D(\mathbf{R}) \\
2 \nabla_x \nabla_y  D(\mathbf{R}) &
\left(\nabla_y^2 - \nabla_x^2\right)  D(\mathbf{R})\\
\end{array}
\right)_{\mathbf{R} \to 0} 
\ne 0
\end{equation}

As a rank two symmetric tensor, gradient covariance has two rotational invariants, the trace and the 
determinant,   or, correspondingly, the determinant of the traceless part. Thus,  in our case we can introduce invariant $I_0$ and $I_2$
\begin{align}
\sigma_{xx} + \sigma_{yy}  = I_0 \\
(\sigma_{xx}-\sigma_{yy})^2 + 4 \sigma_{xy}^2  = I_2^2
 \end{align}
Their dimensionless ratio is important as a measure of anisotropy level
 \begin{equation}
 J_2 \equiv \frac{ (\sigma_{xx}-\sigma_{yy})^2 + 4 \sigma_{xy}^2}{(\sigma_{xx} + \sigma_{yy})^2} = \left( \frac{I_2}{I_0}\right)^2 
 \end{equation}

We now study anisotropy by decomposing the structure function in angular harmonics, which can be done either in Fourier space or real space.

\subsubsection{Fourier space evaluation of gradient covariance tensor}
In Fourier space, where 
\begin{equation}
D(\mathbf{R}) = \int d\mathbf{K} \; P(\mathbf{K}) \left[1 - e^{i \mathbf{K} \cdot \mathbf{R}}\right] ~,
\end{equation}
the angular decomposition is over the dependence of the power spectrum $P(\mathbf{K})$
on the angle of the 2D wave vector $\mathbf{K}$.
Denoting the coordinate angle of $\mathbf{K}$ by $\theta_K$ and that of the projected magnetic field as $\theta_H$, we have for the spectrum
\begin{equation}
P(\mathbf{K}) = \sum_n P_n(K) e^{i n (\theta_H - \theta_K)}
\end{equation}
and for the derivatives of the structure function
\begin{equation}
\nabla_i \nabla_j D(\mathbf{R}) =  \sum_n \int K^3 P_n(K)
\int d\theta_K e^{i n (\theta_H - \theta_K)} 
e^{i K R \cos(\theta_R - \theta_K)} \hat K_i \hat K_j ~,
\end{equation}
where hat designates unit vectors, namely $\hat K_x = \cos \theta_K$ and $\hat K_y = \sin \theta_K$. Performing integration over
$\theta_K$, we obtain the traceless anisotropic part that involves Bessel functions $J_n(K R)$
\begin{equation}
\begin{aligned}
(\nabla_x^2 - \nabla_y^2)  D(\mathbf{R}) & =  \pi \sum_n i^n e^{ i n (\theta_H-\theta_R)} 
\int dK K^3 J_n(k R) \left( P_{n-2}(K) e^{-i 2 \theta_H} + P_{n+2}(K) e^{i 2 \theta_H} \right) \\
\nabla_x \nabla_y D(\mathbf{R}) & = \frac{\pi}{2 i} \sum_n i^{n}
e^{ i n (\theta_H-\theta_R)} \int dK K^3 J_n(k R) \left(-P_{n-2}(K) e^{-i 2 \theta_H} + P_{n+2}(K) e^{i 2 \theta_H} \right)
\end{aligned}
\end{equation}
In the limit $R \to 0$, only $n=0$ term for which $J_0(0)=1$ survives, $\theta_R$ dependence drops out,  and we have
\begin{equation}
\begin{aligned}
(\nabla_x^2 - \nabla_y^2) D(\mathbf{R}) & = 
\left[ 2 \pi  \int dK K^3 P_2(K) \right] \cos 2 \theta_H  \\
2 \nabla_x \nabla_y D(\mathbf{R}) & = \left[
2 \pi \int dK K^3 P_2(K)  \right] \sin 2 \theta_H
\end{aligned}
\end{equation}
We can now identify $I_2$ invariants with the integral over spectrum multipoles
\begin{equation}
I_2 = 2 \pi \int dK K^3 P_2(K)
\end{equation}
Notice that anisotropy of the gradient variance is determined by the quadrupole of the power spectrum as expected.

\subsubsection{Real space approach to evaluating gradient covariance tensor}
Now we demonstrate how gradient convariance tensor
can be evaluated in the real space to link it explicitly to the multipole decompositionstructure-functionfunction that can be performed locally.

We need to compute $R \to 0$ limit of the structure function derivatives
that we represent in polar coordinates as
\begin{equation}
\begin{aligned}
(\nabla_x^2 - \nabla_y^2) & D(R,\theta_R) = \cos2\theta_R 
\left( R \frac{\partial}{\partial R} 
\frac{1}{R} \frac{\partial}{\partial R} - \frac{1}{R^2} \frac{\partial^2}{\partial \theta_R^2}
\right) D(R,\theta_R)
- \sin2\theta_R \frac{\partial}{\partial R} \left(
\frac{2}{R} \frac{\partial}{\partial \theta_R} \right) 
D(R,\theta_R) 
\\
2 \nabla_x \nabla_y & D(R,\theta_R)  = 
\sin2\theta_R \left( R \frac{\partial}{\partial R} 
\frac{1}{R} \frac{\partial}{\partial R} - 
\frac{1}{R^2} \frac{\partial^2}{\partial \theta_R^2}
\right) D(R,\theta_R)
+ \cos2\theta_R \frac{\partial}{\partial R} \left(
\frac{2}{R} \frac{\partial}{\partial \theta_R} \right) 
D(R,\theta_R) \\
(\nabla_x^2 + \nabla_y^2) & D(R,\theta_R)  =  \left( \frac{1}{R} \frac{\partial}{\partial R} 
R \frac{\partial}{\partial R} + 
\frac{1}{R^2} \frac{\partial^2}{\partial \theta_R^2}
\right) D(R,\theta_R)
\end{aligned}
\label{eq:realspacesigma}
\end{equation}
where we added the trace for completeness, but also for the future use.

We present $D(R,\theta_R)$ as an expansion in the eigenfunctions
of the Laplace operator, $\Delta f_{Kn}(R,\theta) = -K^2 f_{Kn}(R,\theta) $,
subject to the boundary condition $D(0,\theta_R)=0$.
Choosing angular directions in respect to symmetry direction $\theta_H$ we have
multipole expansion
\begin{align}
D(R,\theta_R)  & = \sum_n D_n(R) e^{\mathrm{i} n (\theta_R - \theta_H)} 
%\label{eq:DnR}
\end{align}
with radial functions expanded in the Bessel function series
\begin{align}
D_n(R)  = D_{-n}(R) & = 2 \pi \int K dK P_n(K) \left( \delta_{n0}-\mathrm{i}^n J_{|n|}( K R)  \right)
\label{eq:DnR}
\end{align}
where $\delta_{n0}$ term enforces the boundary condition. 
Note, that though it looks very much as a Fourier expansion,  here we can have
a local expansion of the structure function in the  vicinity of some
fixed point in space (from which $R$ separation is measured) instead of global expansion 
into plane waves. Thus, we can
use this expansion in a statistically inhomogeneous situation, one patch at a
time, especially that we are interested in $R \to 0$ limit.

We need to determine the action of three differential operators on
the multipole coefficients $D_n(R)$ which we expect to give
\begin{equation}
\begin{aligned}
\lim_{R \to 0}
\left( R \frac{\partial}{\partial R} 
\frac{1}{R} \frac{\partial}{\partial R} + 
\frac{n^2}{R^2} \right) &  D_n(R) = 
 \frac{1}{2} \left( \delta_{n,2} + \delta_{n,-2} \right) I_2
%\label{eq:lim1}
\\
\lim_{R \to 0}
2 \mathrm{i} \; n \frac{\partial}{\partial R} 
\frac{1}{R} & D_n(R) = 
\frac{\mathrm{i}}{2} \left( \delta_{n,2} - \delta_{n,-2} \right)   I_2
%\label{eq:lim2}
\\
\lim_{R \to 0}
\left( \frac{1}{R} \frac{\partial}{\partial R} 
R \frac{\partial}{\partial R} - 
\frac{n^2}{R^2} \right) & D_n(R)
= \delta_{n0} I_0
\label{eq:lim3}
\end{aligned}
\end{equation}
To prove these relations and determine $I_0$ and $I_2$, we differentiate the Bessel functions as $J_n(x)^\prime = -J_{n+1}(x) + n J_n(x)/x$ 
and, if needed, use the recurrence relation in the direction of eliminating the divergent terms
as we take the limit $x \to 0$.
 Our first combination is
\begin{align}
 \frac{\partial}{\partial R}  \frac{ J_n(K R)}{R} & = \frac{K^2}{x} \left( J_n(x)^\prime - J_n(x)/x\right)
 \nonumber \\
 & =   K^2 \left( -\frac{J_{n+1}(x)}{x}+\frac{n-1}{x^2}  J_n(x)\right) 
 \nonumber \\
 & \to \frac{1}{8} K^2 \delta_{n,2}  - K^2 \delta_{n0} \left( \frac{1}{2}+  \frac{J_0}{x^2} \right)
 \label{eq:deriv1}
 \end{align}
 where divergent at $R \to 0$ $n=0$ term does not contribute to the final result since we multiply
 this derivative by $n$.
 
The other combination is 
\begin{align}
 \left( R \frac{\partial}{\partial R}  
\frac{1}{R} \frac{\partial}{\partial R} + 
\frac{n^2}{R^2} \right)  J_n(K R) & =  K^2\left( x \frac{\partial}{\partial x} \frac{1}{x}
\left( -J_{n+1}(x) + \frac{n}{x} J_n(x)\right) + \frac{n^2}{x^2} J_n(x)
\right)
 \nonumber \\
 & =  K^2 \left(J_{n+2}(x)
 -\frac{2 n}{x}J_{n+1}(x) + \frac{2 n(n-1)}{x^2} J_n(x) \right) 
 \nonumber \\
 & \to  \frac{1}{2} K^2 \delta_{n,2}
 \label{eq:deriv2}
 \end{align}
 
For the last combination (which is the Laplacian)  the answer is already known
\begin{equation}
  \left( \frac{1}{R} \frac{\partial}{\partial R}  
R \frac{\partial}{\partial R} - \frac{n^2}{R^2} \right)  J_n(K R)  =  - K^2 J_n(x) \to -K^2 \delta_{n,0}
\label{eq:deriv3}
 \end{equation}
Using Eqs.~(\ref{eq:deriv1}-\ref{eq:deriv3}) leads to the result Eq.~(\ref{eq:lim3})
with $I_n = 2 \pi \int dK K^3 P_n(K)$.  Taking $R \to 0$ limit in Eq.~(\ref{eq:DnR}) shows that in the real space formulation
the invariants are given by
\begin{align}
I_0 &= 4 \lim_{R \to 0} D_0(R)/R^2 \\
I_2 &= 8 \lim_{R \to 0} D_2(R)/R^2.
\end{align}

Finally, substituting Eqs.~(\ref{eq:lim3}) into Eq.~(\ref{eq:DnR}) and then Eqs.~(\ref{eq:realspacesigma}) 
we confirm that the dependence on the position vector angle cancels in the covariance matrix as it takes the expected form
\begin{equation}
\begin{aligned}
\sigma_{xx}-\sigma_{yy} & =\frac{I_2}{2} \left( \cos2\theta_R
 \left( e^{2 \mathrm{i} (\theta_R -\theta_H)}  +  e^{-2 \mathrm{i} (\theta_R -\theta_H)}
\right)
- \mathrm{i} \sin2\theta_R
\left( e^{2 \mathrm{i} (\theta_R -\theta_H)}  -  e^{-2 \mathrm{i} (\theta_R -\theta_H)}
\right)
\right)
=  I_2 \cos 2 \theta_H \\
2 \sigma_{xy} & = \frac{I_2}{2} 
\left(
\sin2\theta_R  \left( e^{2 \mathrm{i} (\theta_R -\theta_H)}  +  e^{-2 \mathrm{i} (\theta_R -\theta_H)}
\right)
+ \mathrm{i} \cos2\theta_R \left( e^{2 \mathrm{i} (\theta_R -\theta_H)}  -  e^{-2 \mathrm{i} (\theta_R -\theta_H)}
\right)
\right)
= I_2 \sin 2 \theta_H \\
\sigma_{xx}+\sigma_{yy} &  =  I_0
\end{aligned}
\end{equation}

\subsubsection{Eigendirections of the gradient covariance tensor}
The eigendirection of the covariance tensor that corresponds to the largest eigenvalue (``the direction of the gradient'') makes 
an angle $\theta$ with the coordinate x-axis
\begin{equation}
\label{eq:theta_grad}
\tan \theta =
\frac{ 2 \nabla_x \nabla_y D}
{\sqrt{\left(\nabla_x^2 D - \nabla_y^2 D \right)^2 + 4 \left( \nabla_x \nabla_y D \right)^2}+\left( \nabla_x^2 - \nabla_y^2 \right)D}~.
\end{equation}
Substituting expressions for derivatives into this formula
we find that the eigendirection of the gradient variance has the form
\begin{equation}
\tan \theta = \frac{ A \sin 2 \theta_H}{ |A| + A \cos 2 \theta_H} =
\left\{
\begin{array}{rl}
\tan \theta_H & A > 0 \\
-\cot \theta_H & A < 0 
\end{array}
\right.
\end{equation}
and is either parallel or perpendicular to the direction of the magnetic field,
depending on the sign of $A \propto \int dK K^3 P_2(K) $, i.e the sign of the spectral quadrupole $P_2$.

Since the direction of the magnetic field that we aim to track is unsigned,
it is appropriate to describe it as an eigendirection of the rank-2 tensor, rather than a vector.  This naturally leads to the mathematical formalism of Stokes parameters. We can introduce local pseudo-Stokes parameters built from the field gradients
\begin{align}
\widetilde{Q} & \propto (\nabla_x f)^2 - (\nabla_y f)^2 \\
\widetilde{U} & \propto 2 \nabla_x f \nabla_y f 
\end{align}
which averages are directly linked to statistics of the gradients
\begin{align}
\langle \widetilde{Q} \rangle & \propto \sigma_{xx}^2 - \sigma_{yy}^2 = I_2 \cos 2 \theta_H \\
\langle \widetilde{U} \rangle & \propto 2 \sigma_{xy}^2 = I_2 \sin 2 \theta_H
\end{align}
and give the estimate of the direction of the magnetic field
\begin{equation}
\frac{\langle \widetilde{U} \rangle}{\langle \widetilde{Q} \rangle } = \tan 2 \theta_H
\end{equation}

The pseudo Stokes parameters naturally connect the gradient techniques with
polarization studies.
\subsection{Distribution of gradient angles, $J_2$ as a measure of anisotropy}
\label{sec:finetobigpix}

As a model distribution of the gradient angle evaluate in the pixel $p$, $\theta_p = \mathrm{atan}\left[\nabla_y I/\nabla_x  I\right]$, we take the distribution that follows from assuming the gradients to be Gaussian with the covariance matrix $\widetilde{\sigma}_{ij}$ of Eq.\ref{eq:gradcov}. The expression for this distribution can be easily obtained as:
\begin{align}
\label{eq:Gaussian-grads}
P(\theta_p) &= 
\frac{1}{\pi \sqrt{|\sigma_{ij}|}} 
\left[
\left(
\begin{array}{c}
\cos\theta_p \\
\sin\theta_p      
\end{array}
\right) 
{\sigma}_{ij}^{-1}
\left(
\begin{array}{c}
\cos\theta_p \\
\sin\theta_p      
\end{array}
\right)
\right]^{-1} 
\\
%\\
&= \frac{1}{\pi} \times \frac{ 
\sqrt{1 - J_2}}
{1- \sqrt{J_2} \cos2 \left({\theta}_H-\theta_p\right)}
\end{align}
This distribution function has two parameters: $J_2$ and ${\theta}_H$. The first is the rotation invariant ratio of the determinant of the traceless part of the covariance matrix and the (half of) trace of the covariance
\begin{equation}
J_2 \equiv \frac{\left({\sigma}_{xx}-{\sigma}_{yy}\right)^2 + 4 {\sigma}_{xy}^2}
{\left({\sigma}_{xx}+{\sigma}_{yy}\right)^2}=
 4 \left( \lim_{R \to 0} \frac{D_2(R)}{D_0(R)} \right)^2
\end{equation}
and the second is the angle
\begin{equation}
\tan 2 {\theta}_H \equiv \frac{2 {\sigma}_{xy}}{{\sigma}_{xx}-{\sigma}_{yy}}
\end{equation}
The distribution in  Eq.\eqref{eq:Gaussian-grads} is periodic with a period $\pi$ and is normalized to unity on any angular interval of the length of the period, $\int_{\theta_p^*}^{\theta_p^*+\pi} P(\theta_p) d\theta_p = 1$.  Statistically isotropic case
corresponds to ${\sigma}_{xx}={\sigma}_{yy}$ and ${\sigma}_{xy}=0$, i.e $J_2=0$ when
Eq.\eqref{eq:Gaussian-grads} evaluates to uniform distribution.
Any anisotropy leads to non-zero $J_2 > 0$, which on the other hand is bounded
by definition not to exceed unity, $J_2 \le 1$.
Note that fitting angular distribution does not determine the trace of gradient covariance $I_0$.

\section{Additional tools complementary to the current paper}
\label{without}

Most of the current paper describes how to estimate magnetic field within the framework of the Velocity Gradient Technique (VGT). However, the techniques that are developed in the current paper does not require the prediction from the gradient technique to work. In the following, we discuss the complementary techniques that can help us to construct the 3D distribution of magnetic field.

\subsection{Obtaining $M_s$, $M_A$ and the inclination angle $\gamma$}

\noindent{\bf Alfv{\'e}nic Mach number $M_A$}: For instance, the ability of various techniques to estimate the values of $M_A$ and $M_s$ from observational data were studied in literature prior to the time of GT appeared. For instance, the ability of Tsallis statistics to obtain $M_A$ was demonstrated in \cite{2010ApJ...710..125E} and \cite{2011ApJ...736...60T}. In addition, \cite{2005ApJ...631..320E} obtained empirical relations between the anisotropy of the structure function and $M_A$. In fact, the first demonstration of the effect of magnetic field strength on the anisotropy of correlations in the channel maps was presented in \cite{2002ASPC..276..182L}. The analytical relations between $M_A$ the ratio of the quadropole to monopole moments of the structure functions of intensity was obtained in \citetalias{LP12} for synchrotron emission and in \citetalias{KLP17a},\cite{KLP17b} and \cite{KLP18} for the spectroscopic data. The ability of obtaining the mean $M_A$ was demonstrated with polarization data in \cite{dispersion}, while the statistics of $M_A$ was carefully studied in \cite{curvature}.

\noindent{\bf Sonic Mach number $M_s$}: It its turn, $M_s$ was obtained via the analysis of kurtosis and skewness of the emission and polarization observational data (see \citealt{2009ApJ...693..250B,2012ApJ...755L..19B,2013ApJ...771..123B,2014ApJ...790..130B}). Combining thus found $M_s$  with the $M_A$ obtained with the approaches above ,one can successfully obtain the strength of the POS component of the magnetic field.  The results of $M_s$ measurements are very robust, and they are marginally affected either by shear or by large-scale magnetic field curvature. Similarly, the calculation of { $M_A$} is not much influenced by the galactic shear or any other large-scale shear induced by non-turbulent motions.

\noindent{\bf Inclination Angle $\gamma$}: To obtain the full magnetic field strength, an independent way of obtaining $\gamma$ is necessary.  Different measures depend differently on the aforementioned turbulence parameters and $\gamma$. This provides another avenue for obtaining the 3D vector of the magnetic field. Several approaches for obtaining $\gamma$ are suggested and explored in \cite{leakage,2023MNRAS.524.6102M,2024ApJ...965...65M} and \cite{Huetal23II, Huetal23III}. If combined with the value of the magnetic field obtained with the DMA approach, the output provides the full 3D magnetic field vector.

\subsection{\kh Removing density contributions when computing gradients}

As we discussed earlier, MHD turbulence theory is defined in terms of velocity and magnetic field fluctuations (see \citealt{2019tuma.book.....B}). The density can act either as a passive scalar or show its own non-universal scaling. The former is mostly the case of subsonic turbulence and this, as pointed out in \S 3.5 results in the density gradient technique. However, in general, one can get more reliable predictions if the density contributions are removed. 

{\kh Tt is worth emphasizing that the most of the previous predictions on Alfv{\'e}nic Mach number in both the LP-series or the KLP series assume a constant density and negligible thermal broadening effects. While the authors have discussed qualitatively how these effects change the analytical predictions, it is unavoidable to deal with both the density and thermal effects in regimes such as multiphase HI media. To remove the effects of density and thermal broadening from observed channels, \cite{VDA} proposes the {\bf Velocity Decomposition Algorithm (VDA)}, an algorithm that is built upon the foundation of \citetalias{LP00}, that applies to every channel map. The {\it velocity caustics} map $p_v$, which is the channel map without any density fluctuations, is given by:
\begin{equation}
  p_v = p - \left( \langle pI\rangle-\langle p\rangle\langle I \rangle\right)\frac{I-\langle I\rangle}{\sigma_I^2}
\label{eq:VDA_ld2}
\end{equation}
where $p$ is the channel map, $I$ is the column density map and $\langle ..\rangle$ is the averaging operator.  \cite{VDA} shows that the VDA method is extremely accurate reflecting the statistics of velocity fluctuations in sub-sonic media while being fairly accurate in supersonic media. For our current paper's purpose, we need the velocity centroid map with no density effects. The appendix of \cite{VDA} gives a formula for the constant density velocity centroid:
\begin{equation}
  V_{n}({\bf X}) = \int dv v p_v({\bf X},v) 
  \label{eq:C_VDA}
\end{equation}
which is exactly the quantity that is investigated in both \citetalias{KLP16} and \citetalias{KLP17a}. Note, that in this paper we use capital symbols to denote the 2D vectors, i.e. ${\bf X}$ is the 2D Plane of Sky (POS) vector. In below we are going to investigate the gradients of centroid or channels {\it assuming density being constant}. The exact formula in obtaining these observables are given by either Eq.\ref{eq:VDA_ld2} (channels) or Eq.\ref{eq:C_VDA} (centroids).

\subsection{Obtaining 3D magnetic field structures from divergence free condition}

It was discussed in \cite{LY18a} that the 3D distribution of POS magnetic field can be obtained applying velocity gradients to channel maps or reduced centroids and using the galactic rotation curve. This approach was applied in \cite{GL18} to map the POS magnetic field using gradients of the velocity centroids. The corresponding map of 3D distribution of the  POS magnetic field was then successfully tested by comparing the starlight polarization predicted for a number of stars with the known distances with the actual measured polarization. 

The ability of obtaining 3D POS B-field maps provide another way of obtaining the actual distribution of 3D magnetic field vectors. Indeed, the magnetic field is a solenoidal vector three components of which are constrained through the divergence free condition:
$
\nabla \cdot {\bf B}=0,
$
, meaning that that only 2 components in Fourier space are independent. Therefore if we are given 2 components, we can restore the 3d one. The corresponding procedure was suggested in \cite{2021arXiv210705070C} for restoring the z-component of magnetic field when two POS component are obtained using the \cite{LY18a} synchrotron gradient tomography that can restore the 3D POS magnetic field. 

We would like to stress that the same procedure is also applicable to studying magnetic fields using both the POS magnetic fields obtained using the channel maps or reduced centroids in the presence of galactic rotation curve. The caveat for the application of the technique is that the magnetic field is solenoidal in actual 3D space and can lose this property in other coordinates. For instance, the distortions of the 2D slice for which the POS magnetic field is determined can result in errors in determining the LOS component of magnetic field. This means that the POS for the procedure should be sufficiently coarse grained to avoid problems of the mapping.

\section{Numerical simulations employed for testing} 
\label{app:sim}

In Table \ref{tab:sim} we describe the set of numerical simulations that we employ to test our analytical results. 

\begin{table}[h]

\centering
\begin{tabular}{c c c c c | c c c c c}
Model & $M_S$ & $M_A$ & $\beta=2M_A^2/M_S^2$ & Resolution & Model & $M_S$ & $M_A$ & $\beta=2M_A^2/M_S^2$ & Resolution \\ \hline \hline
b11                     & 0.41  & 0.04 & 0.02 & $480^3$ &  huge-0                  & 6.17  & 0.22 & 0.0025 & $792^3$ \\  
b12                     & 0.92  & 0.09 & 0.02 & $480^3$ &  huge-1                  & 5.65  & 0.42 & 0.011 & $792^3$ \\  
b13                     & 1.95  & 0.18 & 0.02 & $480^3$ &  huge-2                  & 5.81  & 0.61 & 0.022 & $792^3$ \\  
b14                     & 3.88  & 0.35 & 0.02 & $480^3$ &  huge-3                  & 5.66  & 0.82 & 0.042 & $792^3$ \\  
b15                     & 7.14  & 0.66 & 0.02 & $480^3$ &  huge-4                  & 5.62  & 1.01 & 0.065 & $792^3$ \\   
b21                     & 0.47  & 0.15 & 0.22 & $480^3$ &  huge-5                  & 5.63  & 1.19 & 0.089 & $792^3$ \\  
b22                     & 0.98  & 0.32 & 0.22 & $480^3$ &  huge-6                  & 5.70  & 1.38 & 0.12 & $792^3$ \\  
b23                     & 1.92  & 0.59 & 0.22 & $480^3$ &  huge-7                  & 5.56  & 1.55 & 0.16 & $792^3$ \\   
b31                     & 0.48  & 0.48 & 2.0 & $480^3$  &  huge-8                  & 5.50  & 1.67 & 0.18 & $792^3$ \\ 
b32                     & 0.93  & 0.94 & 2.0 & $480^3$  &  huge-9                  & 5.39  & 1.71 & 0.20 & $792^3$ \\  
b41                    & 0.16  & 0.49 & 18 & $480^3$   & h0-1200                 & 6.36  & 0.22 & 0.0025 & $1200^3$ \\
b42                     & 0.34  & 1.11 & 18 & $480^3$    & h9-1200                 & 10.79 & 2.29 & 0.098 & $1200^3$ \\
b51                     & 0.05  & 0.52 & 200 & $480^3$   & e5r2                    & 0.13  & 5.99 & 4363 & $1200^3$ \\
b52                     & 0.10  & 1.08 & 200 & $480^3$   & e5r3                    & 0.61  & 0.63 & 2.09 & $1200^3$ \\
Ms0.2Ma0.2              & 0.2   & 0.2  & 2 & $480^3$   & e6r3                    & 5.45  & 0.50 & 0.017 & $1200^3$ \\
Ms0.4Ma0.2              & 0.57   & 0.28  & 0.48 & $480^3$   & e7r3                    & 0.53  & 3.24 & 73.64 & $1200^3$ \\ 
Ms4.0Ma0.2              & 3.81   & 0.18  & 0.00446 & $480^3$   & incompressible & 0 & 0.61 & $\infty$ & $512^3$\\
Ms20.0Ma0.2             & 20.59  & 0.18  & 0.00015 & $480^3$  &&&&& \\  \hline 
\hline
% \hline
\end{tabular}
\caption{\label{tab:sim} Description of MHD simulation cubes {which some of them have been used in the series of papers about VGT \citep{YL17a,YL17b,LY18a,LY18b}}.  $M_s$ and $M_A$ are the R.M.S values at each the snapshots are taken. The incompressible simulation is performed by \cite{2022zndo...8242702H}.}
\end{table}
%\clearpage
\bibliographystyle{aasjournal}

\bibliography{refs.bib}

\begin{thebibliography}{}
\expandafter\ifx\csname natexlab\endcsname\relax\def\natexlab#1{#1}\fi
\providecommand{\url}[1]{\href{#1}{#1}}
\providecommand{\dodoi}[1]{doi:~\href{http://doi.org/#1}{\nolinkurl{#1}}}
\providecommand{\doeprint}[1]{\href{http://ascl.net/#1}{\nolinkurl{http://ascl.net/#1}}}
\providecommand{\doarXiv}[1]{\href{https://arxiv.org/abs/#1}{\nolinkurl{https://arxiv.org/abs/#1}}}

\bibitem[{{Andersson} {et~al.}(2015{\natexlab{a}}){Andersson}, {Lazarian}, \&
  {Vaillancourt}}]{2015ARA&A..53..501A}
{Andersson}, B.~G., {Lazarian}, A., \& {Vaillancourt}, J.~E.
  2015{\natexlab{a}}, \araa, 53, 501,
  \dodoi{10.1146/annurev-astro-082214-122414}

\bibitem[{{Andersson} {et~al.}(2015{\natexlab{b}}){Andersson}, {Lazarian}, \&
  {Vaillancourt}}]{Andersson15}
---. 2015{\natexlab{b}}, \araa, 53, 501,
  \dodoi{10.1146/annurev-astro-082214-122414}

\bibitem[{{Armstrong} {et~al.}(1995){Armstrong}, {Rickett}, \&
  {Spangler}}]{1995ApJ...443..209A}
{Armstrong}, J.~W., {Rickett}, B.~J., \& {Spangler}, S.~R. 1995, \apj, 443,
  209, \dodoi{10.1086/175515}

\bibitem[{{Beattie} {et~al.}(2022){Beattie}, {Krumholz}, {Federrath},
  {Sampson}, \& {Crocker}}]{Bea22}
{Beattie}, J.~R., {Krumholz}, M.~R., {Federrath}, C., {Sampson}, M.~L., \&
  {Crocker}, R.~M. 2022, Frontiers in Astronomy and Space Sciences, 9, 900900,
  \dodoi{10.3389/fspas.2022.900900}

\bibitem[{{Beck}(2015)}]{Beck15}
{Beck}, R. 2015, in Astrophysics and Space Science Library, Vol. 407, Magnetic
  Fields in Diffuse Media, ed. A.~{Lazarian}, E.~M. {de Gouveia Dal Pino}, \&
  C.~{Melioli}, 507, \dodoi{10.1007/978-3-662-44625-6_18}

\bibitem[{Beresnyak \& Lazarian(2019)}]{BeresnyakLazarian+2019}
Beresnyak, A., \& Lazarian, A. 2019, Turbulence in Magnetohydrodynamics
  (Berlin, Boston: De Gruyter), \dodoi{doi:10.1515/9783110263282}

\bibitem[{{Beresnyak} \& {Lazarian}(2019)}]{2019tuma.book.....B}
{Beresnyak}, A., \& {Lazarian}, A. 2019, {Turbulence in Magnetohydrodynamics}

\bibitem[{{Beresnyak} {et~al.}(2005){Beresnyak}, {Lazarian}, \&
  {Cho}}]{2005ApJ...624L..93B}
{Beresnyak}, A., {Lazarian}, A., \& {Cho}, J. 2005, \apjl, 624, L93,
  \dodoi{10.1086/430702}

\bibitem[{{Burkhart} {et~al.}(2009){Burkhart}, {Falceta-Gon{\c{c}}alves},
  {Kowal}, \& {Lazarian}}]{2009ApJ...693..250B}
{Burkhart}, B., {Falceta-Gon{\c{c}}alves}, D., {Kowal}, G., \& {Lazarian}, A.
  2009, \apj, 693, 250, \dodoi{10.1088/0004-637X/693/1/250}

\bibitem[{{Burkhart} \& {Lazarian}(2012)}]{2012ApJ...755L..19B}
{Burkhart}, B., \& {Lazarian}, A. 2012, \apjl, 755, L19,
  \dodoi{10.1088/2041-8205/755/1/L19}

\bibitem[{{Burkhart} {et~al.}(2012){Burkhart}, {Lazarian}, \&
  {Gaensler}}]{2012ApJ...749..145B}
{Burkhart}, B., {Lazarian}, A., \& {Gaensler}, B.~M. 2012, \apj, 749, 145,
  \dodoi{10.1088/0004-637X/749/2/145}

\bibitem[{{Burkhart} {et~al.}(2014){Burkhart}, {Lazarian}, {Le{\~a}o}, {de
  Medeiros}, \& {Esquivel}}]{2014ApJ...790..130B}
{Burkhart}, B., {Lazarian}, A., {Le{\~a}o}, I.~C., {de Medeiros}, J.~R., \&
  {Esquivel}, A. 2014, \apj, 790, 130, \dodoi{10.1088/0004-637X/790/2/130}

\bibitem[{{Burkhart} {et~al.}(2013){Burkhart}, {Lazarian}, {Ossenkopf}, \&
  {Stutzki}}]{2013ApJ...771..123B}
{Burkhart}, B., {Lazarian}, A., {Ossenkopf}, V., \& {Stutzki}, J. 2013, \apj,
  771, 123, \dodoi{10.1088/0004-637X/771/2/123}

\bibitem[{{Carmo} {et~al.}(2020){Carmo}, {Gonz{\'a}lez-Casanova},
  {Falceta-Gon{\c{c}}alves}, {Lazarian}, {Jablonski}, {Zhang}, {Ferreira},
  {Castro}, \& {Yang}}]{2020ApJ...905..130C}
{Carmo}, L., {Gonz{\'a}lez-Casanova}, D.~F., {Falceta-Gon{\c{c}}alves}, D.,
  {et~al.} 2020, \apj, 905, 130, \dodoi{10.3847/1538-4357/abc331}

\bibitem[{{Chandrasekhar} \& {Fermi}(1953)}]{CF53}
{Chandrasekhar}, S., \& {Fermi}, E. 1953, \apj, 118, 113,
  \dodoi{10.1086/145731}

\bibitem[{{Chepurnov}(2021)}]{2021arXiv210705070C}
{Chepurnov}, A. 2021, arXiv e-prints, arXiv:2107.05070,
  \dodoi{10.48550/arXiv.2107.05070}

\bibitem[{{Chepurnov} {et~al.}(2015){Chepurnov}, {Burkhart}, {Lazarian}, \&
  {Stanimirovic}}]{2015ApJ...810...33C}
{Chepurnov}, A., {Burkhart}, B., {Lazarian}, A., \& {Stanimirovic}, S. 2015,
  \apj, 810, 33, \dodoi{10.1088/0004-637X/810/1/33}

\bibitem[{{Chepurnov} {et~al.}(2008){Chepurnov}, {Gordon}, {Lazarian}, \&
  {Stanimirovic}}]{2008ApJ...688.1021C}
{Chepurnov}, A., {Gordon}, J., {Lazarian}, A., \& {Stanimirovic}, S. 2008,
  \apj, 688, 1021, \dodoi{10.1086/591655}

\bibitem[{{Chepurnov} \& {Lazarian}(2009)}]{CL09}
{Chepurnov}, A., \& {Lazarian}, A. 2009, \apj, 693, 1074,
  \dodoi{10.1088/0004-637X/693/2/1074}

\bibitem[{{Chepurnov} \& {Lazarian}(2010)}]{2010ApJ...710..853C}
---. 2010, \apj, 710, 853, \dodoi{10.1088/0004-637X/710/1/853}

\bibitem[{{Cho} \& {Lazarian}(2002)}]{2002PhRvL..88x5001C}
{Cho}, J., \& {Lazarian}, A. 2002, \prl, 88, 245001,
  \dodoi{10.1103/PhysRevLett.88.245001}

\bibitem[{{Cho} \& {Lazarian}(2003)}]{CL03}
---. 2003, \mnras, 345, 325, \dodoi{10.1046/j.1365-8711.2003.06941.x}

\bibitem[{{Cho} {et~al.}(2002){Cho}, {Lazarian}, \&
  {Vishniac}}]{2002ApJ...564..291C}
{Cho}, J., {Lazarian}, A., \& {Vishniac}, E.~T. 2002, \apj, 564, 291,
  \dodoi{10.1086/324186}

\bibitem[{{Cho} \& {Vishniac}(2000)}]{CV00}
{Cho}, J., \& {Vishniac}, E.~T. 2000, \apj, 539, 273, \dodoi{10.1086/309213}

\bibitem[{{Cho} \& {Yoo}(2016)}]{CY16}
{Cho}, J., \& {Yoo}, H. 2016, \apj, 821, 21, \dodoi{10.3847/0004-637X/821/1/21}

\bibitem[{{Clarke}(2010)}]{2010ApJS..187..119C}
{Clarke}, D.~A. 2010, \apjs, 187, 119, \dodoi{10.1088/0067-0049/187/1/119}

\bibitem[{{Crutcher}(2004)}]{2004Ap&SS.292..225C}
{Crutcher}, R.~M. 2004, \apss, 292, 225,
  \dodoi{10.1023/B:ASTR.0000045021.42255.95}

\bibitem[{{Davis}(1951)}]{1951PhRv...81..890D}
{Davis}, L. 1951, Physical Review, 81, 890, \dodoi{10.1103/PhysRev.81.890.2}

\bibitem[{{Draine}(2009)}]{2009ASPC..414..453D}
{Draine}, B.~T. 2009, in Astronomical Society of the Pacific Conference Series,
  Vol. 414, Cosmic Dust - Near and Far, ed. T.~{Henning}, E.~{Gr{\"u}n}, \&
  J.~{Steinacker}, 453.
\newblock \doarXiv{0903.1658}

\bibitem[{{Elmegreen} \& {Scalo}(2004)}]{2004ARA&A..42..211E}
{Elmegreen}, B.~G., \& {Scalo}, J. 2004, \araa, 42, 211,
  \dodoi{10.1146/annurev.astro.41.011802.094859}

\bibitem[{{Esquivel} \& {Lazarian}(2005)}]{2005ApJ...631..320E}
{Esquivel}, A., \& {Lazarian}, A. 2005, \apj, 631, 320, \dodoi{10.1086/432458}

\bibitem[{{Esquivel} \& {Lazarian}(2010)}]{2010ApJ...710..125E}
---. 2010, \apj, 710, 125, \dodoi{10.1088/0004-637X/710/1/125}

\bibitem[{{Falceta-Gon{\c{c}}alves} {et~al.}(2008){Falceta-Gon{\c{c}}alves},
  {Lazarian}, \& {Kowal}}]{Fal08}
{Falceta-Gon{\c{c}}alves}, D., {Lazarian}, A., \& {Kowal}, G. 2008, \apj, 679,
  537, \dodoi{10.1086/587479}

\bibitem[{{Gaensler} {et~al.}(2011){Gaensler}, {Haverkorn}, {Burkhart},
  {Newton-McGee}, {Ekers}, {Lazarian}, {McClure-Griffiths}, {Robishaw},
  {Dickey}, \& {Green}}]{2011Natur.478..214G}
{Gaensler}, B.~M., {Haverkorn}, M., {Burkhart}, B., {et~al.} 2011, \nat, 478,
  214, \dodoi{10.1038/nature10446}

\bibitem[{{Galli} {et~al.}(2006){Galli}, {Lizano}, {Shu}, \&
  {Allen}}]{2006ApJ...647..374G}
{Galli}, D., {Lizano}, S., {Shu}, F.~H., \& {Allen}, A. 2006, \apj, 647, 374,
  \dodoi{10.1086/505257}

\bibitem[{{Galtier} {et~al.}(2000){Galtier}, {Nazarenko}, {Newell}, \&
  {Pouquet}}]{2000JPlPh..63..447G}
{Galtier}, S., {Nazarenko}, S.~V., {Newell}, A.~C., \& {Pouquet}, A. 2000,
  Journal of Plasma Physics, 63, 447, \dodoi{10.1017/S0022377899008284}

\bibitem[{{Girart} {et~al.}(2006){Girart}, {Rao}, \&
  {Marrone}}]{2006Sci...313..812G}
{Girart}, J.~M., {Rao}, R., \& {Marrone}, D.~P. 2006, Science, 313, 812,
  \dodoi{10.1126/science.1129093}

\bibitem[{{Goldreich} \& {Kylafis}(1981)}]{1981ApJ...243L..75G}
{Goldreich}, P., \& {Kylafis}, N.~D. 1981, \apjl, 243, L75,
  \dodoi{10.1086/183446}

\bibitem[{{Goldreich} \& {Sridhar}(1995)}]{GS95}
{Goldreich}, P., \& {Sridhar}, S. 1995, \apj, 438, 763, \dodoi{10.1086/175121}

\bibitem[{{Gonz{\'a}lez-Casanova} \& {Lazarian}(2017)}]{GCL17}
{Gonz{\'a}lez-Casanova}, D.~F., \& {Lazarian}, A. 2017, \apj, 835, 41,
  \dodoi{10.3847/1538-4357/835/1/41}

\bibitem[{{Gonz{\'a}lez-Casanova} \& {Lazarian}(2019)}]{GL18}
---. 2019, \apj, 874, 25, \dodoi{10.3847/1538-4357/ab0552}

\bibitem[{{Habegger} {et~al.}(2024){Habegger}, {Ho}, {Yuen}, \&
  {Zweibel}}]{2024arXiv240307976H}
{Habegger}, R., {Ho}, K.~W., {Yuen}, K.~H., \& {Zweibel}, E.~G. 2024, arXiv
  e-prints, arXiv:2403.07976, \dodoi{10.48550/arXiv.2403.07976}

\bibitem[{{Hayes} {et~al.}(2006){Hayes}, {Norman}, {Fiedler}, {Bordner}, {Li},
  {Clark}, {ud-Doula}, \& {Mac Low}}]{2006ApJS..165..188H}
{Hayes}, J.~C., {Norman}, M.~L., {Fiedler}, R.~A., {et~al.} 2006, \apjs, 165,
  188, \dodoi{10.1086/504594}

\bibitem[{{Heitsch} {et~al.}(2001){Heitsch}, {Zweibel}, {Mac Low}, {Li}, \&
  {Norman}}]{2001ApJ...561..800H}
{Heitsch}, F., {Zweibel}, E.~G., {Mac Low}, M.-M., {Li}, P., \& {Norman}, M.~L.
  2001, \apj, 561, 800, \dodoi{10.1086/323489}

\bibitem[{{Heyer} {et~al.}(2008){Heyer}, {Gong}, {Ostriker}, \&
  {Brunt}}]{2008ApJ...680..420H}
{Heyer}, M., {Gong}, H., {Ostriker}, E., \& {Brunt}, C. 2008, \apj, 680, 420,
  \dodoi{10.1086/587510}

\bibitem[{{Heyer} \& {Brunt}(2004)}]{2004ApJ...615L..45H}
{Heyer}, M.~H., \& {Brunt}, C.~M. 2004, \apjl, 615, L45, \dodoi{10.1086/425978}

\bibitem[{{Ho}(2022)}]{2022zndo...8242702H}
{Ho}, K.~W. 2022, {MHDFlows.jl}, v.0.2.1b,  Zenodo,
  \dodoi{10.5281/zenodo.8242702}

\bibitem[{{Ho} \& {Lazarian}(2021)}]{2021ApJ...911...53H}
{Ho}, K.~W., \& {Lazarian}, A. 2021, \apj, 911, 53,
  \dodoi{10.3847/1538-4357/abe713}

\bibitem[{{Ho} \& {Lazarian}(2023)}]{2023MNRAS.520.3857H}
---. 2023, \mnras, 520, 3857, \dodoi{10.1093/mnras/stad186}

\bibitem[{{Ho} {et~al.}(2023){Ho}, {Yuen}, \& {Lazarian}}]{instability}
{Ho}, K.~W., {Yuen}, K.~H., \& {Lazarian}, A. 2023, \mnras, 521, 230,
  \dodoi{10.1093/mnras/stad481}

\bibitem[{{Ho} {et~al.}(2019){Ho}, {Yuen}, {Leung}, \& {Lazarian}}]{HoL19polar}
{Ho}, K.~W., {Yuen}, K.~H., {Leung}, P.~K., \& {Lazarian}, A. 2019, \apj, 887,
  258, \dodoi{10.3847/1538-4357/ab578c}

\bibitem[{{Houde}(2004)}]{2004ApJ...616L.111H}
{Houde}, M. 2004, \apjl, 616, L111, \dodoi{10.1086/426684}

\bibitem[{{Hu} \& {Lazarian}(2023{\natexlab{a}})}]{Hu_mapping23}
{Hu}, Y., \& {Lazarian}, A. 2023{\natexlab{a}}, \mnras, 524, 2379,
  \dodoi{10.1093/mnras/stad1996}

\bibitem[{{Hu} \& {Lazarian}(2023{\natexlab{b}})}]{HuL23}
---. 2023{\natexlab{b}}, \mnras, 524, 4431, \dodoi{10.1093/mnras/stad2158}

\bibitem[{{Hu} \& {Lazarian}(2024)}]{Huetal23III}
---. 2024, arXiv e-prints, arXiv:2404.07806, \dodoi{10.48550/arXiv.2404.07806}

\bibitem[{{Hu} {et~al.}(2021){Hu}, {Lazarian}, \&
  {Stanimirovi{\'c}}}]{Hu_Snez21}
{Hu}, Y., {Lazarian}, A., \& {Stanimirovi{\'c}}, S. 2021, \apj, 912, 2,
  \dodoi{10.3847/1538-4357/abedb7}

\bibitem[{{Hu} {et~al.}(2024){Hu}, {Lazarian}, {Wu}, \& {Fu}}]{Huetal23II}
{Hu}, Y., {Lazarian}, A., {Wu}, Y., \& {Fu}, C. 2024, \mnras, 527, 11240,
  \dodoi{10.1093/mnras/stad3766}

\bibitem[{{Hu} {et~al.}(2020{\natexlab{a}}){Hu}, {Lazarian}, \&
  {Yuen}}]{Huetal2020}
{Hu}, Y., {Lazarian}, A., \& {Yuen}, K.~H. 2020{\natexlab{a}}, \apj, 897, 123,
  \dodoi{10.3847/1538-4357/ab9948}

\bibitem[{{Hu} {et~al.}(2019{\natexlab{a}}){Hu}, {Yuen}, \&
  {Lazarian}}]{IGvHRO}
{Hu}, Y., {Yuen}, K.~H., \& {Lazarian}, A. 2019{\natexlab{a}}, \apj, 886, 17,
  \dodoi{10.3847/1538-4357/ab4b5e}

\bibitem[{{Hu} {et~al.}(2019{\natexlab{b}}){Hu}, {Yuen}, \&
  {Lazarian}}]{HY19inten}
---. 2019{\natexlab{b}}, \apj, 886, 17, \dodoi{10.3847/1538-4357/ab4b5e}

\bibitem[{{Hu} {et~al.}(2020{\natexlab{b}}){Hu}, {Yuen}, \&
  {Lazarian}}]{2020ApJ...888...96H}
---. 2020{\natexlab{b}}, \apj, 888, 96, \dodoi{10.3847/1538-4357/ab60a5}

\bibitem[{{Hu} {et~al.}(2019{\natexlab{c}}){Hu}, {Yuen}, {Lazarian}, {Fissel},
  {Jones}, \& {Cunningham}}]{velac}
{Hu}, Y., {Yuen}, K.~H., {Lazarian}, A., {et~al.} 2019{\natexlab{c}}, \apj,
  884, 137, \dodoi{10.3847/1538-4357/ab41f2}

\bibitem[{{Hu} {et~al.}(2019{\natexlab{d}}){Hu}, {Yuen}, {Lazarian}, {Ho},
  {Benjamin}, {Hill}, {Lockman}, {Goldsmith}, \& {Lazarian}}]{survey}
{Hu}, Y., {Yuen}, K.~H., {Lazarian}, V., {et~al.} 2019{\natexlab{d}}, Nature
  Astronomy, 3, 776, \dodoi{10.1038/s41550-019-0769-0}

\bibitem[{{Iroshnikov}(1963)}]{1963AZh....40..742I}
{Iroshnikov}, P.~S. 1963, \azh, 40, 742

\bibitem[{{Johns-Krull}(2007)}]{2007IAUS..243...31J}
{Johns-Krull}, C.~M. 2007, in Star-Disk Interaction in Young Stars, ed.
  J.~{Bouvier} \& I.~{Appenzeller}, Vol. 243, 31--42,
  \dodoi{10.1017/S1743921307009398}

\bibitem[{{Kandel} {et~al.}(2016){Kandel}, {Lazarian}, \& {Pogosyan}}]{KLP16}
{Kandel}, D., {Lazarian}, A., \& {Pogosyan}, D. 2016, \mnras, 461, 1227,
  \dodoi{10.1093/mnras/stw1296}

\bibitem[{{Kandel} {et~al.}(2017{\natexlab{a}}){Kandel}, {Lazarian}, \&
  {Pogosyan}}]{KLP17a}
---. 2017{\natexlab{a}}, \mnras, 464, 3617, \dodoi{10.1093/mnras/stw2512}

\bibitem[{{Kandel} {et~al.}(2017{\natexlab{b}}){Kandel}, {Lazarian}, \&
  {Pogosyan}}]{KLP17b}
---. 2017{\natexlab{b}}, \mnras, 470, 3103, \dodoi{10.1093/mnras/stx1358}

\bibitem[{{Kandel} {et~al.}(2018){Kandel}, {Lazarian}, \& {Pogosyan}}]{KLP18}
---. 2018, \mnras, 478, 530, \dodoi{10.1093/mnras/sty1115}

\bibitem[{{Kolmogorov}(1941)}]{1941DoSSR..30..301K}
{Kolmogorov}, A. 1941, Akademiia Nauk SSSR Doklady, 30, 301

\bibitem[{{Kowal} \& {Lazarian}(2010)}]{2010ApJ...720..742K}
{Kowal}, G., \& {Lazarian}, A. 2010, \apj, 720, 742,
  \dodoi{10.1088/0004-637X/720/1/742}

\bibitem[{{Kowal} {et~al.}(2007){Kowal}, {Lazarian}, \&
  {Beresnyak}}]{2007ApJ...658..423K}
{Kowal}, G., {Lazarian}, A., \& {Beresnyak}, A. 2007, \apj, 658, 423,
  \dodoi{10.1086/511515}

\bibitem[{{Kraichnan}(1965)}]{1965PhFl....8.1385K}
{Kraichnan}, R.~H. 1965, Physics of Fluids, 8, 1385, \dodoi{10.1063/1.1761412}

\bibitem[{{Larson}(1981)}]{1981MNRAS.194..809L}
{Larson}, R.~B. 1981, \mnras, 194, 809, \dodoi{10.1093/mnras/194.4.809}

\bibitem[{{Lazarian}(2006)}]{Lazarian06}
{Lazarian}, A. 2006, \apjl, 645, L25, \dodoi{10.1086/505796}

\bibitem[{{Lazarian} \& {Esquivel}(2003)}]{2003ApJ...592L..37L}
{Lazarian}, A., \& {Esquivel}, A. 2003, \apjl, 592, L37, \dodoi{10.1086/377427}

\bibitem[{{Lazarian} {et~al.}(2020){Lazarian}, {Eyink}, {Jafari}, {Kowal},
  {Li}, {Xu}, \& {Vishniac}}]{2020PhPl...27a2305L}
{Lazarian}, A., {Eyink}, G.~L., {Jafari}, A., {et~al.} 2020, Physics of
  Plasmas, 27, 012305, \dodoi{10.1063/1.5110603}

\bibitem[{{Lazarian} {et~al.}(2023){Lazarian}, {HO}, {Yuen}, \&
  {Vishniac}}]{p_turb}
{Lazarian}, A., {HO}, K.~W., {Yuen}, K.~H., \& {Vishniac}, E. 2023, submitted
  to \apj, arXiv:2312.05399, \dodoi{10.48550/arXiv.2312.05399}

\bibitem[{{Lazarian} \& {Hoang}(2007)}]{LazHoang07}
{Lazarian}, A., \& {Hoang}, T. 2007, \mnras, 378, 910,
  \dodoi{10.1111/j.1365-2966.2007.11817.x}

\bibitem[{{Lazarian} \& {Pogosyan}(2000)}]{LP00}
{Lazarian}, A., \& {Pogosyan}, D. 2000, \apj, 537, 720, \dodoi{10.1086/309040}

\bibitem[{{Lazarian} \& {Pogosyan}(2004)}]{LP04}
---. 2004, \apj, 616, 943, \dodoi{10.1086/422462}

\bibitem[{{Lazarian} \& {Pogosyan}(2006)}]{LP06}
---. 2006, \apj, 652, 1348, \dodoi{10.1086/508012}

\bibitem[{{Lazarian} \& {Pogosyan}(2008)}]{LP08}
---. 2008, \apj, 686, 350, \dodoi{10.1086/591238}

\bibitem[{{Lazarian} \& {Pogosyan}(2012)}]{LP12}
---. 2012, \apj, 747, 5, \dodoi{10.1088/0004-637X/747/1/5}

\bibitem[{{Lazarian} \& {Pogosyan}(2016)}]{LP16}
---. 2016, \apj, 818, 178, \dodoi{10.3847/0004-637X/818/2/178}

\bibitem[{{Lazarian} {et~al.}(2002){Lazarian}, {Pogosyan}, \&
  {Esquivel}}]{2002ASPC..276..182L}
{Lazarian}, A., {Pogosyan}, D., \& {Esquivel}, A. 2002, in Astronomical Society
  of the Pacific Conference Series, Vol. 276, Seeing Through the Dust: The
  Detection of HI and the Exploration of the ISM in Galaxies, ed. A.~R.
  {Taylor}, T.~L. {Landecker}, \& A.~G. {Willis}, 182,
  \dodoi{10.48550/arXiv.astro-ph/0112368}

\bibitem[{{Lazarian} \& {Vishniac}(1999)}]{LV99}
{Lazarian}, A., \& {Vishniac}, E.~T. 1999, \apj, 517, 700,
  \dodoi{10.1086/307233}

\bibitem[{{Lazarian} \& {Yuen}(2018{\natexlab{a}})}]{LY18b}
{Lazarian}, A., \& {Yuen}, K.~H. 2018{\natexlab{a}}, \apj, 865, 59,
  \dodoi{10.3847/1538-4357/aad3ca}

\bibitem[{{Lazarian} \& {Yuen}(2018{\natexlab{b}})}]{LY18a}
---. 2018{\natexlab{b}}, \apj, 853, 96, \dodoi{10.3847/1538-4357/aaa241}

\bibitem[{{Lazarian} \& {Yuen}(2018{\natexlab{c}})}]{LY18polar}
---. 2018{\natexlab{c}}, \apj, 865, 59, \dodoi{10.3847/1538-4357/aad3ca}

\bibitem[{{Lazarian} {et~al.}(2018){Lazarian}, {Yuen}, {Ho}, {Chen},
  {Lazarian}, {Lu}, {Yang}, \& {Hu}}]{dispersion}
{Lazarian}, A., {Yuen}, K.~H., {Ho}, K.~W., {et~al.} 2018, \apj, 865, 46,
  \dodoi{10.3847/1538-4357/aad7ff}

\bibitem[{{Lazarian} {et~al.}(2017){Lazarian}, {Yuen}, {Lee}, \& {Cho}}]{LYLC}
{Lazarian}, A., {Yuen}, K.~H., {Lee}, H., \& {Cho}, J. 2017, \apj, 842, 30,
  \dodoi{10.3847/1538-4357/aa74c6}

\bibitem[{{Lazarian} {et~al.}(2022){Lazarian}, {Yuen}, \& {Pogosyan}}]{ch5}
{Lazarian}, A., {Yuen}, K.~H., \& {Pogosyan}, D. 2022, \apj, accepted for
  publication, arXiv: 2204.09731.
\newblock \doarXiv{2002.07996}

\bibitem[{{Lithwick} \& {Goldreich}(2001)}]{2001ApJ...562..279L}
{Lithwick}, Y., \& {Goldreich}, P. 2001, \apj, 562, 279, \dodoi{10.1086/323470}

\bibitem[{{Lu} {et~al.}(2020){Lu}, {Lazarian}, \&
  {Pogosyan}}]{2020MNRAS.496.2868L}
{Lu}, Z., {Lazarian}, A., \& {Pogosyan}, D. 2020, \mnras, 496, 2868,
  \dodoi{10.1093/mnras/staa1570}

\bibitem[{{Makwana} \& {Yan}(2020)}]{2020PhRvX..10c1021M}
{Makwana}, K.~D., \& {Yan}, H. 2020, Physical Review X, 10, 031021,
  \dodoi{10.1103/PhysRevX.10.031021}

\bibitem[{{Malik} {et~al.}(2023){Malik}, {Yuen}, \&
  {Yan}}]{2023MNRAS.524.6102M}
{Malik}, S., {Yuen}, K.~H., \& {Yan}, H. 2023, \mnras, 524, 6102,
  \dodoi{10.1093/mnras/stad2225}

\bibitem[{{Malik} {et~al.}(2024){Malik}, {Yuen}, \&
  {Yan}}]{2024ApJ...965...65M}
---. 2024, \apj, 965, 65, \dodoi{10.3847/1538-4357/ad34d7}

\bibitem[{{Mallet} {et~al.}(2015){Mallet}, {Schekochihin}, \&
  {Chandran}}]{2015MNRAS.449L..77M}
{Mallet}, A., {Schekochihin}, A.~A., \& {Chandran}, B.~D.~G. 2015, \mnras, 449,
  L77, \dodoi{10.1093/mnrasl/slv021}

\bibitem[{{Maron} \& {Goldreich}(2001)}]{MG01}
{Maron}, J., \& {Goldreich}, P. 2001, \apj, 554, 1175, \dodoi{10.1086/321413}

\bibitem[{{McKee} \& {Ostriker}(2007)}]{MO07}
{McKee}, C.~F., \& {Ostriker}, E.~C. 2007, \araa, 45, 565,
  \dodoi{10.1146/annurev.astro.45.051806.110602}

\bibitem[{{Mestel} \& {Spitzer}(1956)}]{1956MNRAS.116..503M}
{Mestel}, L., \& {Spitzer}, L., J. 1956, \mnras, 116, 503,
  \dodoi{10.1093/mnras/116.5.503}

\bibitem[{{Meyrand} {et~al.}(2018){Meyrand}, {Kiyani}, {G{\"u}rcan}, \&
  {Galtier}}]{2018PhRvX...8c1066M}
{Meyrand}, R., {Kiyani}, K.~H., {G{\"u}rcan}, {\"O}.~D., \& {Galtier}, S. 2018,
  Physical Review X, 8, 031066, \dodoi{10.1103/PhysRevX.8.031066}

\bibitem[{{Montgomery} \& {Turner}(1981)}]{1981PhFl...24..825M}
{Montgomery}, D., \& {Turner}, L. 1981, Physics of Fluids, 24, 825,
  \dodoi{10.1063/1.863455}

\bibitem[{{Mouschovias} {et~al.}(2006){Mouschovias}, {Tassis}, \&
  {Kunz}}]{2006ApJ...646.1043M}
{Mouschovias}, T.~C., {Tassis}, K., \& {Kunz}, M.~W. 2006, \apj, 646, 1043,
  \dodoi{10.1086/500125}

\bibitem[{Oughton {et~al.}(1997)Oughton, R\"adler, \&
  Matthaeus}]{PhysRevE.56.2875}
Oughton, S., R\"adler, K.-H., \& Matthaeus, W.~H. 1997, Phys. Rev. E, 56, 2875,
  \dodoi{10.1103/PhysRevE.56.2875}

\bibitem[{{Pavaskar} {et~al.}(2024){Pavaskar}, {Yuen}, {Yan}, \&
  {Malik}}]{2024arXiv240517985P}
{Pavaskar}, P., {Yuen}, K.~H., {Yan}, H., \& {Malik}, S. 2024, arXiv e-prints,
  arXiv:2405.17985.
\newblock \doarXiv{2405.17985}

\bibitem[{{Seta} {et~al.}(2018){Seta}, {Shukurov}, {Wood}, {Bushby}, \&
  {Snodin}}]{2018MNRAS.473.4544S}
{Seta}, A., {Shukurov}, A., {Wood}, T.~S., {Bushby}, P.~J., \& {Snodin}, A.~P.
  2018, \mnras, 473, 4544, \dodoi{10.1093/mnras/stx2606}

\bibitem[{{Skalidis} \& {Tassis}(2021)}]{2021A&A...647A.186S}
{Skalidis}, R., \& {Tassis}, K. 2021, \aap, 647, A186,
  \dodoi{10.1051/0004-6361/202039779}

\bibitem[{{Stone} {et~al.}(2020){Stone}, {Tomida}, {White}, \&
  {Felker}}]{2020ApJS..249....4S}
{Stone}, J.~M., {Tomida}, K., {White}, C.~J., \& {Felker}, K.~G. 2020, \apjs,
  249, 4, \dodoi{10.3847/1538-4365/ab929b}

\bibitem[{{Tofflemire} {et~al.}(2011){Tofflemire}, {Burkhart}, \&
  {Lazarian}}]{2011ApJ...736...60T}
{Tofflemire}, B.~M., {Burkhart}, B., \& {Lazarian}, A. 2011, \apj, 736, 60,
  \dodoi{10.1088/0004-637X/736/1/60}

\bibitem[{{Yan} \& {Lazarian}(2006)}]{YanL06}
{Yan}, H., \& {Lazarian}, A. 2006, \apj, 653, 1292, \dodoi{10.1086/508704}

\bibitem[{{Yan} \& {Lazarian}(2007)}]{YanL07}
---. 2007, \apj, 657, 618, \dodoi{10.1086/510847}

\bibitem[{{Yan} \& {Lazarian}(2008)}]{YanL08}
---. 2008, \apj, 673, 942, \dodoi{10.1086/524771}

\bibitem[{{Yuen} {et~al.}(2024){Yuen}, {Ho}, {Law}, \&
  {Chen}}]{filament_review}
{Yuen}, K.~H., {Ho}, K.~W., {Law}, C.~Y., \& {Chen}, A. 2024, Reviews of Modern
  Plasma Physics, {\it invited review}, arXiv:2404.19101,
  \dodoi{10.1007/s41614-024-00156-5}

\bibitem[{{Yuen} {et~al.}(2022){Yuen}, {Ho}, {Law}, {Chen}, \&
  {Lazarian}}]{spectrum}
{Yuen}, K.~H., {Ho}, K.~W., {Law}, C.~Y., {Chen}, A., \& {Lazarian}, A. 2022,
  \apjl, submitted, arXiv:2204.13760

\bibitem[{{Yuen} {et~al.}(2021){Yuen}, {Ho}, \& {Lazarian}}]{VDA}
{Yuen}, K.~H., {Ho}, K.~W., \& {Lazarian}, A. 2021, \apj, 910, 161,
  \dodoi{10.3847/1538-4357/abe4d4}

\bibitem[{{Yuen} \& {Lazarian}(2017{\natexlab{a}})}]{YL17a}
{Yuen}, K.~H., \& {Lazarian}, A. 2017{\natexlab{a}}, \apjl, 837, L24,
  \dodoi{10.3847/2041-8213/aa6255}

\bibitem[{{Yuen} \& {Lazarian}(2017{\natexlab{b}})}]{YL17b}
---. 2017{\natexlab{b}}, arXiv e-prints, arXiv:1703.03026.
\newblock \doarXiv{1703.03026}

\bibitem[{{Yuen} \& {Lazarian}(2020{\natexlab{a}})}]{YL20}
---. 2020{\natexlab{a}}, \apj, 898, 65, \dodoi{10.3847/1538-4357/ab9307}

\bibitem[{{Yuen} \& {Lazarian}(2020{\natexlab{b}})}]{curvature}
---. 2020{\natexlab{b}}, \apj, 898, 66, \dodoi{10.3847/1538-4357/ab9360}

\bibitem[{{Yuen} {et~al.}(2023){Yuen}, {Yan}, \& {Lazarian}}]{leakage}
{Yuen}, K.~H., {Yan}, H., \& {Lazarian}, A. 2023, \mnras, 521, 530,
  \dodoi{10.1093/mnras/stad287}

\bibitem[{{Zhang} {et~al.}(2020{\natexlab{a}}){Zhang}, {Chepurnov}, {Yan},
  {Makwana}, {Santos-Lima}, \& {Appleby}}]{2020NatAs...4.1001Z}
{Zhang}, H., {Chepurnov}, A., {Yan}, H., {et~al.} 2020{\natexlab{a}}, Nature
  Astronomy, 4, 1001, \dodoi{10.1038/s41550-020-1093-4}

\bibitem[{{Zhang} {et~al.}(2020{\natexlab{b}}){Zhang}, {Hu}, {Cho}, \&
  {Lazarian}}]{ZhangL20}
{Zhang}, J.-F., {Hu}, K., {Cho}, J., \& {Lazarian}, A. 2020{\natexlab{b}},
  \apj, 895, 20, \dodoi{10.3847/1538-4357/ab88ac}

\bibitem[{{Zhang} {et~al.}(2019){Zhang}, {Lazarian}, {Ho}, {Yuen}, {Yang}, \&
  {Hu}}]{ZhangL19}
{Zhang}, J.-F., {Lazarian}, A., {Ho}, K.~W., {et~al.} 2019, \mnras, 486, 4813,
  \dodoi{10.1093/mnras/stz1176}

\bibitem[{{Zhao} {et~al.}(2024{\natexlab{a}}){Zhao}, {Yan}, {Liu}, {Yuen}, \&
  {Shi}}]{2024ApJ...962...89Z}
{Zhao}, S., {Yan}, H., {Liu}, T.~Z., {Yuen}, K.~H., \& {Shi}, M.
  2024{\natexlab{a}}, \apj, 962, 89, \dodoi{10.3847/1538-4357/ad132e}

\bibitem[{{Zhao} {et~al.}(2024{\natexlab{b}}){Zhao}, {Yan}, {Liu}, {Yuen}, \&
  {Wang}}]{2024NatAs.tmp...71Z}
{Zhao}, S., {Yan}, H., {Liu}, T.~Z., {Yuen}, K.~H., \& {Wang}, H.
  2024{\natexlab{b}}, Nature Astronomy, \dodoi{10.1038/s41550-024-02249-0}

\end{thebibliography}
%\bibliography{ch9_submit.bbl}
\label{lastpage}

% No idea why it goes wrong.
\end{document}